\documentclass[ DIV=10,
                onecolumn
              ]{scrartcl}

\usepackage{hyperref}
\usepackage{color} 
\usepackage{psfrag}
\usepackage{subfig}
\usepackage{amsmath}
\usepackage{graphics}
\usepackage{pstool} 
\usepackage{todonotes}

\usepackage{pgfplots}
\usepackage{pgfplotstable}
\usepackage{filecontents}

\usepackage{tikz}
\usepackage{tikzscale}
\usepackage{tikz-uml}
\usetikzlibrary{plotmarks}
\definecolor{darkblue}{rgb}{0.0,0.0,0.5}
\definecolor{darkred}{rgb}{0.5,0.0,0.0}
\definecolor{darkgreen}{rgb}{0.0,0.5,0.0}
\definecolor{lightgreen}{rgb}{0.0,0.8,0.0}

\usepackage{booktabs}   
\usepackage{makecell} 
\usepackage{colortbl}   

\usepackage{xspace}
\usepackage{color}     
\usepackage{setspace}   
 
\usepackage{soul}
\usepackage{ulem}
\usepackage{currfile}
\usepackage{import}

\usepackage{authoraftertitle} 
\usepackage{authblk} 

\usepackage[square,sort,comma,numbers]{natbib}
\usepackage{algorithm}
\usepackage[]{algpseudocode}
\usepackage{graphicx}
\usepackage{cleveref}
\usepackage{txfonts}
\usepackage{amsmath,amssymb}
\usepackage{float}
\usepackage{xfrac}

\usepackage{enumitem}

\usepackage{amssymb}
\usepackage{pifont}
\usepackage{bm}
\usepackage{url}

\usepackage[utf8]{inputenc}

\pgfplotsset{compat=1.8}

\newcommand{\findmax}[3]{
    \pgfplotstablesort[sort key={#2},sort cmp={float >}]{\sorted}{#1}%
    \pgfplotstablegetelem{0}{#2}\of{\sorted}%
    \let #3=\pgfplotsretval%
}

\pgfplotscreateplotcyclelist{myCycleListBlackAndWhite}
{%
  {black, mark=o},
  {black, mark=square},
  {black, mark=triangle},
  {black, mark=diamond}, 
  {black, mark=pentagon}, 
  {black, mark=*},
  {black, mark=square*},
  {black, mark=triangle*},
  {black, mark=diamond*}, 
  {black, mark=pentagon*}, 
}

\definecolor{darkgreen}{rgb}{0,0.4,0} 
\definecolor{darkbrown}{rgb}{0.5, 0.396, 0.09}

\definecolor{c1}{rgb}{0.0, 0.4196078431372549, 0.6431372549019608}
\definecolor{c2}{rgb}{1.0, 0.5019607843137255, 0.054901960784313725}
\definecolor{c3}{rgb}{0.6705882352941176, 0.6705882352941176,
0.6705882352941176} \definecolor{c}{rgb}{0.34901960784313724, 0.34901960784313724, 0.34901960784313724}
\definecolor{c4}{rgb}{0.37254901960784315, 0.6196078431372549,
0.8196078431372549} \definecolor{c}{rgb}{0.7843137254901961, 0.3215686274509804, 0.0}
\definecolor{c5}{rgb}{0.5372549019607843, 0.5372549019607843,
0.5372549019607843} \definecolor{c}{rgb}{0.6352941176470588, 0.7843137254901961, 0.9254901960784314}
\definecolor{c6}{rgb}{1.0, 0.7372549019607844, 0.4745098039215686}
\definecolor{c7}{rgb}{0.8117647058823529, 0.8117647058823529,
0.8117647058823529}

\pgfplotscreateplotcyclelist{myCycleListColor}
{%
  {blue, mark=o},
  {red, mark=square},
  {darkbrown, mark=triangle},
  {darkgreen, mark=diamond},
  {black, mark=pentagon},
  {blue, mark=*},
  {red, mark=square*},
  {darkbrown, mark=triangle*},
  {darkgreen, mark=diamond*},
  {black, mark=pentagon*},
}

\pgfplotsset{every axis/.append style= 
              {
                font=\small,
                mark size=2,
                line width = 0.1,
                legend style={font=\small, mark size=3, draw=none, fill=none},
                legend cell align=left,
                cycle list name=myCycleListColor,
              }
            }
            
\pgfplotstableset
{
  every odd row/.style={before row={\rowcolor[gray]{0.9}}},
  every head row/.style=
  {
    before row=
    {%
      \toprule
    },
    after row=\midrule
  },
  every last row/.style=
  {
    after row=\bottomrule
  }
}

\newif\ifdrawboundingbox

\usetikzlibrary{external} 
\tikzset{external/system call={pdflatex \tikzexternalcheckshellescape
-halt-on-error -interaction=batchmode -jobname "\image" "\texsource"}} 

\pgfkeys
{ 
  /pgf/number format/.cd,
  fixed,
  set thousands separator={\,},
  1000 sep in fractionals,
}

\newcolumntype{C}[1]{>{\centering\arraybackslash}m{#1}}
\newcolumntype{R}[1]{>{\raggedright\arraybackslash}m{#1}}
\newcolumntype{L}[1]{>{\raggedleft\arraybackslash}m{#1}}

\newcommand{\delete}[1]{\xspace}

\title{\vspace{-2em}Integrating CAD and numerical analysis: ‘Dirty geometry’ handling using the Finite Cell Method}

\author[1]{Benjamin Wassermann\thanks{benjamin.wassermann@tum.de, Corresponding Author}}
\author[1]{Stefan Kollmannsberger\thanks{stefan.kollmannsberger@tum.de}}
\author[1,2]{Shuohui Yin\thanks{yinsh@xtu.edu.cn}}
\author[1]{L\'{a}szl\'{o} Kudela\thanks{laszlo.kudela@tum.de}}
\author[1,3]{Ernst Rank\thanks{ernst.rank@tum.de}}

\affil[1]{Chair for Computation in Engineering,
Technical University of Munich,
Arcisstr. 21, 80333 M\"unchen, Germany}

\affil[2]{School of Mechanical Engineering, Xiangtan University, Hunan 411105, PR China}

\affil[3]{Institute for Advanced Study, 
Technical University of Munich, 
Lichtenbergstr. 2a, 85748 Garching, Germany}

\begin{document}
\normalem
\maketitle

\hrule
\subsubsection*{Abstract}

This paper proposes a computational methodology for the integration of Computer Aided Design (CAD) and the Finite Cell Method (FCM) for models with ``dirty geometries''. FCM, being a fictitious domain approach based on higher order finite elements, embeds the physical model into a fictitious domain, which can be discretized without having to take into account the boundary of the physical domain. The true geometry is captured by a precise numerical integration of elements cut by the boundary. Thus, an effective Point Membership Classification algorithm that determines the inside--outside state of an integration point with respect to the physical domain is a core operation in FCM. To treat also ``dirty geometries'', i.e. imprecise or flawed  geometric models, a combination of a segment-triangle intersection algorithm and a flood fill algorithm being insensitive to most CAD model flaws is proposed to identify the affiliation of the integration points. The present  method thus allows direct computations on geometrically and topologically flawed models. The potential and merit for practical applications of the proposed method is demonstrated by several numerical examples.
\vspace*{0.2cm}

\noindent \textit{Keywords:} Computer-Aided Design, Dirty geometry, Finite Cell Method, 
Flood Fill, Point Membership Classification, Flawed geometry
\vspace*{0.2cm}
\hrule
\vspace*{0.2cm}
\noindent\textcopyright 2019. This manuscript version is made available under the CC-BY-NC-ND 4.0 license.\\
Published in Computer Methods in Applied Mechanics and Engineering\\
\href{https://www.sciencedirect.com/science/article/pii/S0045782519302208?via\%3Dihub}{https://www.sciencedirect.com/science/article/pii/S0045782519302208?via\%3Dihub}\\
\href{https://doi.org/10.1016/j.cma.2019.04.017}{DOI: 10.1016/j.cma.2019.04.017}

\newpage
\tableofcontents
\newpage

\section{Introduction}
Product development in the scope of Computer Aided Engineering (CAE) typically involves Computer Aided Design (CAD) and numerical analyses. The life cycle of almost every complex mechanical product starts with the creation of a CAD model which is then converted into a suitable format for downstream CAE applications such as Finite Element Analysis, Rapid Prototyping, or automated manufacturing. However, a truly smooth transition from a geometric to a computational model is still challenging. This is especially the case for numerical simulations like the Finite Element Method. Very often, complex and time-consuming model preparation and pre-processing steps are necessary to obtain a decent numerical model that is suitable for analysis purposes. For complex CAD models, this transition process can take up to 80\% of the overall analysis time~\citep{Cottrell2009}.

For this reason, various alternative numerical approaches have been developed which seek to avoid or shorten this costly transition process (e.g., meshing). 
Isogeometric analysis (IGA) as the most prominent example aims at easing the transition from CAD to computational analysis by using the same spline basis functions for geometric modeling and numerical simulation~\citep{Hughes2005,Cottrell2009}. In a related earlier approach Cirak and Scott~\citep{Cirak2002} presented an integrated design process based on Subdivision Surfaces. Kagan and Fischer~\citep{Kagan2000} used B-spline finite elements in an effort to join design and analysis.

\begin{figure}[ht]
	\centering
	\includegraphics[width=0.8\textwidth]{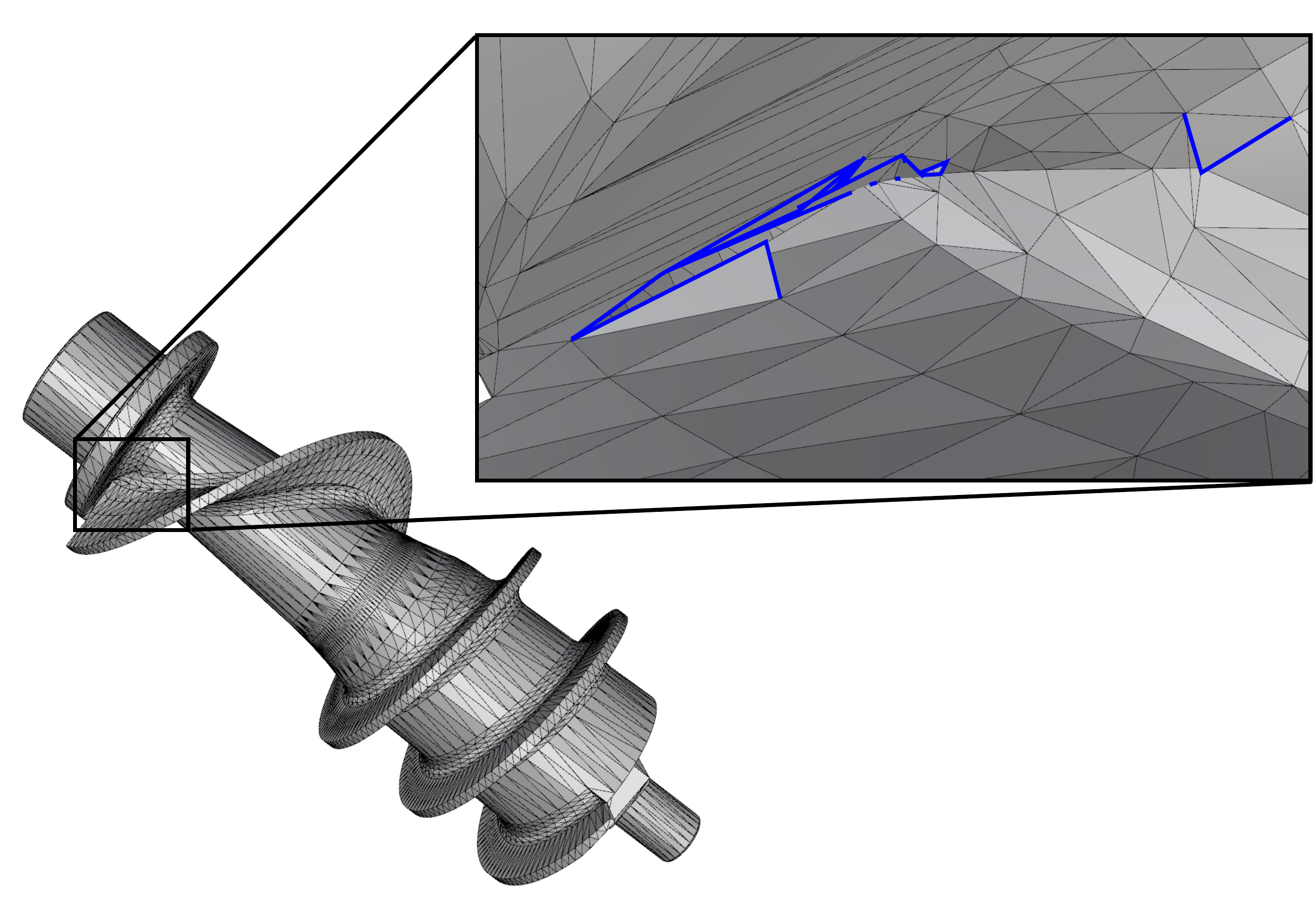}
	\caption{Cad model of a screw with flaws. Free edges are highlighted in blue. }
	\label{fig:flaws}	
\end{figure}

However, independent of the respective numerical approach, flaws may appear in the CAD model during the design--analysis cycle -- such as double entities, gaps, overlaps, intersections, and slivers -- as shown in Fig.~\ref{fig:flaws}. They are mainly due to data loss while the model is exchanged between different CAD and/or CAE systems, to inappropriate operations by the designer, or to approximation steps resulting in incompatible geometries. These model flaws, also called \emph{dirty topologies} or \emph{dirty geometries}, may be extremely small or even unapparent. While they are of no particular importance to a CAD engineer they may, however, cause serious problems for structural analyses. In the best case, they merely generate excessively fine meshes in some regions which are not relevant to structural analysis (e.g., at fillets, etc.) but drive up computational time unnecessarily. In the worst case, computations fail completely because no finite element mesh can be created. This is due to the fact that neither classical finite element approaches nor the newly developed methods mentioned above are designed to handle dirty topologies and geometries. 
Thus, extra effort is necessary to repair, heal, or reconstruct the model into an analysis-suitable geometry~\cite{Yang2006}, even if the affected region is not of special interest to the structural analyst. This is also a major obstacle for IGA, which heavily relies on flawless geometries.

Solid CAD modeling systems mainly rely on two different representation techniques: Boundary Representation (B-Rep) and Constructive Solid Geometry (CSG)\citep{Mantyla1988} which is often extended to a so-called procedural modeling. In CSG, a volume is described by volumetric primitives, whereas in B-Rep it is described via its surfaces. Consequently, B-Rep models provide direct and easy access to the explicit boundaries. However, B-Rep models are not necessarily valid, meaning that it might in some cases not be possible to determine whether a point lies inside or outside (Point Membership Classification). In contrast, CSG models are inherently watertight. Problems such as non-manifolds, dangling faces, or lines, or Boolean operations on disjoint objects need to be handled accordingly by the respective CAD system. A novel representation technique -- V-Rep (volumetric representation) -- was recently proposed by Elber et al.\citep{Massarwi2016} and implemented into the IRIT solid modeler\footnote{http://www.cs.technion.ac.il/~gershon/GuIrit/}. V-Reps are constructed of volumetric, non-singular B-Spline primitives, thus, providing both an explicit volume and explicit surface description. As the V-Rep models follow the CSG idea of combining valid primitives, this approach can help to overcome several pitfalls in solid modeling. Within this paper, we focus on flawed or 'dirty' B-Rep models. 
The most direct way to address CAD model flaws is to heal or repair the model before meshing. The healing process involves identifying the type of model errors and fixing them individually. Butlin and Stops~\cite{Butlin1996} listed topological and geometrical inconsistencies. The geometrical inconsistencies relate to their positions in space, while the topological inconsistencies relate to the connections or relationships among entities. 
Gu et al.~\cite{Gu2001} presented a visual catalog of potential flaws. 
Petersson and Chand~\cite{Petersson2001} developed a suite of tools for the preparation of CAD geometries that are imported from IGES files and stored in the boundary representation for mesh generation; the algorithm can identify gross flaws and remove them automatically. 
Yang et al.~\cite{Yang2005a} classified topological and geometrical flaws in CAD models and proposed a procedural method to verify 19 flaw types in STEP format and 12 types in the IGES format. 
 Yang and Han~\cite{Yang2006} conducted a case study to investigate the typical nature of CAD model flaws. They reported the classification and frequency of each of the six most common error types that significantly increase the lead times, and they proposed a repair method based on the design history. Healing methods act either on the CAD model or on the mesh~\cite{Yang2006}. 
According to their approach, these methods can be classified into surface~\cite
{Chong2007}, volumetric~\cite{Nooruddin2003} and hybrid~\cite{Bischoff2005a} types. Surface-based geometry repair methods perform local modifications merging and fixing incorrect surface patches. Volumetric techniques are used to reconstruct a new global shape without flaws. However, this approach typically leads to information loss, especially at sharp features such as kinks. Hybrid methods combine the advantages of local surface healing and global volumetric healing. To this end, flaws are detected and a volumetric reconstruction is performed only in their vicinity. These methods have been used for CAD models that are represented in typical B-Rep formats (e.g., STEP and IGES) as well as for polygonal meshes~\cite{Busaryev2009}.

Although healing and repair methods have been applied successfully in recent years, healing can still be very labor intensive and time consuming in the scope of product development. As a remedy, mesh generation techniques have been developed which have the potential to generate meshes from flawed geometric models. In this line of research, Wang and Srinivasan~\cite{Wang2002} proposed an adaptive Cartesian mesh generation method. Herein, the computational grid is created inside the domain, which then connects to the boundary. Another technique -- the Cartesian shrink-wrapping technique -- was presented in~\cite{Lee2010} to generate triangular surface meshes automatically for 3D flawed geometries without healing. However, to generate a mesh, an initial watertight shell (called wrapper surface) needs to be constructed.
Another line of research proposed by Gasparini et al.~\cite{Gasparini2013} is an approach to analyze geometrically imperfect models based on a geometrically adaptive integration technique that uses different model representations, i.e. space decomposition, B-Rep, and distance fields. This approach relies on a method that was first introduced by Kantorovich~\cite{Kantorovich1958} and that has recently been commercialized~\cite{IntactSolutions2013}. Furthermore, this approach requires computation of a well-defined distance function to the boundaries -- which is non-trivial for dirty geometries, as the orientation of boundary surfaces might be incorrect or the location of the boundaries is anticipated incorrectly, e.g., due to spurious entities, or intersections.
However, two main issues arise applying geometry healing: (i) In the case that the geometry is healed locally, i.e. each flaw on its own, it is almost impossible to heal all flaws. Hence, a subsequent volumetric meshing is likely to fail. (ii) If the model is healed in a volumetric sense, i.e. the model is entirely reconstructed, a valid model can be obtained. However, typically sharp features, such as edges, corners or small details are lost. The automatic assumptions which are made during the volumetric healing lead to a changed model which is likely to be not in the designer's intent.

In this work, we present an alternative computational methodology which aims at dealing \emph{robustly} with dirty topologies and geometries. At its core, it utilizes the Finite Cell Method (FCM)~\cite{Parvizian2007, Duster2008}, a fictitious domain method which uses classical linear, or higher-order finite elements. The FCM embeds the physical model into a fictitious domain which is then discretized by a simple, often axis-aligned grid. This grid does not have to conform to the boundary of the physical domain. Instead, the physical domain is recovered on the level of integration of element matrices and load vectors. A Point Membership Classification (PMC) test is carried out at each integration point to determine whether it lies inside or outside the physical domain. Hence, the only information needed from the CAD model is a reliable and robust PMC, which strongly reduces the geometrical and topological requirements on the validity of the geometric model. This observation allows for a new paradigm in the computational analysis: not to create an analysis-suitable model and/or to derive a mesh or distance field, but rather \textit{to directly compute on geometrically and/or topologically flawed models by a flaw-insensitive computational method}. Thus, it is neither required to heal the flawed geometry nor to construct conforming meshes or distance fields. Instead, a PMC is constructed which is robust w.r.t. to a large number of model flaws. The Point Membership Classification test can then be evaluated with a certainty at least up to a geometric magnitude of the defect itself (as, e.g., in the case of gaps). This is important because a subsequent computational analysis can then directly be carried out without healing. Moreover, the computational analysis may still deliver the necessary accuracy on those flawed models as their effect on the results of the computation remains local to the flaw itself. Only, if the local flaw lies directly in the region of interest it must be fixed. This is, however, only necessary to achieve higher accuracy -- an analysis can be carried out either way. 

The Finite Cell Method is a widely applicable method itself. While the original publications concerning the FCM treated linear elasticity in 2D and 3D~\cite{Duster2008}, the scope of application was extended to various fields, such as elastoplasticity~\cite{Abedian2014}, constructive solid geometric models~\cite{Wassermann2017}, topology optimization~\cite{Cai2014,Groen2016}, local enrichment for material interfaces~\cite{Joulaian2013}, elastodynamics and wave propagation~\cite{Joulaian2014, Duczek2014,Elhaddad2015}, and contact problems~\cite{Bog2017, Mongeau2015}. Further developments include weakly enforced essential boundary conditions~\cite{Kollmannsberger2015}, local refinement schemes~\cite{Zander2015}, and efficient integration techniques~\cite{Kudela2016, Fries2015, Joulaian2016, Hubrich2017}. Furthermore, the concept of the FCM is independent of the underlying approximation method.
It does not have to be based on hierarchical Legendre shape functions but can also be built on a spline-based approximation like in Isogeometric Analysis, or spectral shape functions~\cite{Giraldo2017}. In this case, the fictitious domain approach is an adequate method for trimming Isogeometric Analysis, as presented and analyzed, e.g., in~\citep{Schillinger2012b, Rank2012, Ruess2014}. In \cite{dePrenter2017a}, an efficient method to overcome the inherent problem of bad condition numbers based on precondition is presented. Approaches very similar to the FCM have been presented more recently, like the cutFEM method \cite{Burman2015a}, which builds on earlier publications of Hansbo et al. \cite{Burman2010}. Therein, small elements are explicitly stabilized by controlling the gradients across embedded boundaries connected neighboring cells in the fictitious domain. This is different to FCM where a stabilization is achieved to a certain extent by a small but non-zero stiffness in the fictitious domain.

In this contribution, the FCM is extended in order to directly simulate a CAD model with flaws. The paper is structured as follows: Section \ref{sec:DirtyGeometry} provides a brief overview over geometrical and topological flaws. The basic formulation of the FCM and the requirements of a numerical simulation on flawed geometric CAD models are given in Section \ref{sec:FCM}. A robust algorithm for Point Membership Classification on dirty geometries is presented in Section \ref{sec:Method}. Several numerical examples for the proposed methodology are presented and discussed in Section \ref{sec:Examples}. Finally, conclusions are drawn in Section \ref{sec:Conclusion}.

\section{Dirty Topology/Geometry} \label{sec:DirtyGeometry}
In this section, we provide a very short general overview of Boundary Representation (B-Rep) models (sec.~\ref{sec:bRep}) and necessary conditions for their validity (sec.~\ref{sec:conditions}). By implication, 'dirty' geometries, or topologies are models which do not meet these requirements and are therefore mathematically invalid. To describe the wide variety of different flaws (sec:~\ref{sec:ModelFlaws}), we define mathematical operators (sec.~\ref{sec:operators}) and apply them to a valid B-Rep model, thereby transforming a 'valid' into a 'dirty' B-Rep model (sec.~\ref{sec:ApplicationOperators}). \\
Several of these flaw operators allow introducing a control parameter $\varepsilon$, indicating a geometric size of the respective flaws. Applying a sequence of flaw operators maps a flawless model to exactly one resulting flawed model. It is obvious that, given some flawed model, it is not possible to determine on which flawless model it could be based meaning that a class of equivalent flawless models can be associated to one 'dirty' model. Our conceptual approach therefore only \textit{assumes} the existence of a flawless model that is expected to be 'close' to the 'dirty' one. This is used as the geometric basis for analysis. Further, it is to be noted that no explicit knowledge of this flawless model is required.    

\subsection{Boundary Representation Models} \label{sec:bRep}
B-Rep objects are described by their boundaries. A model $\Omega$ can consist of several sub-domains, which all describe a separate closed volumetric body $B_i$. 
\begin{equation}
\Omega = \lbrace\, B_i \; \mid \; i \in \{1,...,n\} \,\rbrace
\end{equation}
with $n$ being the number of volumetric bodies. For simplicity of presentation, we assume that a B-Rep model consists only of one domain $\Omega = B$. A B-Rep body consists of topology $T$ and geometry $G$~\citep{Mantyla1988}:
\begin{equation}
B\,(\,T, G \,)
\end{equation}
The topology $T$ describes the relations or logical location of all entities (\ref{sec:topology}), whereas the geometry $G$ provides the physical location of points, consequently defining the actual shape of the model  (\ref{sec:geometry}).

\subsubsection{Topology}\label{sec:topology}
The topology $T(\,t, r^{int}, r^{ext}\,)$ provides the logical internal $r^{int} = \{r^{int}_i\}$ and external relations $r^{ext} = \{r^{ext}_i\}$ between the topological entities $t = \{t_i\}$, i.e. vertices $v_i$, edges $e_j$, and faces $f_k$. Thereby, each topological entity $t_i$ has its own local, internal relation $r_i^{int}$, defining how and from which underlying topological entities it is constructed.
Topological entities are typically represented by sets:
\begin{align}
	& V = \lbrace\, v_i \; \mid \; i \in \{1,...,n\} \,\rbrace \\
	& E = \lbrace\, e_i \; \mid \; i \in \{1,...,m\} \, ,\, e_i = (v_{\alpha}, v_{\beta}) \, ,\,v_{\alpha},v_{\beta} \in V \,\rbrace \label{eq:edgeSet}\\
	& F = \lbrace\, f_i \; \mid \; i \in \{1,...,o\} \, ,\,f_i = \left(\,(e_{\kappa})_{\kappa \in \{\alpha,...,\psi \}}\, , \bm{n}_i \,\right) \, ,\,e_{\kappa} \in E \,\rbrace \label{eq:faceSet}
\end{align}

\noindent with $n,m,o$ being the number of vertices, edges, and faces, respectively. The ordered pair of vertices $(v_{\alpha}, v_{\beta})$ contains the bounding vertices of an edge. A face $f_i$ is described by an ordered pair containing: (i) the boundary edges, denoted by an ordered $n$-tuple $(e_{\kappa})$, with $n$ being the number of boundary edges, and (ii) the respective normal vector $\bm{n}_i$. In some cases, the normal vector is provided implicitly by the order of the boundary edges $(e_{\kappa})$. 
The external relations $r^{ext}$ describe the global adjacency relations between the particular entities (e.g., which faces are neighbors to each other). There are various possible methods to represent the internal and external adjacency relations,  or a combination of both, such as the \textit{winged edge model} or the \textit{double connected edge list} \citep{Mantyla1988}. Thereby, the adjacency relations can be represented by graphs. \Cref{fig:AdjacenyMatrix} shows an exemplary detail of a topology consisting of three triangles. The pure external relations can, for example, be represented by the adjacency matrix  $r^{ext}_{FF}$ (see equation~(\ref{eq:adjacency_1})). The adjacency matrices for faces and edges $r_{FE}$ (see equation~(\ref{eq:adjacency_2})) and for edges and vertices $r_{VE}$ (see equation~(\ref{eq:adjacency_3})) represent a combination of internal and external relations \footnote{\textit{Please note:} An entry $1$ in the adjacency matrix shows which entity (row) is connected to which other entity (column). An entry $0$ indicates that no direct adjacency exists.}.

\begin{minipage}{.42\textwidth}
	\begin{figure}[H]
  		\centering
  		\includegraphics[width=\textwidth]{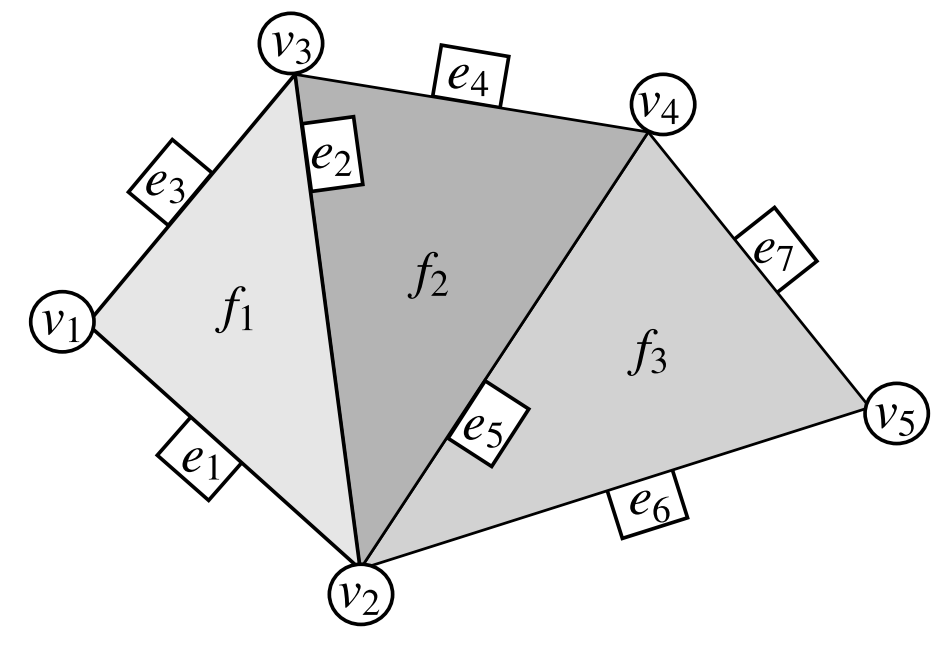}
  		\caption{Example topology with $n=5$ vertices, $m=7$ edges, and $o = 3$ faces.}
		\label{fig:AdjacenyMatrix}
	\end{figure}
\end{minipage}
\hfill
\begin{minipage}{.53\textwidth}
	\begin{equation}
		r^{ext}_{FF} \subset F \times F = \begin{bmatrix} 
			0 & 1 & 0 \\ 
			1 & 0 & 1 \\ 
			0 & 1 & 0 
		\end{bmatrix}
	\label{eq:adjacency_1}
	\end{equation}
	\begin{equation}
		r_{FE} \subset F \times E = \begin{bmatrix} 
			1 & 1 & 1 & 0 & 0 & 0 & 0\\ 
			0 & 1 & 0 & 1 & 1 & 0 & 0\\ 
			0 & 0 & 0 & 0 & 1 & 1 & 1
		\end{bmatrix}
	\label{eq:adjacency_2}
	\end{equation}
	\begin{equation}
		r_{VE} \subset V \times E = \begin{bmatrix} 
			1 & 0 & 1 & 0 & 0 & 0 & 0 \\ 
			1 & 1 & 0 & 0 & 1 & 1 & 0 \\
			0 & 1 & 1 & 1 & 0 & 0 & 0 \\
			0 & 0 & 0 & 1 & 1 & 0 & 1 \\
			0 & 0 & 0 & 0 & 0 & 1 & 1
		\end{bmatrix}
	\label{eq:adjacency_3}
	\end{equation}	
\end{minipage}

\subsubsection{Geometry}\label{sec:geometry}
\begin{figure}[H]
  	\centering
  	\includegraphics[width=0.7\textwidth]{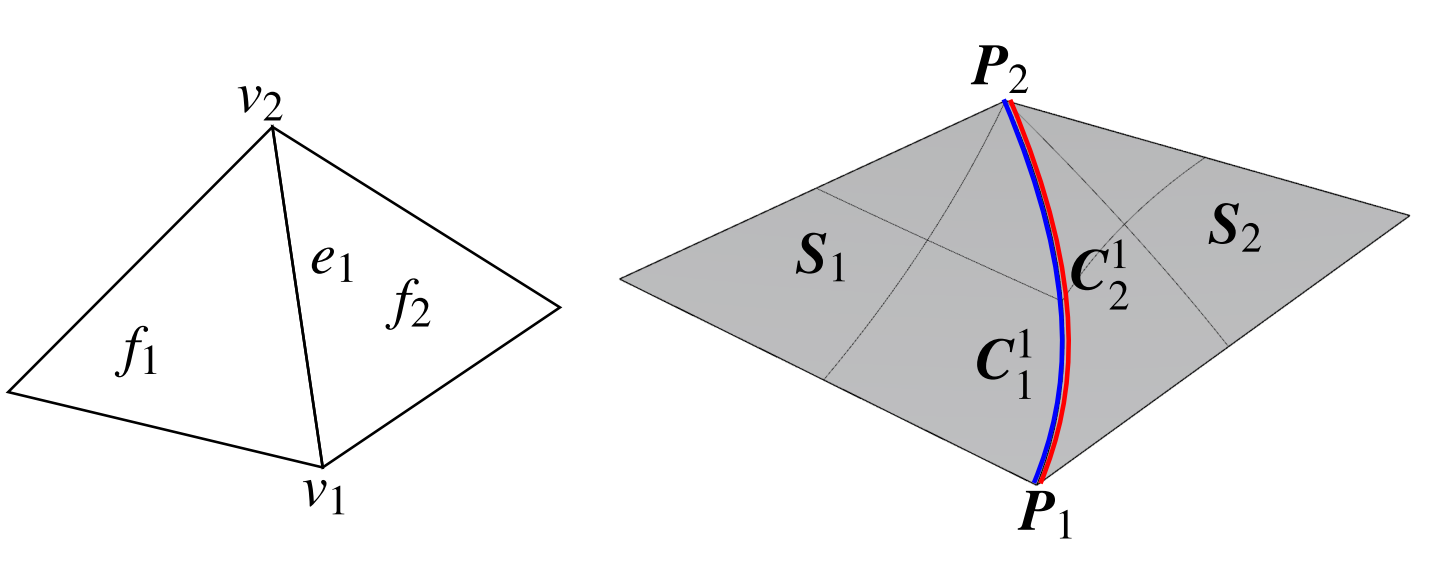}
  	\caption{At each edge $e_i$ two boundary curves $\bm{C}_{S_j}^{e_i}$ and $\bm{C}_{S_k}^{e_i}$ meet.}
	\label{fig:GeometryTopology}
\end{figure}

The geometry $G(\{g_i\})$ contains the geometric entities $g_i$, i.e. the points $\bm{P}_i$, curves $\bm{C}_j(\xi)$, and surfaces $\bm{S}_k(\xi,\eta)$, which describe the actual physical location of the boundary and, thus, the shape of the geometry. Curves and surfaces are often expressed in parametric representation:
\begin{align}
	& \bm{P}_i =  \left( \, x_i, \, y_i, \, z_i \, \right)^\mathrm{T}\\
	& \bm{C}_i(\xi) = \begin{pmatrix} x(\,\xi\,) \\ y(\,\xi\,) \\ z(\,\xi\,) \end{pmatrix} \quad e.g.,  \quad \bm{C}_i(\xi) = \sum_{j}^{n_{Q_j}} N_{j}(\,\xi\,) \cdot \bm{Q}_j  \\
	& \bm{S}_i(\bm{\xi}) = \begin{pmatrix} x(\,\bm{\xi}\,) \\ y(\,\bm{\xi}\,) \\ z(\,\bm{\xi}\,) \end{pmatrix} \quad e.g., \quad \bm{S}_i(\bm{\xi}) = \sum_j^{n_{Q_j}}\sum_k^{n_{Q_k}} N_j(\,\xi\,) \cdot N_k(\,\eta\,) \cdot \bm{Q}_{j,k}
\end{align}
with $\xi \in \mathbb{R}$ and $\bm{\xi} = (\xi,\eta) \in \mathbb{R}^2$. $N_i(\xi)$ denote shape functions (such as Lagrange or Legendre polynomials, B-Splines, NURBS, etc.) and $\bm{Q}_i$ the associated (control-)points, which can, depending on the curve description, coincide with the geometrical points $\bm{P}_i$. \\
 Analogous to the topology, the geometry $G$ can be represented by sets:
\begin{align}
	& P = \lbrace\, \bm{P}_i \;\mid\; i =  \{1,...,n\}  \,\rbrace \\
	& C = \lbrace\, \bm{C}_i \;\mid\; i =  \{1,...,2\cdot m\} \,\rbrace  \\
	& S = \lbrace\, \bm{S}_i \;\mid\; i =  \{1,...,o\} \,\rbrace 
\end{align}

\noindent where the number of points and surfaces equals the number of vertices $n$ and faces $o$, respectively. A special case are curves, where at each edge two adjoined faces meet, whose underlying surfaces have each their own boundary curves. Consequently, the number of curves is $2\cdot m$ (see Fig.~\ref{fig:GeometryTopology}).

\subsubsection{Minimal B-Rep and the STL format}\label{sec:stl}
The most commonly used B-Rep exchange format between CAD and analysis is STL (STereoLithography, or more expressive Standard Tessellation Language). STL can be interpreted as a minimal B-Rep format, as it provides only the least amount of necessary information. Additionally, no explicit separation between topology and geometry is made. STL consists of independent triangles, which are defined by their three corner points. As geometric information in form of point coordinates is provided explicitly only for vertices, curves and surfaces are linearly interpolated. No adjacency, or 'consistency' information is provided, which makes STL quite flexible -- but also particularly prone to a variety of potential flaws. The relation between faces and vertices reads:
\begin{equation}
	 F^{STL} = \lbrace\, f_i \; \mid \; i \in \{1,...,o\} \, ,\,f_i = \left((v_{\alpha},\,v_{\beta},\,v_{\gamma}),\,\bm{t}_i \right) \, ,\,v_{\kappa} \in V \,,\,\lvert V \rvert = 3\cdot o \,\rbrace 
\end{equation}
Note that -- due to the multiple definition of vertices -- STL models are, strictly speaking, topologically not valid. Furthermore, the redundancy of point definitions and normal vectors, which could be derived from the orientation of the face has an eminent impact on the required memory for storage.

\subsection{Conditions for valid B-Rep models}\label{sec:conditions}
Although intuitively quite apparent, it is not straightforward to define a valid B-Rep model. Patrikalakis et al. \citep{Patrikalakis2000} provided a definition: 
"A B-Rep model is valid if its faces form an orientable 2-manifold without boundary."
From this, several requirements can be derived, some of which are also mentioned by M\"antyl\"a \citep{Mantyla1988} and Hoffmann \citep{Hoffmann1989}. 

\begin{figure}[H]
  \centering
  	\includegraphics[width=0.7\textwidth]{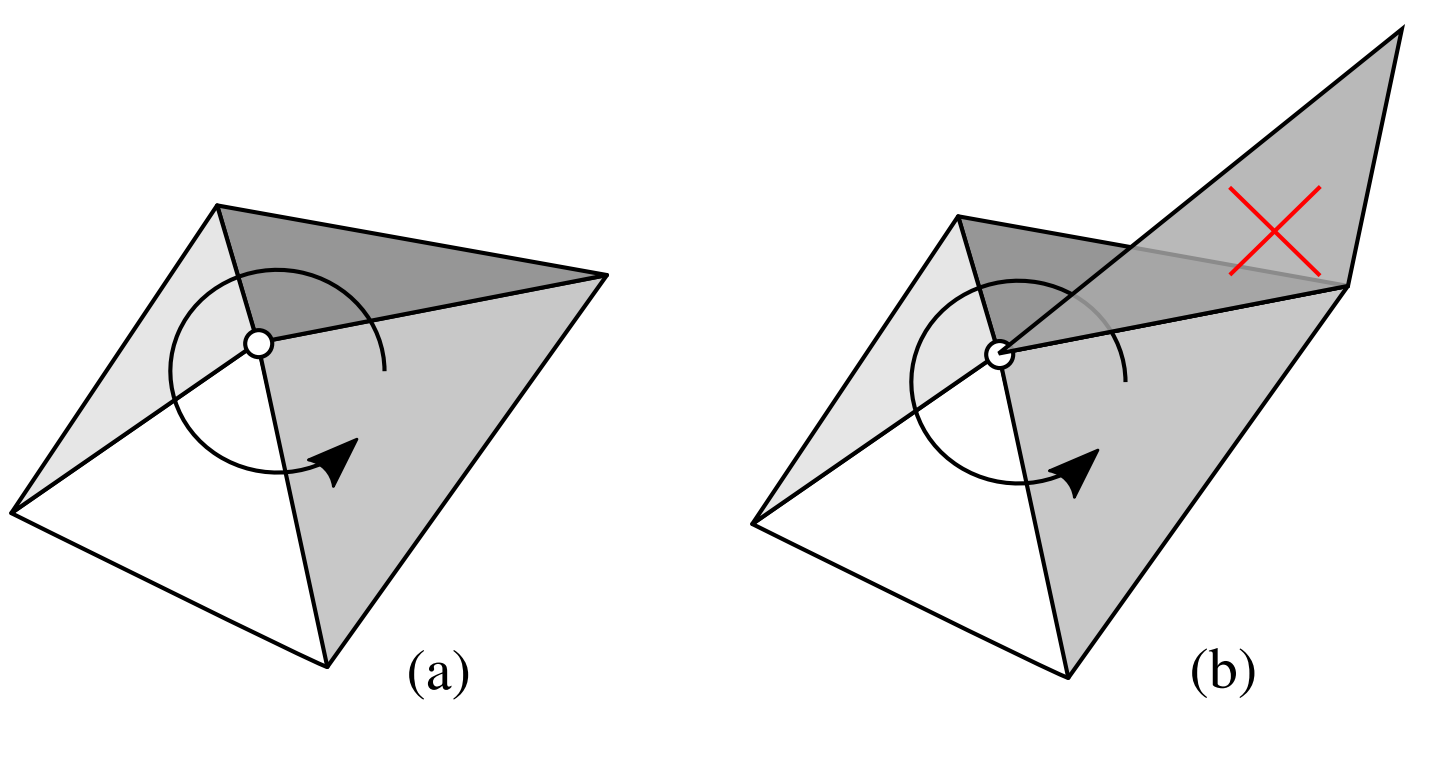}
  \caption{Vertex with adjoined edges and faces: (a) It is possible to cycle through the faces, passing each adjoined edge once. Hence, all faces belong to the same surface. (b) Not all faces belong to the same surface.}
  \label{fig:CycleFaces}
\end{figure}

\noindent{Topology:}
\begin{enumerate}
	\item \label{condition:VertexCoords} Different vertices do have different  coordinates (see Fig. \ref{fig:doubleVertex}). 
	\item \label{condition:Edge2Faces} One edge is shared by exactly two faces (see Fig. \ref{fig:CycleFaces}). 
	\item \label{condition:SameHull} Faces at one vertex belong to one surface, i.e. at a vertex it is possible to cycle through all adjacent faces such that all of the vertex' edges are crossed exactly once (see Fig. \ref{fig:CycleFaces}). 
	\item \label{condition:Orientation} The orientation of faces must follow \textit{Moebius' Rule}, i.e. inside and outside must be distinguishable from each other (see Fig. \ref{fig:orientation}). 
\end{enumerate}

\noindent{Geometry:}
\begin{enumerate}
\setcounter{enumi}{4}
    \item \label{condition:curveOnSurf} A curve must lie on the respective surface whose partial boundary it forms. 
	\item \label{condition:coincide} Both boundary curves at one edge must coincide (see Fig. \ref{fig:coincide}). 
	\item \label{condition:selfIntersect} Surfaces must not self-intersect. From this -- and from \ref{condition:curveOnSurf} -- it follows that curves do not self-intersect either (see Fig. \ref{fig:selfIntersect}).
    \item \label{condition:intersect} Surfaces must not touch or intersect with other surfaces except at common edges (see Fig. \ref{fig:surfaceIntersect}).
\end{enumerate}

\subsection{CAD model flaws}\label{sec:ModelFlaws}
\begin{figure}[H]
  	\centering
	\includegraphics[width=0.7\textwidth]{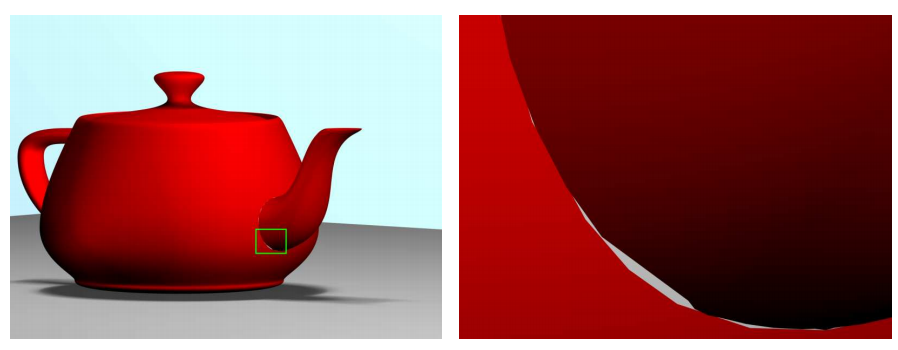}
  \caption{Gaps between trimmed NURBS patches of the Utah teapot (Picture taken from~\cite{Sederberg2008})}
	\label{fig:utahTeapot}	
\end{figure}

\begin{figure}[!htbp]
  \centering
	\subfloat[Double vertices]{
  		\includegraphics[width=0.3\textwidth]{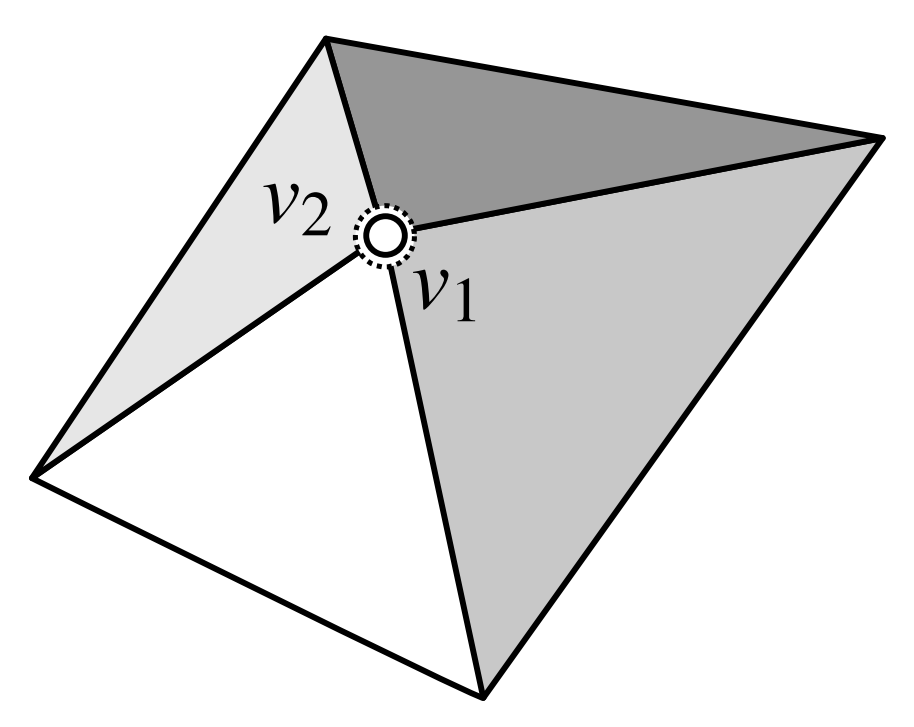}\label{fig:doubleVertex}}
	\quad
	\subfloat[Double edges]{
  		\includegraphics[width=0.3\textwidth]{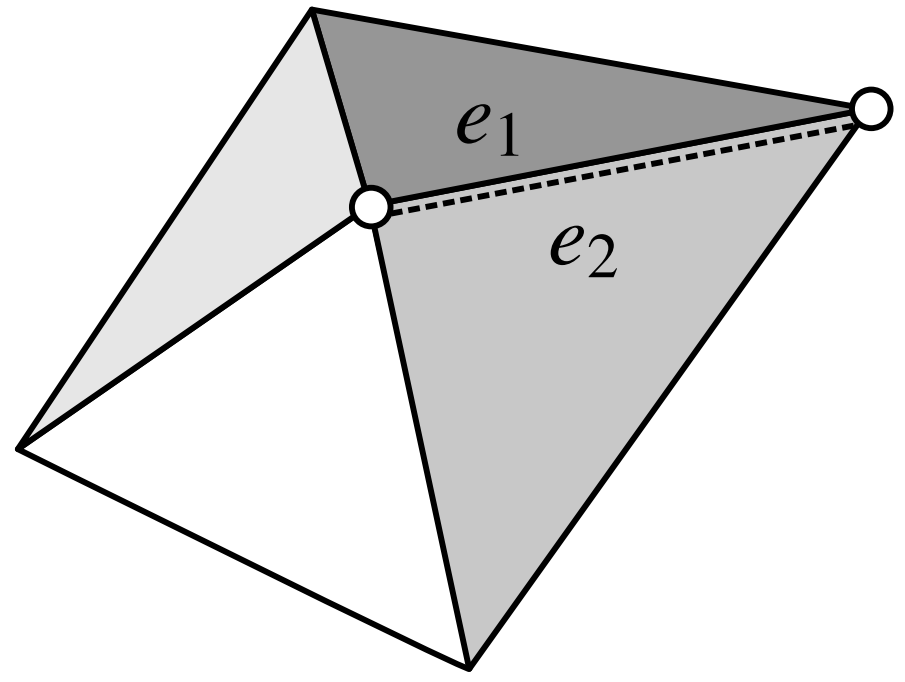}\label{fig:doubleEdge}}
	\quad
	\subfloat[Double faces]{
  		\includegraphics[width=0.3\textwidth]{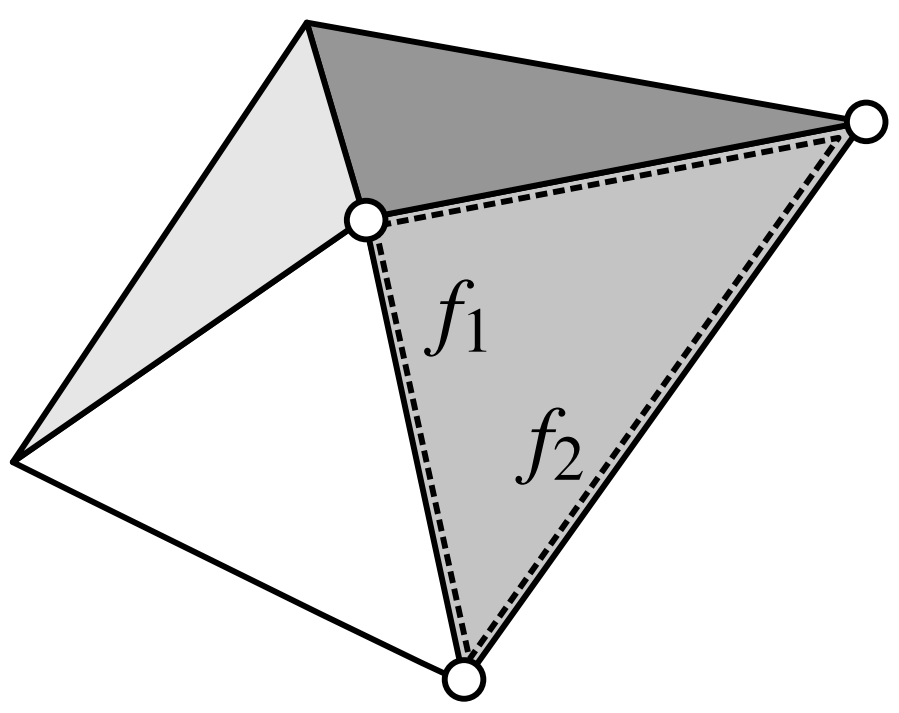}\label{fig:doubleFace}}
	\quad
	\subfloat[Wrong orientations]{
  		\includegraphics[width=0.3\textwidth]{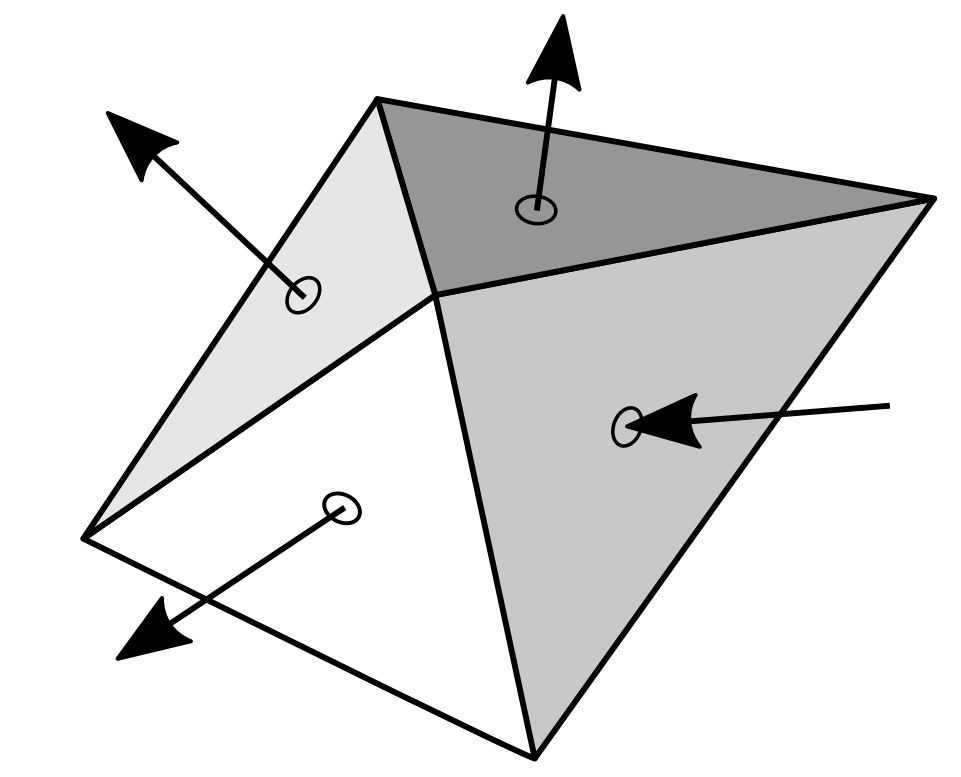}\label{fig:orientation}}
	\quad
	\subfloat[Missing faces]{
  		\includegraphics[width=0.3\textwidth]{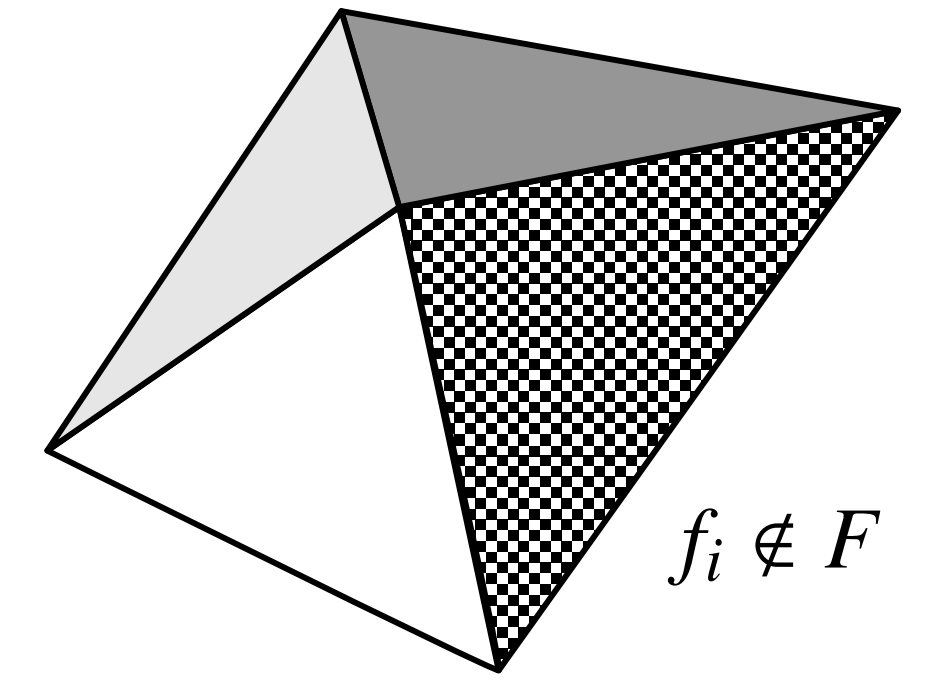}\label{fig:missingFace}}
	\quad
	\caption{Topological flaws}\label{fig:topoFlaws}
\end{figure}

\begin{figure}[!htbp]
  \centering
	\subfloat[Curves at common edge do not coincide]{
  		\includegraphics[width=0.3\textwidth]{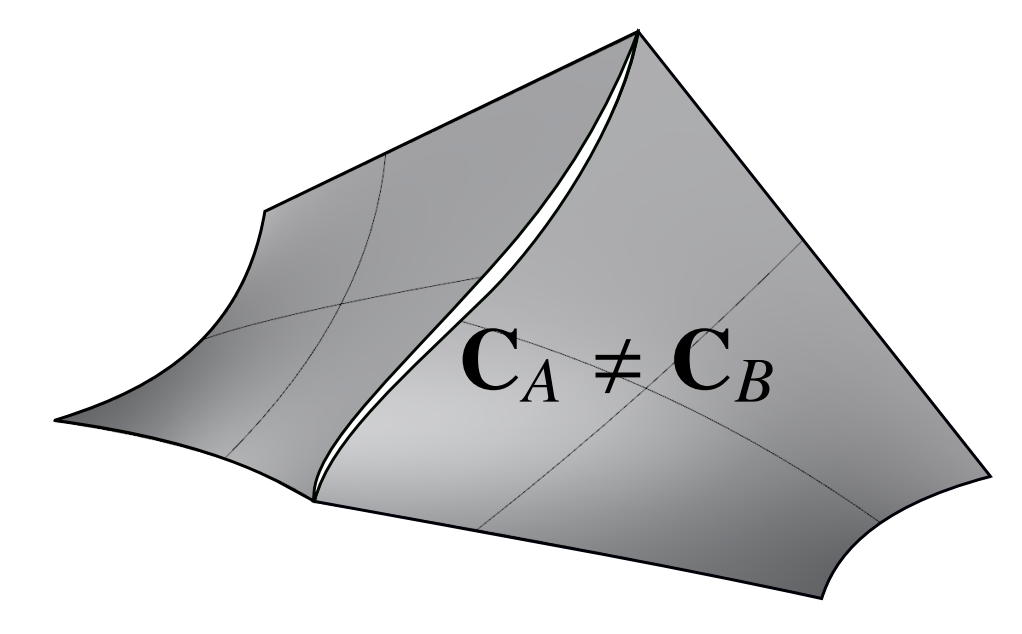}\label{fig:coincide}}
	\quad	
	\subfloat[Surface and boundary curve self-intersect]{
  		\includegraphics[width=0.3\textwidth]{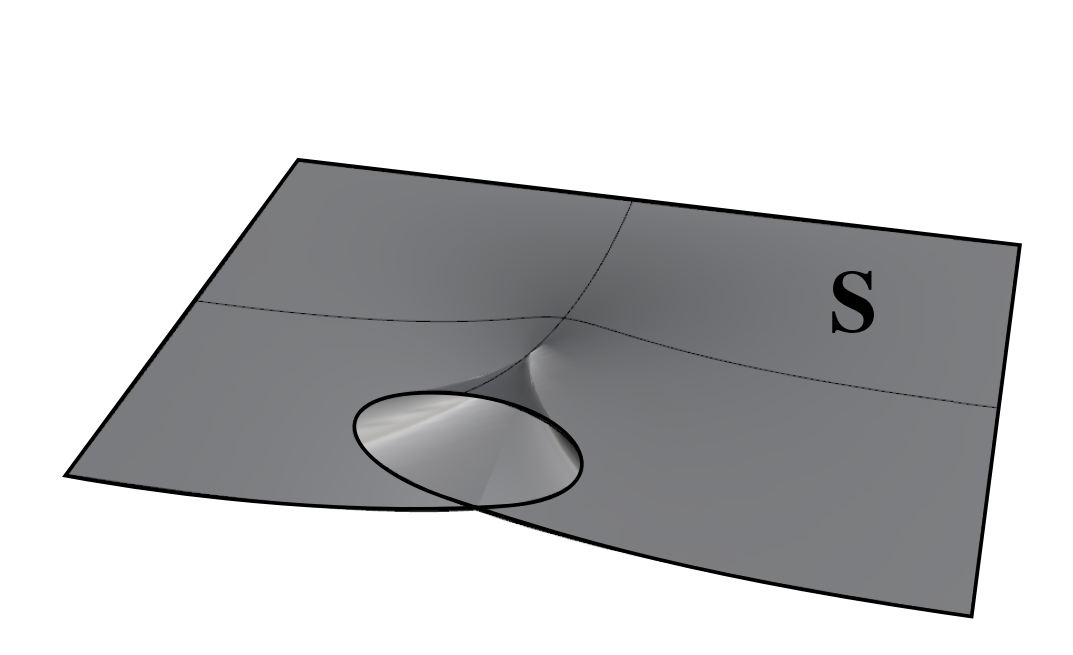}\label{fig:selfIntersect}}
	\quad	
	\subfloat[Surface intersects with another surface]{
  		\includegraphics[width=0.3\textwidth]{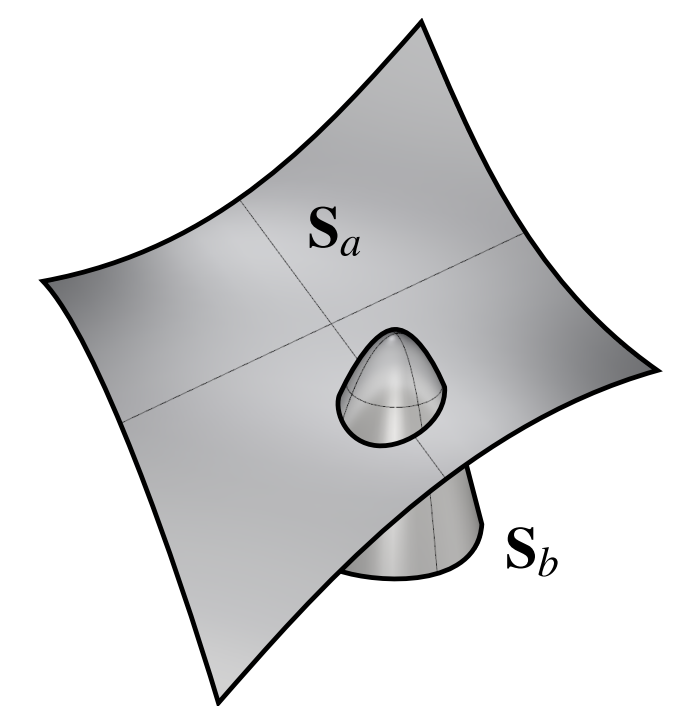}\label{fig:surfaceIntersect}}
	\quad
	\caption{Geometrical flaws}\label{fig:geomFlaws}
\end{figure}

\begin{figure}[!htbp]
  \centering
  	\subfloat[Gaps]{
  		\includegraphics[width=0.3\textwidth]{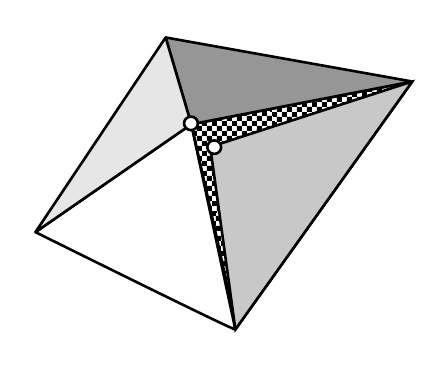}\label{fig:gap}}
	\quad
	\subfloat[Overlaps]{
  		\includegraphics[width=0.3\textwidth]{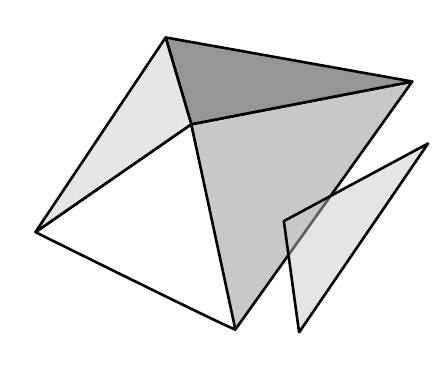}\label{fig:overlap}}
	\quad
	\subfloat[Intersections]{
  		\includegraphics[width=0.3\textwidth]{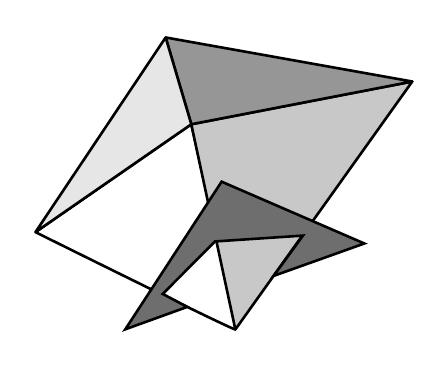}\label{fig:intersection}}
	\quad
	\subfloat[Artifacts]{
  		\includegraphics[width=0.3\textwidth]{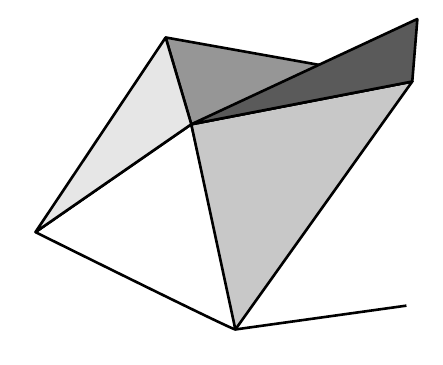}\label{fig:spurious}}
	\quad
	\subfloat[Offsets]{
  		\includegraphics[width=0.3\textwidth]{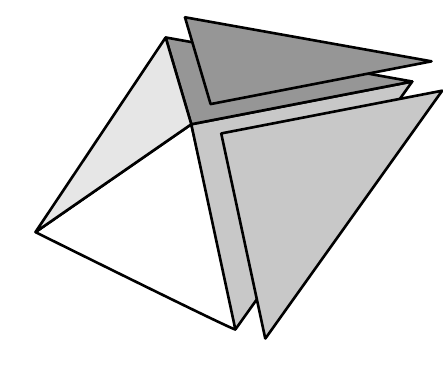}\label{fig:offset}}
	\quad	
	\caption{Hybrid flaws which consist of topological and geometrical components}\label{fig:topoGeomFlaws}
\end{figure}

Model flaws can originate from different sources, such as mathematical inaccuracies, data conversion problems between different software systems, mistakes by designers, different design goals, etc. The probably most famous example of mathematical inaccuracies is the `leaking teapot' model, as depicted in Fig.~\ref{fig:utahTeapot}. The gap between spout and body of the teapot could only be avoided by more complex spline types (see, e.g., T-splines \cite{Bazilevs2010a}), or unreasonably high polynomial degrees. The simplification results in a non-watertight geometry, a major obstacle for the interoperability between CAD and CAE. Figures \ref{fig:topoFlaws}, \ref{fig:geomFlaws}  and \ref{fig:topoGeomFlaws} provide an overview over the most common topological and geometrical modeling flaws. 

\newpage
\subsection{Flaw operators}\label{sec:operators}
\noindent In the following,  we will introduce several operators that perform transformations on a valid B-Rep model, allowing for a controlled imposition of different flaws. To measure the size of the flaws, we introduce an error parameter $\varepsilon$, indicating the 'dirtiness' or inaccuracy of the model.

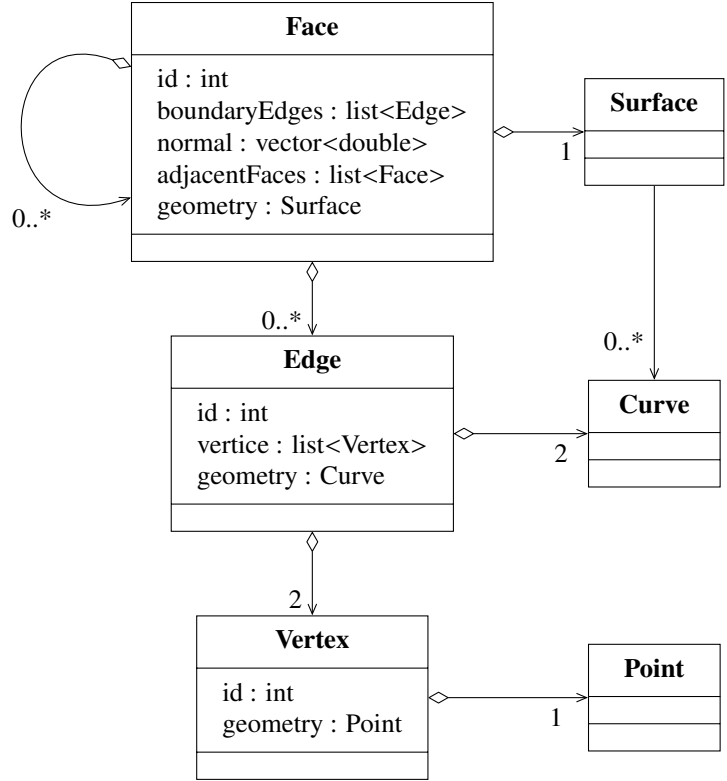
\begin{figure}
\centering
\begin{tikzpicture}
\umlclass[fill=white]{Face}{
id : int \\
boundaryEdges : list$<$Edge$>$\\
normal : vector$<$double$>$ \\
adjacentFaces : list$<$Face$>$\\
geometry : Surface}{}
\umlclass[fill=white,y=-4]{Edge}{
id : int \\
vertice : list$<$Vertex$>$\\
geometry : Curve}{}
\umlclass[fill=white,y=-7.5]{Vertex}{
id : int \\
geometry : Point}{}
\umlemptyclass[fill=white,x=4.5]{Surface}
\umlemptyclass[fill=white,x=4.5, y=-4]{Curve}
\umlemptyclass[fill=white,x=4.5, y=-7.5]{Point}
\umluniaggreg [mult2=1] {Face}{Surface}
\umluniaggreg [mult2=2] {Edge}{Curve}
\umluniaggreg [mult2=1] {Vertex}{Point}
\umluniaggreg [mult2=0..*] {Face}{Edge}
\umluniaggreg [mult2=2] {Edge}{Vertex}
\umluniassoc [mult2=0..*] {Surface}{Curve}
\umluniaggreg [mult=0..*, angle1=160, angle2=200,loopsize=2cm ] {Face}{Face}
\end{tikzpicture}
\caption{UML-diagram of a possible object-oriented B-Rep implementation}
\label{fig:UML}
\end{figure}

To provide an easily understandable formulation of the operators, we consider an object-oriented B-Rep data structure. Thereby, the implementation must allow a distinction between internal and external/adjacency relations. \Cref{fig:UML} provides a UML diagram of a possible hierarchical implementation. For an introduction to the notation of the UML (Unified Modeling Language) see, e.g., \citep{Rumpe2016}. Here, the external adjacency relations $r^{ext}$ are realized at the faces, where the adjacent faces are stored in the field: \textit{adjacentFaces}. All other external adjacency relations (e.g., which edges are adjacent) can be derived from this and from the respective internal relations $r^{int}$.

\begin{figure}[H]
  	\centering
  	\includegraphics[width=0.9\textwidth]{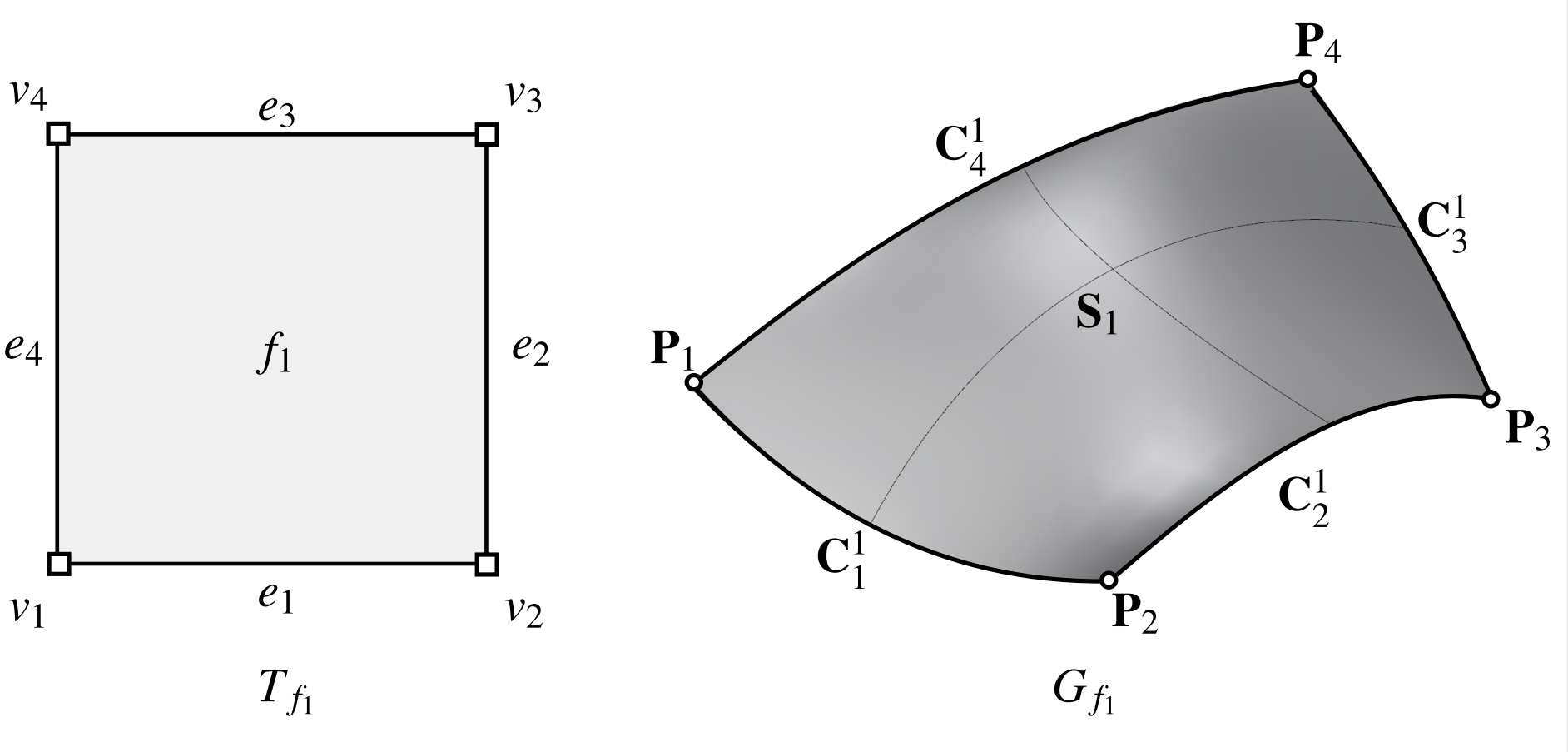}
  	\caption{B-Rep sub-part $\omega^{f_1}$, which corresponds to face $f_1$ and consists of the topology $T^{f_1}$ and the corresponding geometry $G^{f_1}$.}
	\label{fig:BRepSegment}
\end{figure}

Let $\omega^{t_i}$ be the B-Rep sub-part, or segment, which corresponds to a topological entity $t_i$, e.g., a face, an edge, or a vertex. The segment $\omega^{t_i}$ consists of all information that is needed to visualize $t_i$. Hence, it must contain $t_i$ and, recursively, all underlying sub-topologies and geometries that are related by respective internal adjacency relations $r^{int}$ (see Fig.~\ref{fig:BRepSegment}).
\begin{equation}
\omega^{t_i}(T^{t_i}, G^{t_i}) \subset B 
\end{equation}
with $T^{t_i}\left(\, \tau, \rho^{int} \,\right)$ and $G^{t_i}(\gamma)$ being the respective topology and geometry, where $\tau =\{\tau_i\}$ and $\gamma = \{\gamma_i\}$ denote the sets of those topological and geometrical entities which are recursively related by the internal relations $\rho^{int}=\{\rho^{int}_i\}$. Consequently, three different segments are possible: 1) vertex segments, b) edge segments, and c) face segments. As topological entities $\{\tau_i\}$, a face segment, for example, contains the face itself and all associated edges and vertices. As geometric entities $\{\gamma_j\}$, it holds the corresponding surface with its boundary curves and corner points. Additionally, all internal relations $\{\rho^{int}_i\}$ are contained, i.e. the relations among face, boundary edges, and vertices, as well as the relations to the geometric entities. Not contained are \textit{external} adjacency relations, i.e. those to neighboring faces or edges.

\noindent In the following, $\tilde{a}$ denotes the object $a$ after the transformation and let $\mathrm{dist}\,(a,b) = \mathrm{inf}\{\,\lVert a,b \rVert_2\,\}$ be the minimum Euclidean distance between two objects, e.g., the distance between the closest points on two different surfaces.

\begin{enumerate}

\item Let $O^{select}$ be an extraction operator, which selects for a topological entity $t_i$ the corresponding segment $\omega^{t_i}$ from the body $B$.
\begin{equation}
	 O^{select} \, \left( B \, ,\, t_i \right) \mapsto \omega^{t_i} \label{eq:extract}
\end{equation}
\noindent Note that the external relations are not extracted. Hence, the segment forgets about its \textit{logical} location in the body.

\item Let $O^{join}$ be a join operator that adds a segment $\omega^{t_i}$ to the body $B$.
\begin{equation}
	 O^{join} \, \left( B \, ,\, \omega^{t_i} \right) \mapsto \tilde{B}\,,\;\; \mathrm{where} \;\;\tilde{B} =B \, \cup \,\omega^{t_i} \label{eq:join}
\end{equation}

\item Let $O^{shallowCopy}$ be a shallow copy operator that copies an arbitrary entity $a_i$. For a more detailed description of the object-oriented concept of 'shallow' and 'deep' copying see, e.g., \citep{Goldberg1983}. $a_i$ can be a topological entity $t_i$, a geometrical entity $g_i$, or an internal $r_i^{int}$ or external relation $r_i^{ext}$.
\begin{equation}
	 O^{shallowCopy} \, \left(a_i \right) \mapsto \tilde{a}_i\,,\;\; \mathrm{where} \;\; \tilde{a}_i := a_i \label{eq:shallowCopy}
\end{equation}
The '$:=$' in (\ref{eq:shallowCopy}) is to be understood as the shallow copy assignment, according to~\citep{Goldberg1983}. Note that the new object is distinguishable from the old object, e.g., by an updated id, or, in the context of object-oriented programming, by a different memory address. Yet, it still uses the same references to other objects as the original segment.

\item Let $O^{deepCopy}$ be an internal deep copy operator \citep{Goldberg1983} that performs a deep copy operation on a segment $\omega^{t_i}$. To this end, a shallow copy operation is carried out on all corresponding topological and geometrical entities, as well as the internal adjacency relations.
\begin{equation}
\begin{aligned}
	 O^{deepCopy} &\, \left( \omega^{t_i} \right)  \mapsto \tilde{\omega}^{t_i}(\tilde{T}^{t_i}, \tilde{G}^{t_i}) \,,\;\; \mathrm{with} \;\; \tilde{T}^{t_i}\left(\,\tilde{\tau},\tilde{\rho}^{int}\,\right)\, , \, \tilde{\tau}_{i} := O^{shallowCopy}(\tau_i)\,,\\
	 & \quad \tilde{\rho}^{int}_i := O^{shallowCopy}(\rho^{int}_i) \;\;\forall \tau_i, \rho^{int}_i \in {T}^{t_i} \;, \\
	 & \mathrm{and} \\
	 & \quad \tilde{G}^{t_i}\left(\,\tilde{\gamma}\right) \; , \;  \tilde{\gamma}_i := O^{shallowCopy}(\gamma_i) \;\; \forall \gamma_{i} \in G^{t_i}
\end{aligned}\label{eq:deepCopy}
\end{equation}
Note that the deep copied segment $\omega^{t_i}$ has no information about its \textit{logical} location in $B$, i.e. it has no external adjacency relations, and that all internal relations are updated to reference the new topological and geometrical entities.

\item Let $O^{delete}$ be a deletion operator that deletes a face $f_i$ and its related geometry $g^{f_i} \subset G$ consisting of the underlying surface $\bm{S}_i$ and the corresponding boundary curves $\{\bm{C}^{\bm{S}_i}_j\}$. Thereby, the characteristic size of the resulting opening must not exceed a given, e.g., user-defined minimal accuracy $\varepsilon$. Let $\delta$ be the diameter of the largest possible inscribed sphere of the surface $\bm{S}_i$ to be deleted.
\begin{align}
\begin{aligned}
	O^{delete}\, \left(B, f_i, g^{f_i}  \right) \mapsto \tilde{B}\,(\tilde{T},\tilde{G})\,,\;\; \mathrm{where} \;\; \tilde{F} = F \setminus f_i \; ;\; \tilde{G} = G \setminus g^{f_i} \; ,\; \delta < \varepsilon
\end{aligned}\label{eq:delete}
\end{align}
Note that, as a deletion of an edge, or vertex would lead to an uncontrollable cascade of deletions of superior entities, only a face deletion is allowed in this context. A B-Rep model with a deleted edge or node without deletion of referencing faces would not even be readable and is not considered in our investigation.

\item Let $O^{explode}$ be an operator that removes all external relations $r^{ext}$ from a body $B$. This can be achieved by extracting (\ref{eq:extract}), copying (\ref{eq:deepCopy}), and joining (\ref{eq:join}) all face segments $\omega_{f_i}\; \forall f_i \in F$. The resulting body $\tilde{B}$ is then described by independent topological sup-parts/segments $\omega_{f_i}$.
\begin{align}
\begin{aligned}
	O^{explode}\,(B) = O^{join}\,\left(B^\circ,\, O^{deepCopy}\,\left(O^{extract}\left(B,\,F\right)\right)\right) \mapsto  \tilde{B} \;,\;\; \mathrm{where} \;\; \tilde{r}^{ext} = \emptyset
\end{aligned}\label{eq:explode}
\end{align}
where $B^\circ$ is an empty body.

\item Let $O^{flip}$ be a topological flip operator that flips the normal $\bm{n}_i$ of a face  $f_i$.
\begin{align}	
	& O^{flip} \,\bigl(f_i \bigr) \mapsto \tilde{f_i} \;,\;\; \mathrm{where} \;\;\tilde{f_i} = \left( (e_{\kappa}), \tilde{\bm{n}}_i = -1\cdot \bm{n}_i \right) \label{eq:flip}
\end{align}

\item Let $O^{move}$ be a geometric move operation that moves the point $\bm{P}_i$ within the range $\varepsilon$. Additionally, all adjoined surfaces and curves are adapted consistently such that they form a 2-manifold without boundaries after the operation. This involves the following adaptions to the adjoined surfaces $S^{\bm{P}_i} = \{\bm{S}_{i}^{\bm{P}_i}\}$ and curves $C^{\bm{P}_i} =\{\bm{C}_{i}^{\bm{P}_i}\}$:
\begin{itemize}
 \item The resulting point $\bm{\tilde{P}}_i$ must again lie on the altered surfaces $\tilde{S}^{\bm{\tilde{P}}_i} $ and curves $\tilde{C}^{\bm{\tilde{P}}_i} $ .
 \item All pairs of the resulting adjoined surfaces $(\bm{\tilde{S}}_A^{e_k},\,\bm{\tilde{S}}_B^{e_k})$ must again meet at their common edge/curve $e_k$. Consequently, the two respective boundary curves $(\bm{\tilde{C}}_A^{e_k},\,\bm{\tilde{C}}_B^{e_k})$ must coincide.
\end{itemize} 
The latter condition is omitted in the case of an already broken topology, where an edge no longer has two adjoined faces/surfaces.
\begin{align}
\begin{aligned}
O^{move}\,&(\bm{P}_i,\,G ) \mapsto \tilde{G}  \;,\;\; \mathrm{where}\;\; 0 < \mathrm{dist}\left( \bm{P}_i,\tilde{\bm{P}}_i \right) < \varepsilon\,, \\
& \mathrm{and} \\
& \quad \exists\; \bm{\xi} : \bm{\tilde{S}}_i^{\bm{\tilde{P}}_i}(\bm{\xi}) = \bm{\tilde{P}}_i \;\; \forall \, \bm{\tilde{S}}_i^{\bm{\tilde{P}}_i}\; , \; \exists\; \zeta : \bm{\tilde{C}}_i(\zeta) = \bm{\tilde{P}}_i \;\; \forall  \,\bm{\tilde{C}}_i^{\bm{\tilde{P}}_i} \; , \\
& \mathrm{and} \\
& \quad \mathrm{dist}\left( \bm{\tilde{C}}_A^{e_k},\,\bm{\tilde{C}}_B^{e_k}\right) = 0 \; \;\forall \,(\bm{\tilde{S}}_A^{e_k},\,\bm{\tilde{S}}_B^{e_k}) \;\mathrm{at}\; v_i 
\end{aligned}\label{eq:move}
\end{align}

\item Let $O^{detach}$ be a geometrical operator that detaches two adjacent surfaces, $\bm{S}_i$ and $\bm{S}_j$, which meet at the edge $e_k$ (see~\cref{fig:coincide}). To this end, one surface $\bm{S}_i$  and its respective boundary curve $\bm{C}_{\bm{S}_i}^{e_k}$ at $e_k$ are changed. Again, the characteristic size of the potentially resulting opening must not exceed $\varepsilon$.
\begin{align}
\begin{aligned}
O^{detach}\,&\left(G, e_k\right) \mapsto \; \tilde{G} \;,\;\; \mathrm{where} \;\;  \mathrm{dist}\left(\bm{\tilde{S}}_i, \bm{\tilde{C}}_{\bm{S}_i}^{e_k}(\xi) \right) = 0\; ,\\
& 0 \leq \mathrm{dist}\left(\bm{C}_{\bm{S}_i}^{e_k}, \bm{\tilde{C}}_{\bm{S}_i}^{e_k}(\xi)\right) < \varepsilon \;\; \forall \xi \in [\xi_a, \xi_b]
\end{aligned} 
 \label{eq:detach}
\end{align}

with $[\xi_a, \xi_b]$ being the respective interval on which the boundary curve is defined.

\item Let $O^{intersect}$ be a geometric operator that alters a surface $\bm{S}_i(\bm{\xi})$ such that it touches or intersects with another surface $\bm{S}_j(\bm{\eta})$ apart from common edges. Note that we assume that there is no intersection in the original model, according to the definition of a valid B-Rep model.
\begin{equation}
\begin{aligned}
	 & O^{intersect}\,\left(\bm{S}_i(\bm{\xi})\right) \mapsto \;\tilde{\bm{S}}_i(\bm{\xi})\;, \;\; \mathrm{where} \\
	 & \quad \exists \;(\bm{\xi}, \bm{\eta})\;:\; \mathrm{dist}\left( \bm{\tilde{S}}_i(\bm{\xi}),\bm{S}_j (\bm{\eta}) \, \right) = 0 \; \wedge \; \mathrm{dist}\left(\bm{\tilde{S}}_i(\bm{\xi}),\bm{\tilde{C}}^{\bm{\tilde{S}}_i}_k\right) > 0 \;\\
	 & \quad  \forall \,\bm{\tilde{C}}^{\bm{\tilde{S}}_i}_k \in \Gamma^{\tilde{\bm{S}}_i} \;,\; i \neq j
\end{aligned} \label{eq:intersect}
\end{equation}
with $\Gamma^{\tilde{\bm{S}}_i}$ being the set of boundary curves of $\bm{\tilde{S}}_i$. 

A special case of intersections are self-intersections:
\begin{equation}
\begin{aligned}
	 O^{selfIntersect}\,\left(\bm{S}_i(\bm{\xi})\right) \mapsto  \;\tilde{\bm{S}}_i(\bm{\xi})\;, \;\; \mathrm{where} \;\;  \exists \;(\bm{\xi}, \bm{\eta})\;:\; \mathrm{dist}\left( \bm{\tilde{S}}_i(\bm{\xi}),\bm{S}_i (\bm{\eta}) \, \right) = 0 \,,\; \bm{\xi} \neq \bm{\eta}
\end{aligned} \label{eq:selfintersect}
\end{equation}

\end{enumerate}

\subsection{Application of flaw operators}\label{sec:ApplicationOperators}
We now continue with the definition of a flawed model. To this end, we apply the flaw operators defined in \cref{sec:operators} onto a valid B-Rep model. The 'dirtiness' of the model is then defined by $\varepsilon$. It should be mentioned that, for models that are drafted by a real-life CAD system, flaws do not necessarily originate from these operators, yet most flawed models can equivalently be created by a sequence of these operators.

Let $B(T,G)$ be a valid flawless B-Rep body. Note that operators acting on the body is to be understood as acting on a segment $\omega^{t_i}$, or single topological, or geometrical entity, or relation.

\begin{enumerate}

\item Single topological entities $t_i$ and their corresponding segments $\omega_{t_i}$ can be copied and added to $B$ with a combination of the extraction (\ref{eq:extract}), the deep copying (\ref{eq:deepCopy}), and the joining (\ref{eq:join}) operator:
\begin{equation}
\tilde{B}\,(\tilde{T},\tilde{G})  := O^{join}\,\left(B\,(T,G),\, O^{deepCopy}\,\left(O^{extract}\left(B\,(T,G),\,t_i\right)\right)\right)
\end{equation}
The resulting B-Rep model is invalid as it has multiple entities (refer to Figs.~\ref{fig:doubleVertex}, \ref{fig:doubleEdge}, and \ref{fig:doubleFace}), which violates condition \ref{condition:VertexCoords}. As an example, consider the STL format where each triangle (re-)defines its corner points. Also, multiply defined faces/surfaces appear frequently in free form CAD models, which leads to a touching/intersection of the surfaces (refer to Fig. \ref{fig:intersection}).
 
\item Application of the deletion operator (\ref{eq:delete}) on a face $f_i$:
\begin{equation}
\tilde{B}\,(\tilde{T},\tilde{G}) :=  O^{delete}\, \left(B, f_i, g_{f_i}\right)
\end{equation}
The deletion of a face violates condition~\ref{condition:Edge2Faces} (see Fig. \ref{fig:missingFace}). Thereby, the size of the resulting opening restricted to be smaller than $\varepsilon$.

\item Application of the explosion operator (\ref{eq:explode}):
\begin{equation}
\tilde{B}\;(\tilde{T},\tilde{G}) :=  O^{explode}(B(T,G))
\end{equation}
Most B-Rep models are constructed from independent surfaces, which are later joined into a (hopefully) valid B-Rep model. This join operation corresponds to the inverse of the explosion operation. It is yet well known that a strict 'join'-operation is not necessarily possible (or maybe not feasible) e.g., in case of an intersection of two NURBS surfaces~\citep{Sederberg2008}. Also, STL models are constructed by independent triangles. Such models violate the topological conditions \ref{condition:VertexCoords}, \ref{condition:Edge2Faces}, and~\ref{condition:SameHull}. Geometrically, they can still form a closed 2-manifold without boundaries. However, these models are very prone to a variety of different flaws, as no external adjacency relations are provided explicitly.

\item Application of the flip operator (\ref{eq:flip}):
\begin{equation}
\tilde{B}\;(\tilde{T},G) :=  O^{flip}(B(T,G))
\end{equation}
The resulting B-Rep model does not fulfill \textit{Moebius' Rule} anymore~(see condition~\ref{condition:Orientation}). This flaw usually appears if the normal is defined implicitly by the order of the boundary edges (see Fig.~\ref{fig:orientation}). However, this error also appears quite frequently if the normal is given explicitly, e.g., in the case of STL.
 
\item Application of the move operator (\ref{eq:move}):
\begin{equation}
\tilde{B}\;(T,\tilde{G}) := O^{move}(B(T,G))
\end{equation}
Applied on a valid B-Rep body, the move operator preserves a geometric \textit{2-manifold, without boundary}. However, the orientability can be lost (see condition:\ref{condition:Orientation}). As an example, consider a point $\bm{P}_i$ on surface $\bm{S}_j$, which is close to surface $\bm{S}_k$ with distance $\mathrm{dist}\left(\bm{P}_i,\,\bm{S}_k\right) < \varepsilon$. A movement then can lead to an intersection of the two surfaces. This violates condition~\ref{condition:intersect} (see Fig. \ref{fig:surfaceIntersect}). 

\item Application of the detach operator (\ref{eq:detach}):
\begin{equation}
\tilde{B}\;(T,\tilde{G}) := O^{detach}(B(T,G))
\end{equation}
The resulting model violates condition~\ref{condition:coincide}. This is likely to happen at the intersection of free-form surfaces. The boundary curves would require unreasonably high polynomial degrees to perfectly coincide. Possible flaws can e.g., be openings or intersections (see Figs.~\ref{fig:coincide} and \ref{fig:surfaceIntersect}). As an example, consider the leaking Utah teapot. 

\item Application of the intersection operator (\ref{eq:intersect}):
\begin{equation}
\tilde{B}\;(T,\tilde{G}) := O^{intersect}(B(T,G))
\end{equation}
The resulting model may violate conditions~\ref{condition:selfIntersect} or ~\ref{condition:intersect}. Apart from gaps, intersections frequently appear at patch boundaries as well (see Fig.~\ref{fig:coincide}). Intersections can also occur if two surfaces are too close to each other. In this case, they additionally violate \textit{Moebius' Rule}~\ref{condition:Orientation} (see Figs.~\ref{fig:surfaceIntersect} and ~\ref{fig:selfIntersect}). A special case are overlaps, where two surfaces touch each other (see Fig. \ref{fig:overlap}).

\item Application of the copy (\ref{eq:deepCopy}) and the move operators (\ref{eq:move}) to a single face $f_i$ :
\begin{align}
\begin{aligned}
& \tilde{\omega}^{f_i} := O^{deepCopy}\left(O^{extract}\left( B,\,f_i\right)\right)\\
& \breve{\omega}^{f_i} := O^{move}(\bm{P}_j \in \tilde{\omega}^{f_i},\, \tilde{G}^{t_i}) \\
& \tilde{B}\;(\tilde{T},\tilde{G}) := O^{join}\left(B,\,\breve{\omega}^{f_i} \right) \\
\end{aligned}
\end{align}

This chain of operations allows to create offsets and artifacts, i.e. entities which do not belong to the outer hull and lead to a violation of the conditions~\ref{condition:SameHull} and ~\ref{condition:Edge2Faces} (see Figs.~\ref{fig:spurious} and \ref{fig:offset}).

\item Application of the explosion (\ref{eq:explode}) and move operators (\ref{eq:move}):
\begin{align}
\begin{aligned}
& \tilde{B}(\tilde{T},\tilde{G}) := O^{explode}(B)\\
& \breve{B}(\tilde{T},\breve{G}) := O^{move}(\bm{\tilde{P}}_i \in \tilde{G},\, \tilde{G})
\end{aligned}
\end{align}
Starting from an exploded model, moving one or more points can lead to various common flaws -- such as gaps, intersections, or overlaps (see Figs.~\ref{fig:gap},~\ref{fig:intersection}, ~\ref{fig:overlap}). As many B-Rep modeling tools work with exploded models, i.e. with independent surfaces, these flaws appear very commonly, particularly at patch boundaries. Also, the STL format stores a body with independent triangles.

\end{enumerate}

\noindent Note that, independent of the performed operations, it is imperative for the presented method that the size of all openings and gaps is restricted to be smaller than a pre-defined $\varepsilon$. This is required not only for each individual flaw operation but also for the resulting model after a sequence of flaw operations, e.g., a sequence of individual moves of a segment.

The resulting flawed models are invalid in a mathematical sense, which renders a subsequent conversion into a simulation model either impossible or invalid. The necessity to heal the flaws can neither be circumvented by meshing, as in the classical FEM, nor by a direct simulation as in IGA. It is, however, possible to compute 'dirty' models directly with an \emph{embedded domain method} such as the Finite Cell Method (Section \ref{sec:FCM}). To this end, we construct a specially adapted Point Membership Classification test (Section \ref{sec:Method}) which is blind to flaws up to a characteristic size $\varepsilon$.

\section{Finite Cell Method} \label{sec:FCM}
The Finite Cell Method is a higher order fictitious domain method. However, the approach presented within this paper does not rely on higher-order elements. Hence, it can be also applicable for linear fictitious domain methods. FCM offers simple meshing of potentially complex domains into a structured grid of, e.g., cuboid cells without compromising the accuracy of the underlying numerical method. For completeness of this paper, the basic concepts are briefly introduced in this section. We restrict ourselves to linear elasticity -- emphasizing however, that the FCM has been extended to more general partial differential equations \cite{Schillinger2012, Zander2012, Elhaddad2018, Kollmannsberger2018}.

\subsection{Basic formulation}
\begin{figure*}[ht]
  \centering
	\includegraphics[width=0.98\textwidth]{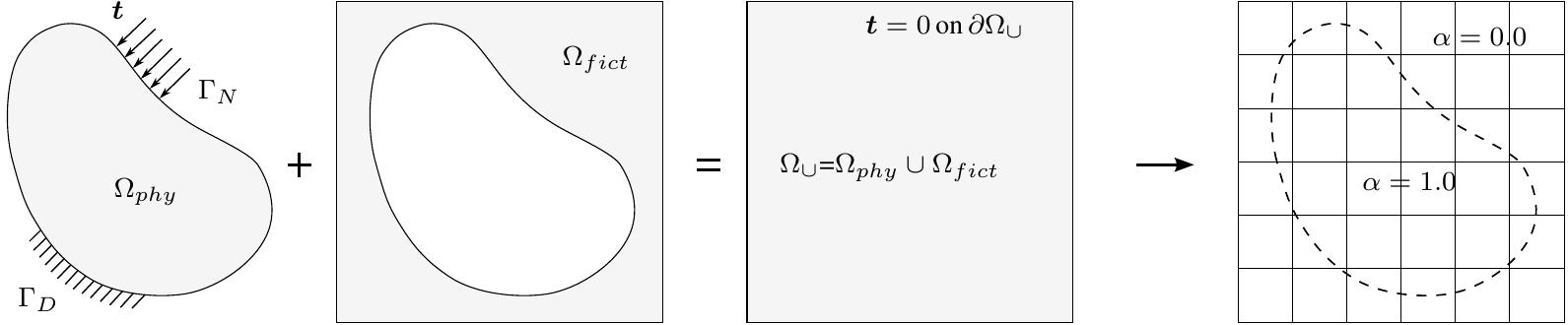}
  \caption{The concept of the Finite Cell Method~\cite{Duster2017}}	
\label{fig:fictdomain}
\end{figure*}

In the Finite Cell Method, an n-dimensional open and bounded physical domain $\Omega_{{phy}}$ is embedded 
in a fictitious domain $\Omega_{{fict}}$ to form an extended 
domain $\Omega_{\cup}$, as illustrated in Fig.~\ref{fig:fictdomain} in two dimensions. The 
resulting domain $\Omega_{\cup}$ has a simple shape which can be meshed easily, without conforming to the boundary of $\Omega_{{phy}}$. 

The weak form of the equilibrium equation for the extended domain $\Omega_{\cup}$ is defined as
\begin{align}
  \int _{\Omega_{\cup}}
 	[\mathbf{L} \mathbf{v}]^T \alpha \mathbf{C}  [\mathbf{L} \mathbf{u}] 
 	~\mathrm{d}\Omega
	&= 	\int _{\Omega_{\cup}}
	   \mathbf{v}^T \alpha \mathbf{f} 
 	  ~\mathrm{d}\Omega
 	+ \int_{\Gamma_N}
	\mathbf{v}^T \mathbf{\overline{t}}
 	~\mathrm{d}\Gamma \quad ,
	\label{eq:VirtualWork}
\end{align}
where $\mathbf{u}$ is a displacement function, $\mathbf{v}$ a test function, $\mathbf{L}$ is the linear strain operator, and $\mathbf{C}$ denotes the elasticity matrix of the physical domain $\Omega_{{phy}}$, yet extended to $\Omega_{\cup}$. $\mathbf{f}$ and $\mathbf{\overline{t}}$ denote the body load and the prescribed tractions on the Neumann boundary, respectively. The indicator function $\alpha$ is defined as
\begin{align}
\alpha(\mathbf{x})&=\begin{cases}
1 & \forall \mathbf{x} \in \Omega_{phy} \\
10^{-q} & \forall\mathbf{x} \in \Omega_{fict}
\end{cases} \quad ,
	\label{eq:alpha}
\end{align}
In the limiting case of $q\xrightarrow{} \infty$, the standard weak form for an elasticity problem on $\Omega_{phys}$ is obtained. In practical applications, a sufficiently large  $q = 6..10$ (see~\cite{Parvizian2007,Duster2008}) is chosen, introducing a modeling error to the formulation~\cite{Dauge2015}, which yet stabilizes the numerical scheme and controls the conditioning number of the discrete equation system -- see~\cite{dePrenter2017a} for a detailed analysis. $\Omega_{fict}$ is then discretized in 'finite cells' of simple shape (rectangles or cuboids). In the context of this paper, we assume for simplicity a uniform grid of finite cells, yet note that generalizations to locally refined grids \cite{Schillinger2011, Zander2016} and unstructured meshes ~\cite{Varduhn2016, Kamensky2015, Duczek2016} have been studied extensively.

\subsection{Geometry treatment}
In FCM, the physical domain $\Omega_{phys}$ (i.e. the geometry) is recovered by the discontinuous scalar field $\alpha$. Consequently, the complexity of the geometry is shifted from the finite elements to the integration of the element matrices and load vectors, which imposes less geometrical requirements on the model. It is in fact sufficient to provide a robust Point Membership Classification (PMC), i.e. for every point $x \in \mathbb{R}^n$, it must be possible to decide whether it is inside or outside of ${\Omega}_{phys}$. This implies that $\Omega_{phys}$ must have a mathematically valid description. Due to the discontinuity of $\alpha$, the integrands in cut cells need to be computed by specially constructed quadrature rules, see, e.g.,~\cite{Abedian2013,Kudela2016,Hubrich2017} for a recent overview of possible schemes. To perform a suitable integration, the domain is approximated by a space-tree $T\!R_{int}$. The leaves of $T\!R_{int}$ are called integration leaves $c_{int}$.
Additional information, such as explicit surface descriptions are only needed for the application of boundary conditions as well as for post-processing (see Sec.~\ref{sec:boundaryConditions}).

\subsection{Boundary conditions} \label{sec:boundaryConditions}
Neumann boundary conditions are applied according to equation \eqref{eq:VirtualWork} in an integral sense on the boundary $\Gamma_{N}$. Homogeneous Neumann conditions (i.e. zero traction) require no treatment, as they are automatically satisfied by setting $\alpha=0$ or, in an approximate sense, to a small value in $ \Omega_{fict}$. As the boundary of the physical model typically does not coincide with the edges/faces of the finite cell mesh, Dirichlet boundary conditions need to be enforced also in a weak sense. To this end, several methods have been adopted, such as the penalty method, Nitsche's method, or Lagrange Multipliers~\cite{Ruess2012,Ruess2013, Kollmannsberger2015, Guo2015a}. \\
For the integration of Dirichlet and inhomogeneous Neumann boundary conditions, an explicit surface description is needed. This can be of poor quality. For the enforcement of Neumann boundary conditions, however, a surface without multiple faces/surfaces or large overlaps is required, as these flaws would introduce physically modified boundary conditions (i.e. additional loads, heat sources, etc.). To this end, we propose the following automatable method to convert a 'dirty' surface into a surface without multiple entities or overlaps: 
\begin{enumerate}
	\item Triangulate the respective surface (if not already provided, e.g., with STL).

	\item Get the intersection points between the surface mesh and the element boundaries. 

	\item Create an element-wise point cloud from the intersection points and respective triangle corner points.

	\item Perform an element-wise Delaunay triangulation on the respective point cloud.
\end{enumerate}
The resulting element-wise triangular meshes are used only for integration and can consequently be independent of each other. Note that the requirements to these local surface meshes are by far less restrictive than they would be for a surface mesh as a starting point for volume mesh generation. Note that a potential triangulation of the surface will cause an approximation error.

\section{Robust Point Membership Classification for flawed CAD models} \label{sec:Method}
As explained in Section~\ref{sec:FCM}, the only geometric information required to setup the system matrices for the Finite Cell Method is an unambiguous statement about the location of a point, i.e whether it lies inside or outside of the domain of computation. Considering flawed CAD models (e.g., with undesired openings), the concept of `inside' or `outside' is fuzzy -- at least up to the characteristic size of the flaw $\varepsilon$. In this section, we present a robust Point Membership Classification method for 'dirty' STL B-Rep models. The presented approach is, however, not restricted to STL models, and it can easily be extended to other boundary representations.

\subsection{Point Membership Classification for valid CAD models}\label{sec:PMCvalid}
PMC algorithms are fundamental and extensively used operations, e.g., in computer graphics, computer games, and in geoinformatics~\cite{Preparata1985}. For different geometric representations, various PMC algorithms exist. For CSG models, a point is classified against all the underlying primitives and the resulting Boolean expressions (see \cite{Wassermann2017}). Ray casting~\cite{Preparata1985} is often used for boundary representation models. Further variants are approximation-tree-based algorithms ~\cite{Zalik2001}, point cloud methods~\cite{Sitek2006}, sign of offset~\cite{Taylor1994}, and the swath method ~\cite{Salomon1978}. As the space-tree based approximation and the ray-casting are needed in the following, these aspects will be explained in more detail:

\begin{itemize}
\item \textit{Ray casting:} The ray casting method is an efficient and suitable algorithm for general polytopes, and it is extensively used in computational graphics, e.g., for depth maps. To classify a given point with respect to a geometric model, a ray is shot in an arbitrary direction and the intersections with the boundary are counted. The parity (even, or odd) of intersections then provides information on whether the point lies inside or outside. For flawless models, ray-casting is accurate. For flawed CAD models, however, ray casting delivers no reliable statement about the point's domain membership, as almost all flaws influence the parity of intersections

\item \textit{Space-tree based PMC:} For the tree-based PMC, the domain is discretized by a space-tree $T\!R_{int}$, with leaves $c_{geo}$. Leaves intersected by the surface are marked as cut. Subsequently, a flood-fill algorithm is applied to the leaves $c_{geo}$. Starting from a seed point, whose domain membership is known, all connected leaves are marked as inside or outside, respectively. A challenge in this methodology is posed only by undesired openings or unintentional gaps. In these cases, a too fine approximation with leaves smaller than the size of the flaws would cause the flood-fill algorithm to mark the entire domain as inside or outside. Furthermore, despite its robustness against most flaws, the octree $T\!R_{int}$ gives only a coarse step-wise approximation of the geometry.

\end{itemize}

\subsection{General approach for flawed models}
The presented PMC method combines the robustness of space-tree approximation with the accuracy of ray-casting. The general approach works as follows:

\begin{enumerate}
\item The CAD model is approximated by a watertight space-tree $T\!R_{geo}$. Watertightness is imperative to ensure that the subsequent flood-fill can distinguish between inside and outside.

\item A flood fill algorithm is applied on $T\!R_{geo}$ to mark all connected points as inside and outside, respectively. This yields a filled space tree $\widehat{T\!R}_{geo}$. Remark: For all points that are not on cut leaves, the approximation tree $\widehat{T\!R}_{geo}$ can be used as fast, efficient, and accurate PMC.

\item An additional ray-casting is only carried out for points lying inside the cut boundary leaves -- in order to approximate the structure more precisely.
\end{enumerate}

Step 1 and Step 3 will now be described in more detail. For a description of the well-known flood fill algorithm in step 2 we refer to, e.g., \cite{Foley1997}.  

\subsection{Watertight space tree approximation} \label{sec:SurfaceRecon}
To ensure that the approximation space-tree $\widehat{T\!R}_{geo}$ is watertight, the size of the smallest leafs $d_{c_{geo}}$ must not undercut the characteristic size of the largest gap/opening $\varepsilon_{gap}$.
\begin{equation}
	d_{c_{geo}} > \varepsilon_{gap} \label{eq:maximumCellSize}
\end{equation}

$\varepsilon_{gap}$ is typically not known apriori and is determined by an iterative decrease of the cell size, until the subsequent fill algorithm fills the entire domain. 

From this, it follows that the maximal partitioning depth $n_{max}$ of $\widehat{T\!R}_{geo}$ is bounded by the ratio of domain size $d_{domain}$ of the tree $\widehat{T\!R}_{geo}$ to the dimension of the gaps/openings $\varepsilon_{gap}$:
\begin{equation}
	n_{max} < \mathrm{log}_2\left(\frac{d_{domain}}{\varepsilon_{gap}}\right)
\end{equation}
This limitation might allow, depending on the size of the gaps/openings only a very coarse approximation of the true geometry (see Figure~\ref{fig:failedFloodfill}). Concerning all other types of considered flaws, a test using the space tree $\widehat{T\!R}_{geo}$ is robust. Note that the reconstruction tree can be set up for an arbitrary flaw size $\varepsilon_{gap}$, as long as at least one inner cell can be detected.  The quality of the result will then only be dependent on the secondary PMC test (see Section~\ref{sec:PMC}).

Note that, generally, the space-trees $T\!R_{int}$ and $\widehat{T\!R}_{geo}$ are distinct. While $T\!R_{int}$ is constructed in order to numerically integrate the discontinuous element matrices for finite cells (see Section~\ref{sec:FCM}), the purpose of $\widehat{T\!R}_{geo}$ is merely to support the Point Membership Classification of the integration points. 

\begin{figure}[H]
	\centering
        \includegraphics[width=0.6\textwidth]{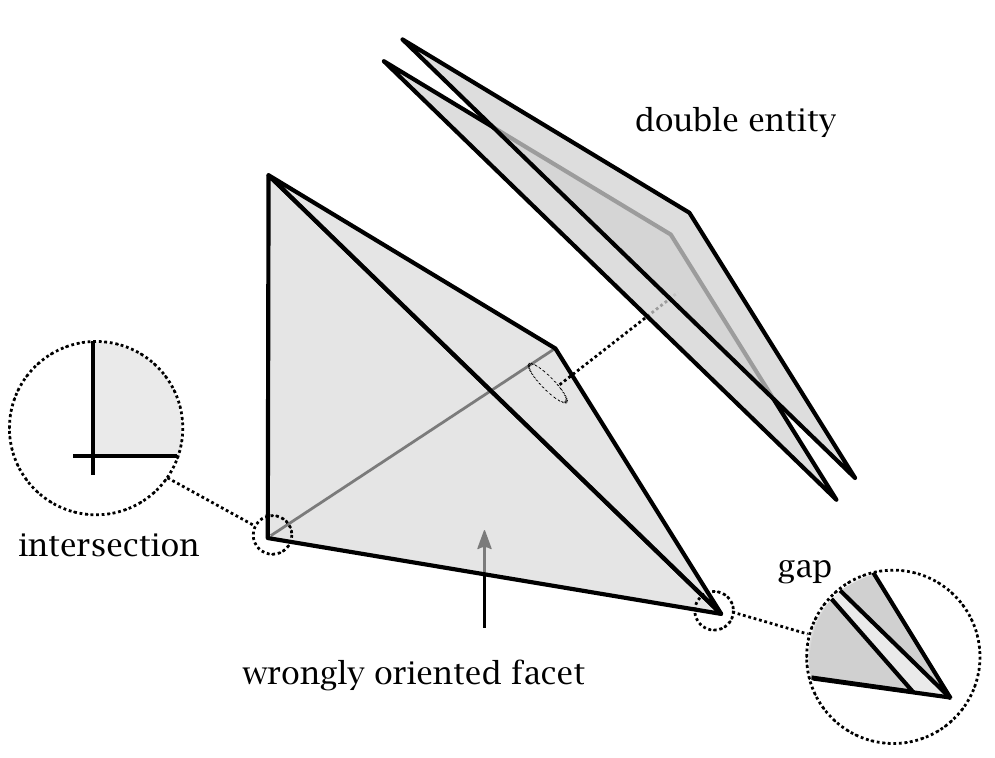}
    \caption{Example of an STL model with typical flaws.}
	\label{fig:SimpleExample}
\end{figure}

\begin{figure}[H]
	\centering
	\subfloat[Part of a 3D octree approximation]{
		\includegraphics[width=0.45\textwidth]{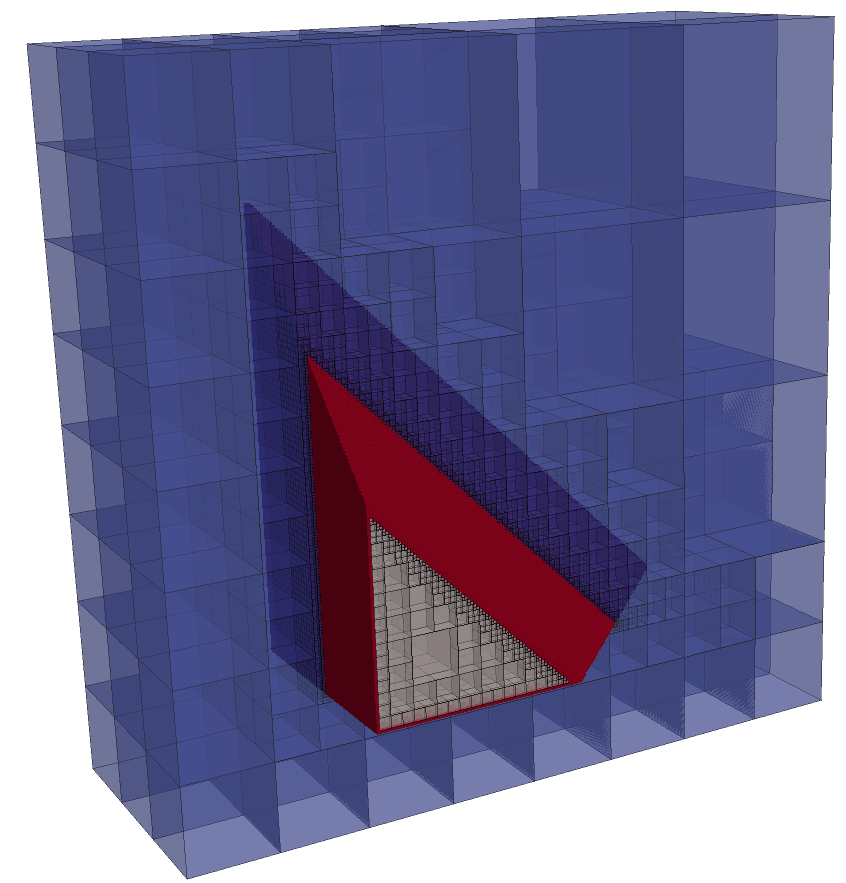}}
		\quad
	\subfloat[2D slice]{
        \includegraphics[width=0.45\textwidth]{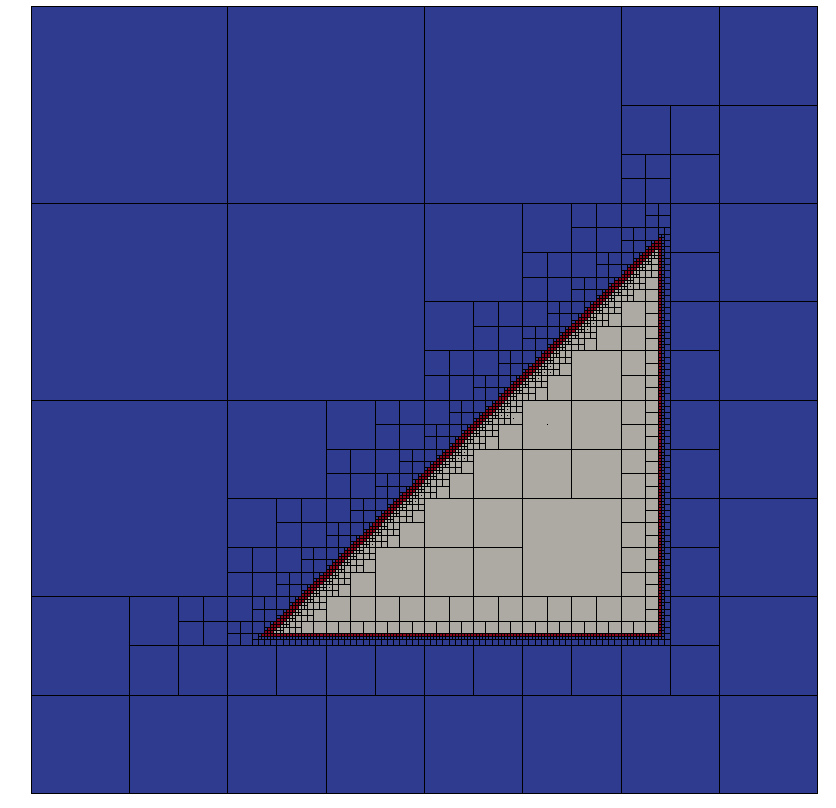}}
    \caption{Octree approximation of the embedded tetrahedral domain. The outer domain (blue) is separated by the cut leaves (red) from the inner domain (gray). The subdivision level is $n_{max}=7$.}
	\label{fig:FloodFillResult}
\end{figure}

After the surface is approximated by the space tree, the flood fill algorithm~\citep{Foley1997} can be applied to mark connected regions.
Figure~\ref{fig:FloodFillResult} shows the octree approximation of a simple example (Fig.~\ref{fig:SimpleExample}), which has several typical flaws. The size of the opening $\varepsilon_{gap}$ allows a maximum subdivision level of $n_{max} = 7$. Hence, the ratio of the largest gap to overall size is in the range of:
\begin{equation}
	\frac{1}{256} < \frac{\varepsilon_{gap}}{d_{domain}} < \frac{1}{128} \quad. 
\end{equation} 

\subsection{Point Membership Classification on cut leaves} \label{sec:PMC}
\begin{figure}[H]
  \centering
	\subfloat[Flawless model]{
	\includegraphics[width=0.3\textwidth]{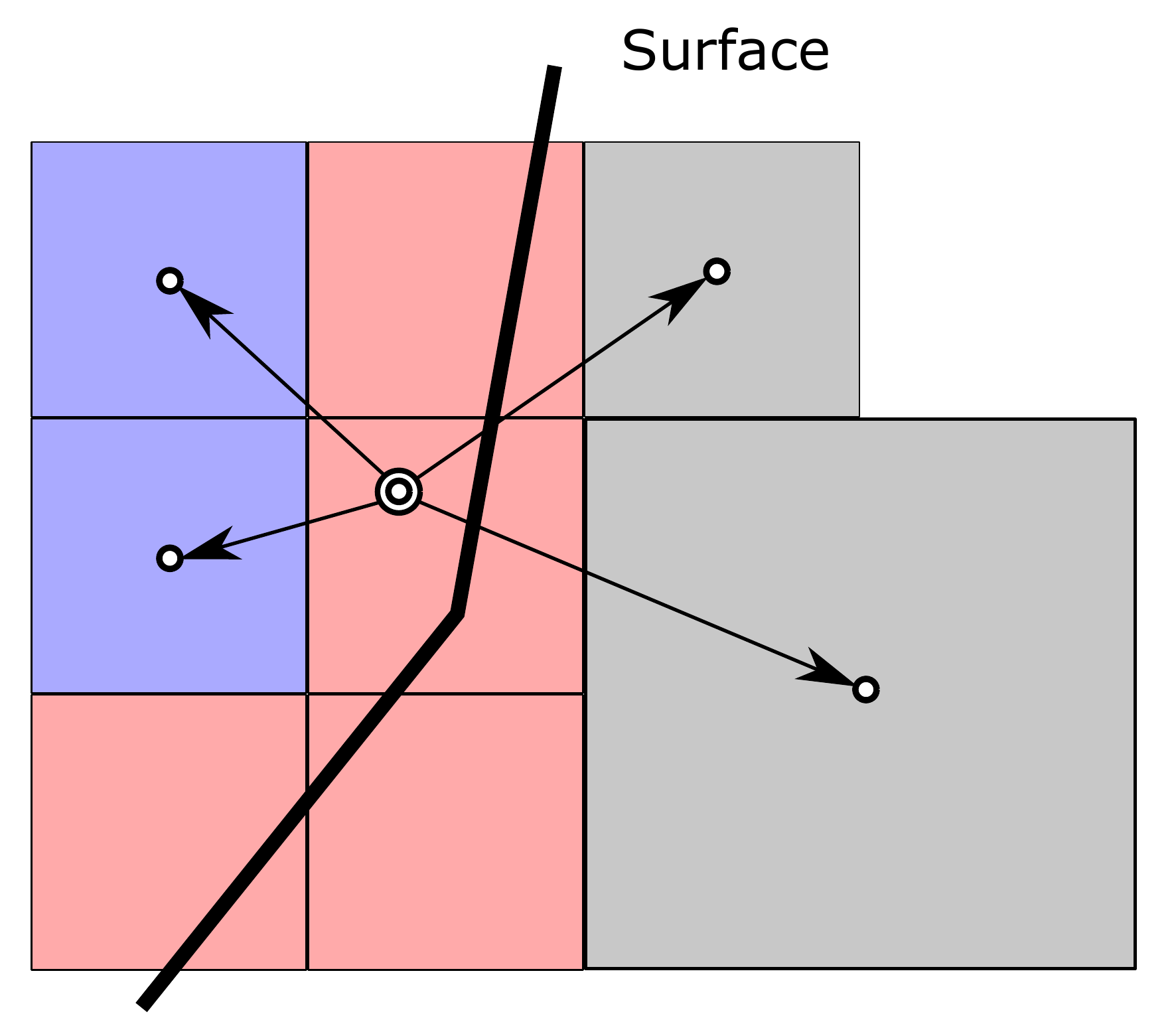}}
	\quad
	\subfloat[Gap, opening, missing entity]{
	\includegraphics[width=0.3\textwidth]{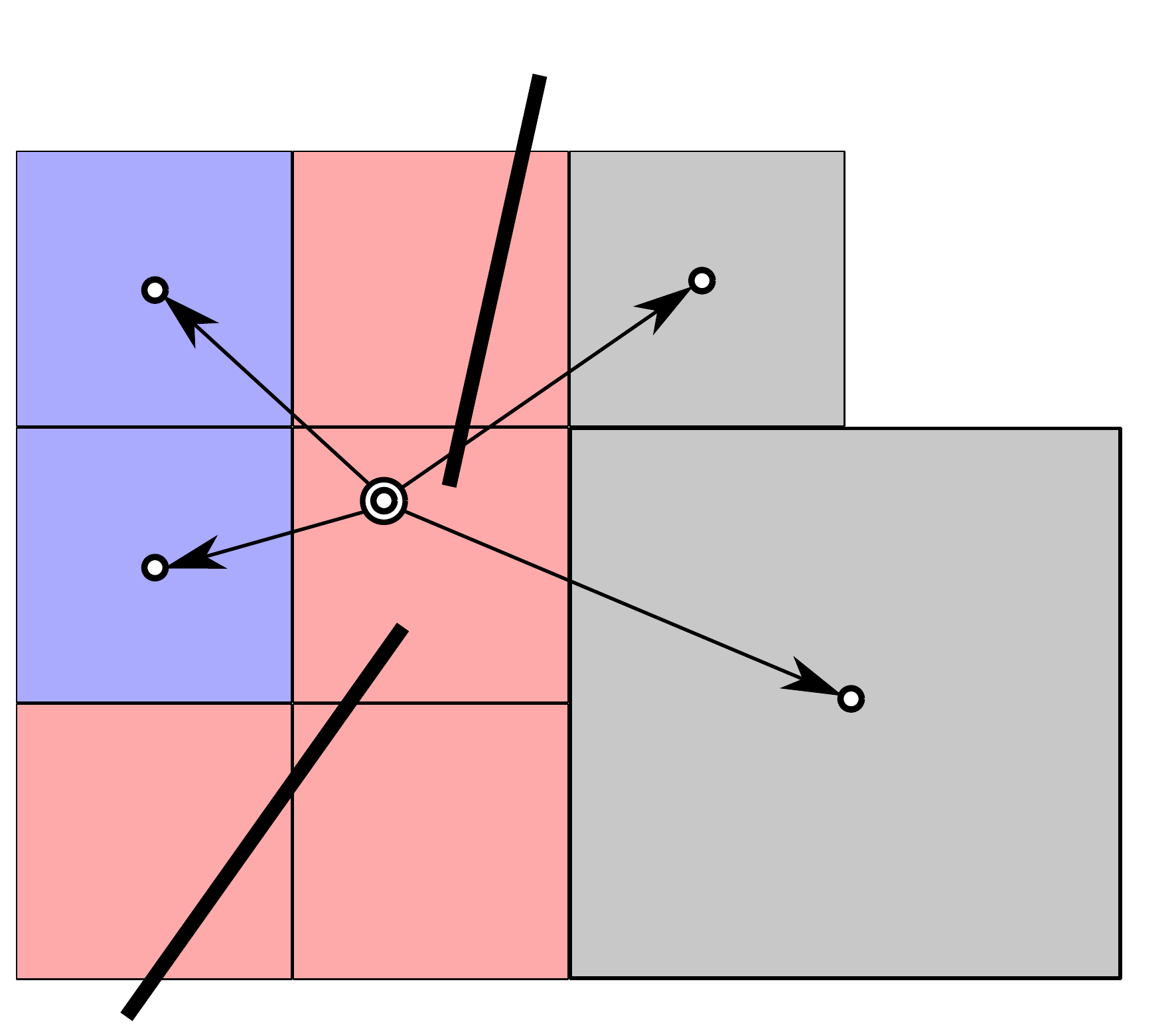}}
	\quad
	\subfloat[Double entity]{
	\includegraphics[width=0.3\textwidth]{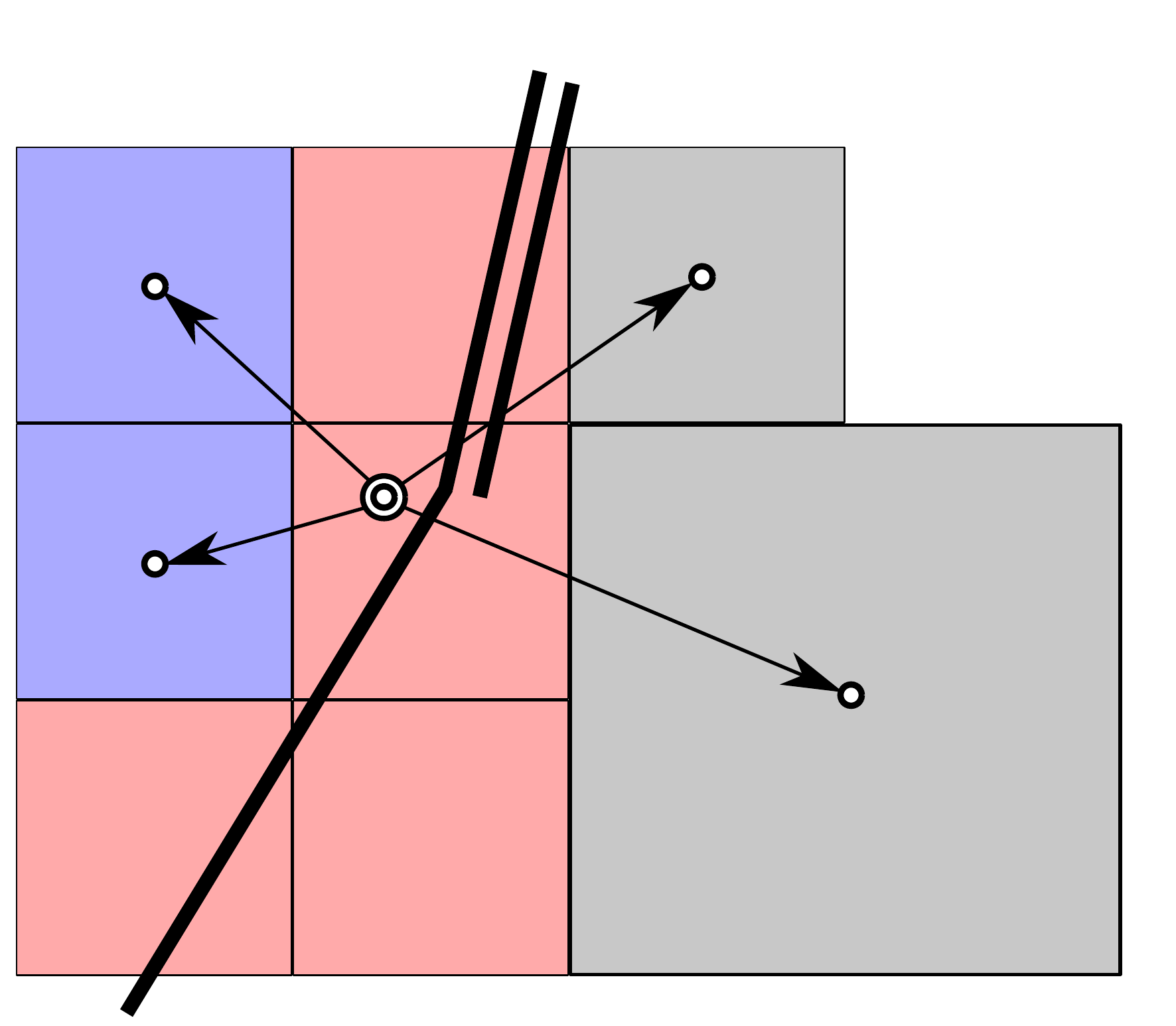}}
	\quad
	\subfloat[Several double entities]{
	\includegraphics[width=0.3\textwidth]{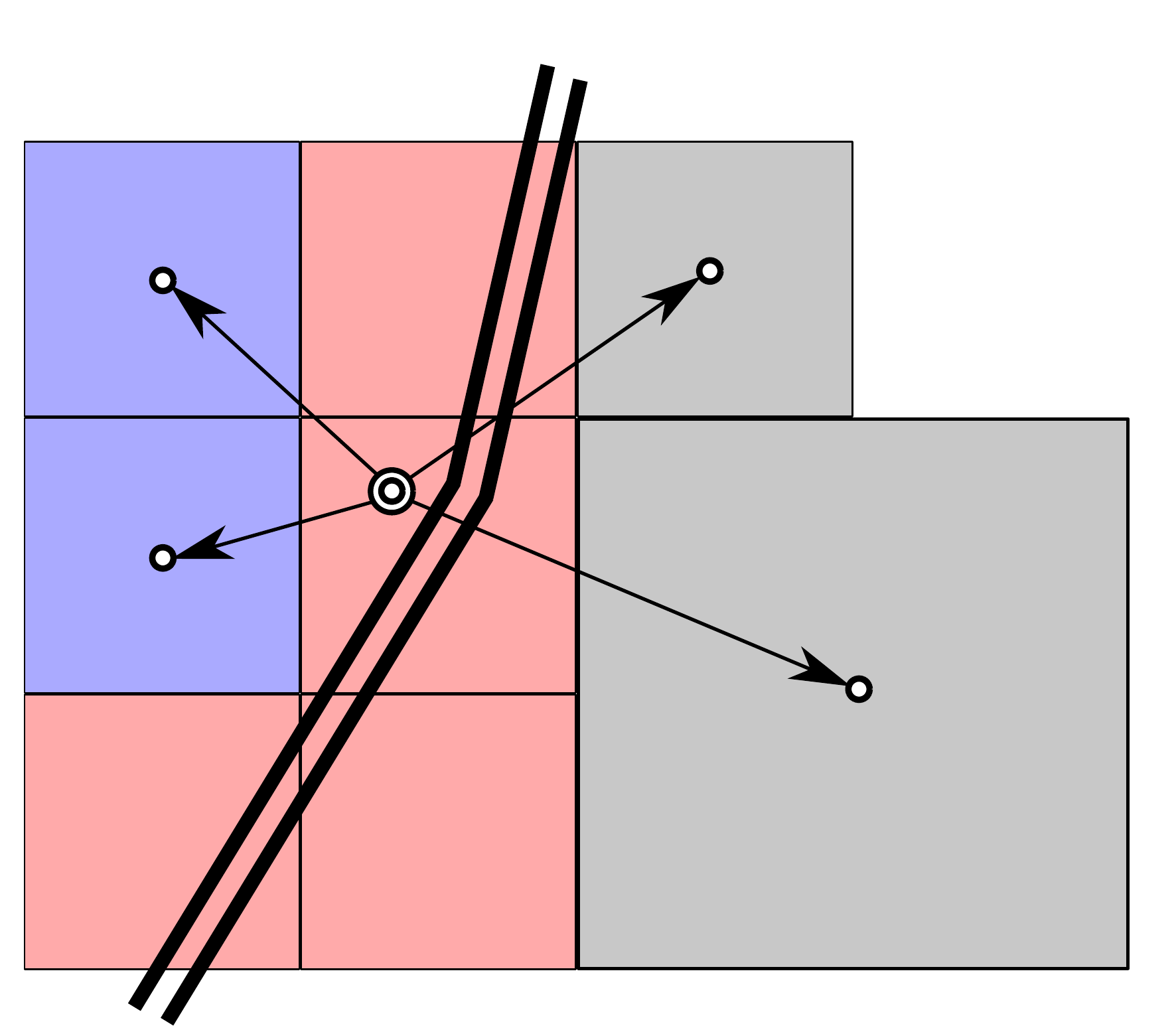}}
	\quad
	\subfloat[Spurious entity]{
	\includegraphics[width=0.3\textwidth]{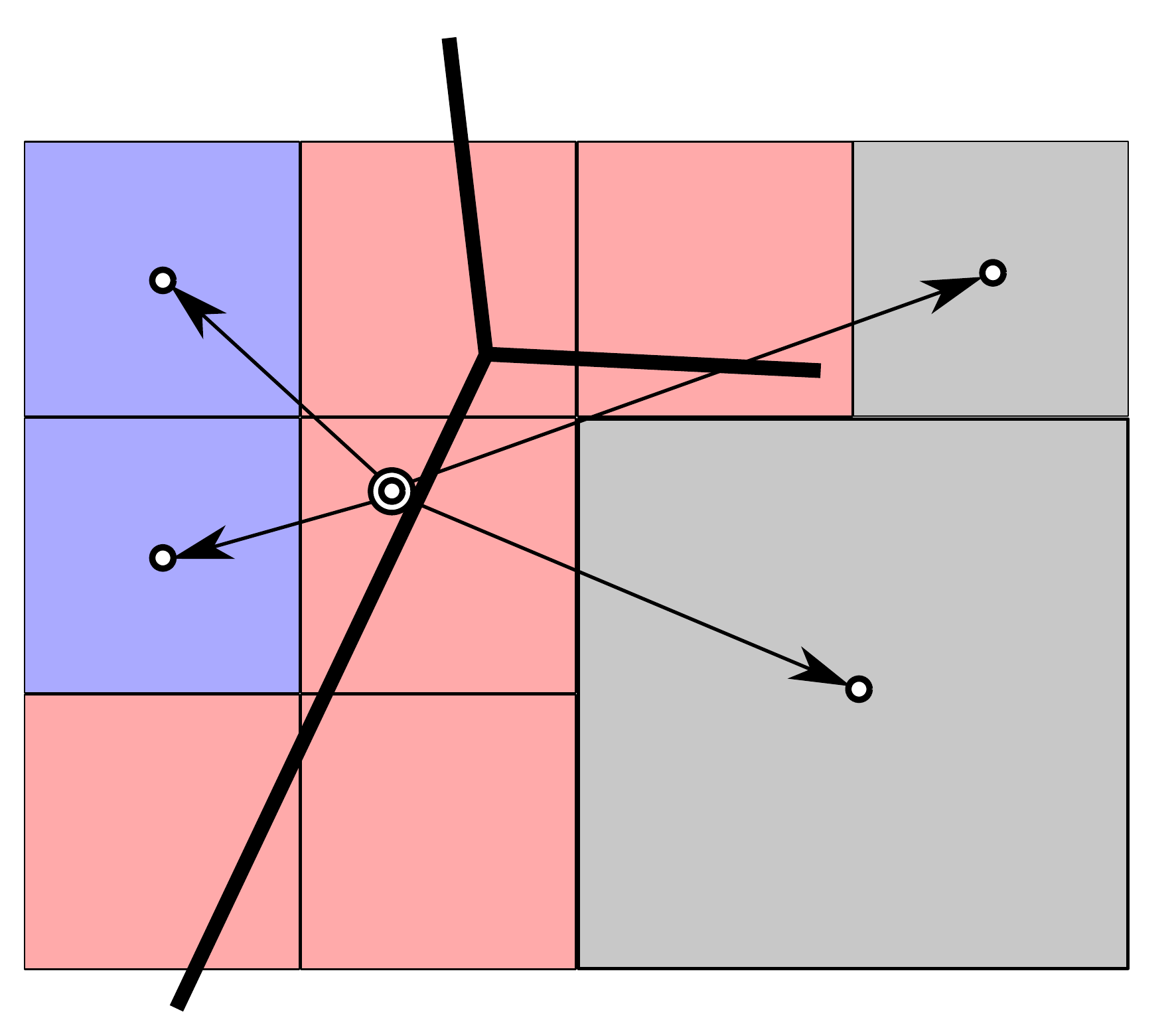}}		
	\quad
	\subfloat[Intersection]{
	\includegraphics[width=0.3\textwidth]{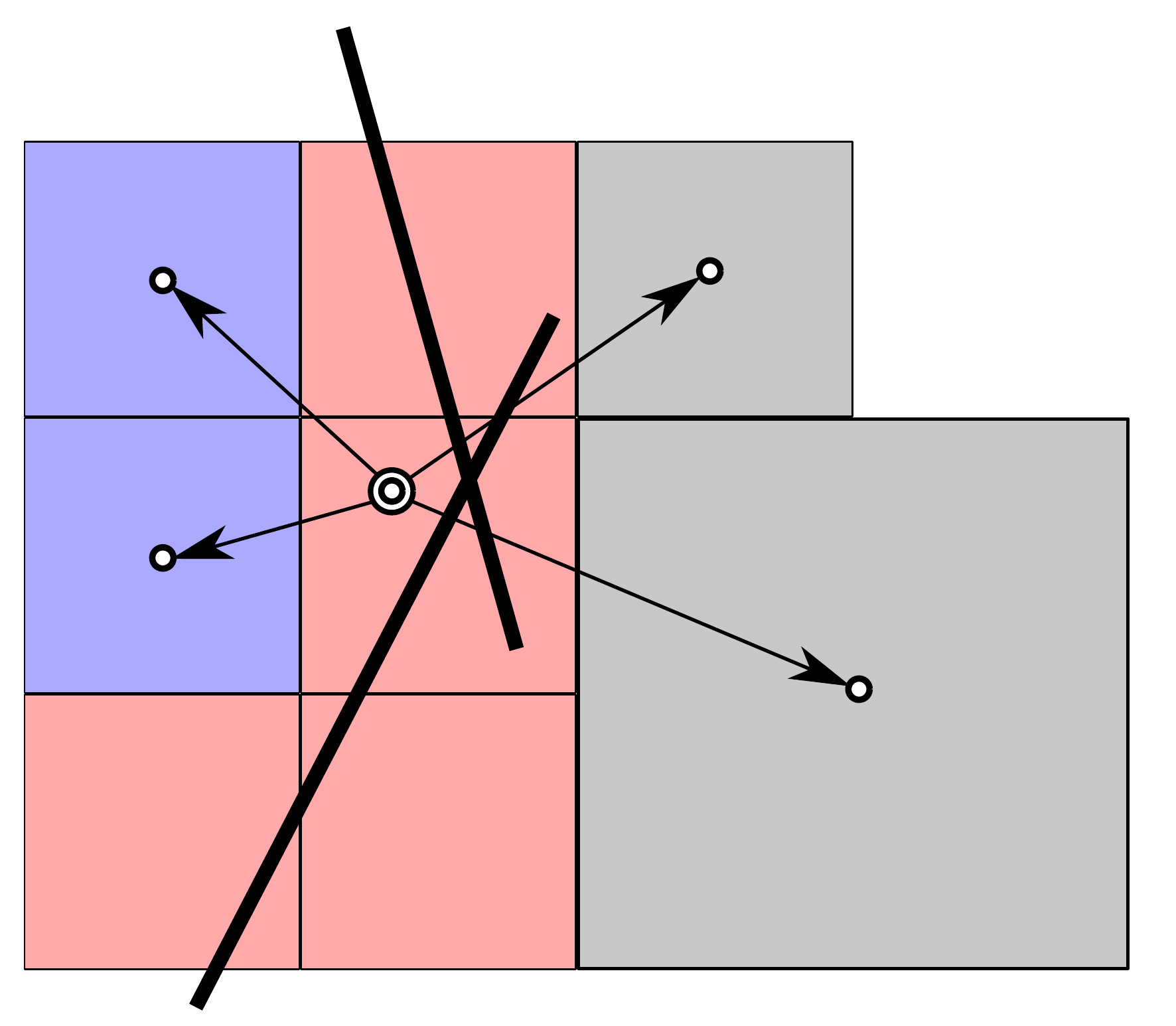}}
	\caption{Multiple ray casting for different flaws (red: cut leafs, gray: inside, blue: outside). In these examples d) and f) would lead to indifferent results.}
	\label{fig:RayCasting}
\end{figure}

The space-tree $\widehat{T\!R}_{geo}$ represents the surface only very roughly and, thus, cannot be used for a precise numerical analysis. Hence, in order to improve the representation of the boundary, an additional PMC using ray casting is carried out on cut leaves. Let us first assume that the model is flawless (see Fig.~\ref{fig:RayCasting}a). Then, the ray test for any integration point in an integration leaf $c_{int}$ yields a unique result without ambiguity, independent of the direction of the ray. In case of a flawed surface, the result may be ambiguous, depending on the selected direction of the ray (Figs.~\ref{fig:RayCasting}b-f). To handle this problem, we test rays in different directions, more precisely to the midpoints of all neighboring non-cut cells, which restricts the intersection tests to be carried out in the vicinity of the integration point and guarantees that various directions are queried. Hence, the probability for a correct result is increased. The PMC is then decided 'following the vote of the majority'. Clearly, this 'vote' can be wrong w.r.t. the (in general unknown) flawless model. This wrong decision results in an integration error for the computation of element matrices. In a mathematical sense, we are performing a 'variational crime' (see, e.g.,~\citep{Strang1973}). For geometrically small flaws, the smallness of this integration error can be readily assumed -- as, by construction of the two-stage PMC, it can only occur in the smallest leaf $c_{geo}$ cut by the surface.

We can even bound this error by bracketing, i.e. by solving the elasticity problem (23) -- once under the assumption that all ambiguous integration points are inside, and once assuming them outside of the domain of computation (see Section~\ref{sec:ParameterStudy} and Example~\ref{sec:Example1}), thus ensuring that the approximation quality of the method is not corrupted. 

\noindent Note that also other possibilities for the secondary PMC test can be applied, such as ray-casting in only a few, or just one direction, which will lead to a significant speedup but increases the probability of wrong results. Another possibility lies in the combination with a PMC test based on point clouds, as this test is sensitive to other types of flaws, such as wrongly oriented normals, or intersections.

\subsection{Parameter study on the influence of the gap size} \label{sec:ParameterStudy}
\begin{figure}[H]
  	\centering
  	\includegraphics[width=\textwidth]{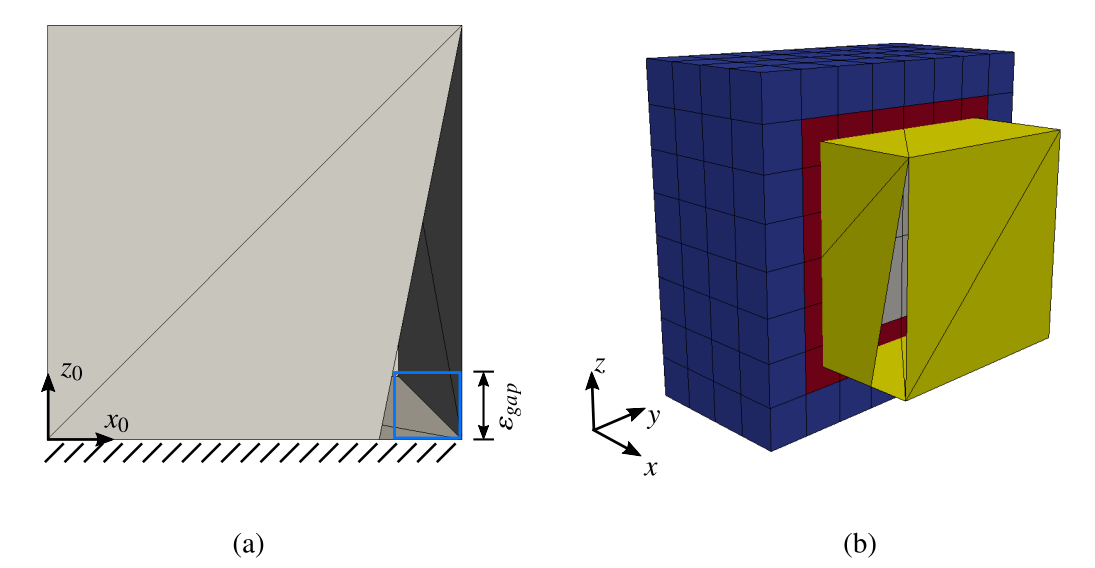}
  	\caption{Parameter study on a unit cube: a) Characteristic gap size $\varepsilon_{gap}$. b) Reconstruction tree on the flawed geometry.}
	\label{fig:IncreasingGap}
\end{figure}

As stated in Section~\ref{sec:ApplicationOperators} a flawed model has, in general, no mathematically valid solution. Therefore it is not possible to define an 'error' of the computed approximation w.r.t. an exact solution. Yet, in order to judge the quality we compare for a simple example energies of approximate and reference solution in dependence of the size of a flaw in the B-Rep model.  In particular, we investigate the influence of the largest gap size $\varepsilon_{gap}$ on the internal strain energy for a cube with the dimensions $1\times 1\times 1$ loaded under self-weight. The cube is clamped at the bottom. It is embedded in $9\times 9\times 9$ elements employing integrated Legendre polynomials of degree $p=3$. The B-Rep model of the cube consists of twelve triangles. One triangle is not properly connected to two of its neighbors resulting in a flawed model with a gap of characteristic size $\varepsilon^i_{gap}$ (see Figure~\ref{fig:IncreasingGap}). The size of the gap limits the maximum subdivision depth of the reconstruction tree, meaning that more refined trees would lead to a non-watertight boundary of the tree (see Section~\ref{sec:SurfaceRecon}). 

\begin{figure}[H]
  	\centering
  	\includegraphics[width=\textwidth]{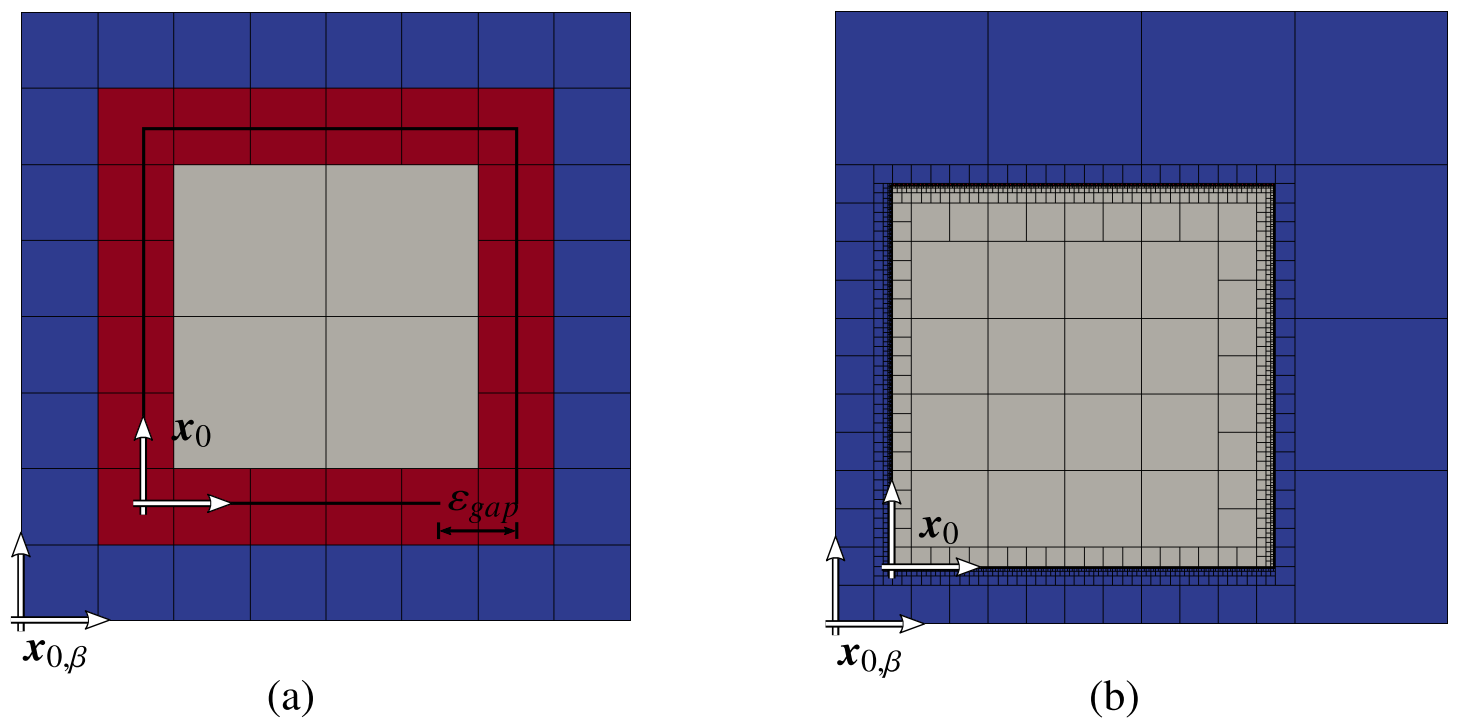}
  	\caption{
Cut through two reconstruction trees for different gap sizes, consequently maximum subdivision depths and for different origin positions: a) $\varepsilon_{gap}=0.2$, $n_{max}=3$, shift by $\beta=0$. b) $\varepsilon_{gap}=0.0031$, $n_{max}=9$, shift by $\beta=3$.}
	\label{fig:ReconstructionUnitCube}
\end{figure}

\noindent The embedding domain used for the reconstruction tree has the dimension $1.6 \times 1.6\times 1.6$. The quality of reconstruction not only depends on the depth of the tree but also on the relative position of the domain of computation (the cube) and the tree. This influence is studied by gradually 'shifting' the origin $x_{0,beta}$ of the cube along a diagonal in space:
\begin{equation}
\bm{x}_{0,\beta} =  \begin{bmatrix} -0.3 \\ -0.3 \\ -0.3 \end{bmatrix} + \beta	\cdot \begin{bmatrix} 0.05 \\ 0.05 \\ 0.05 \end{bmatrix} \quad ,
\end{equation} 

with $\beta = 0...3$. Figure~\ref{fig:ReconstructionUnitCube} shows two different reconstruction trees for different origin positions.

\begin{figure}[H]
  \centering
  	\subfloat[$\varepsilon_{gap}=0$, $n_{max}=\infty$]{
		\includegraphics[width=0.3\textwidth]{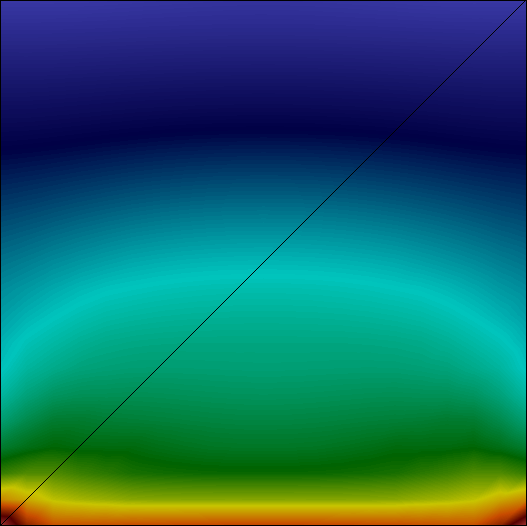}}
	\quad
	\subfloat[$\varepsilon_{gap}=0.025$, $n_{max}=6$]{
		\includegraphics[width=0.3\textwidth]{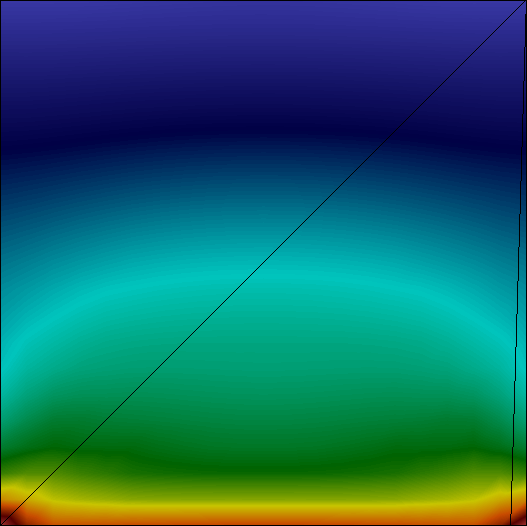}}
	\quad
	\subfloat[$\varepsilon_{gap}=0.2$, $n_{max}=3$]{
		\includegraphics[width=0.3\textwidth]{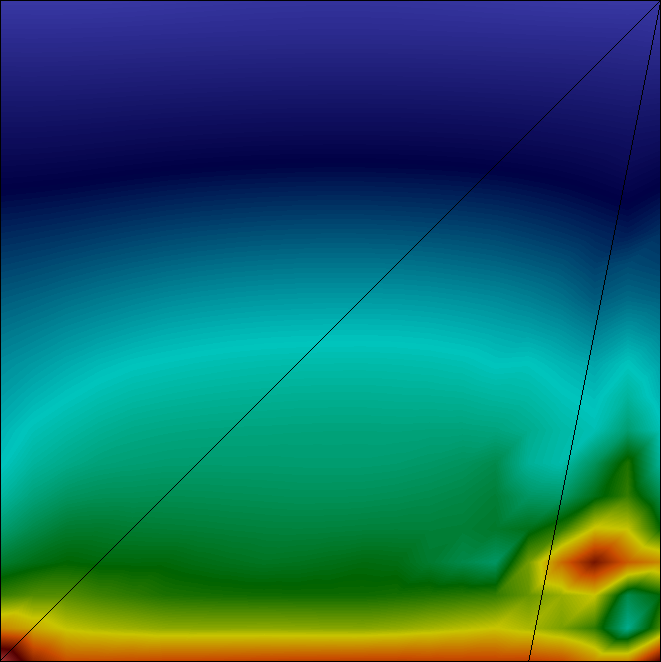}}
	\caption{Principal stresses for the flawless model (a) and gap sizes of $\varepsilon_{gap} = 2.5 \%$ (b) and $\varepsilon_{gap}  = 20 \%$ (c).}
	\label{fig:parameterStudyResults}
\end{figure}

\noindent Figure~\ref{fig:GapToErrorPMC} shows the influence of the characteristic size of the gap $\varepsilon_{gap}$ on the error in the internal energy. The abscissa depicts the characteristic size of the gap compared to the unit length of the cube in percent. The values correspond to the respective maximum subdivision depths $n_{max}^i = 9...3$ resulting from gap sizes $\varepsilon_{gap}^i = \frac{1.6}{2^{n_{max}^i}}$ from left to right. The ordinate shows the deviation of the internal strain energy $U$ to the reference energy $U_{ref}$ in percent. The reference energy $U_{ref}$ is computed on a flawless model.
Accurate results in energy are obtained even for large gap sizes of up to $20\%$ of the domain length. The quality of the solution is confirmed by Figure~\ref{fig:parameterStudyResults}, showing a plot of principle stresses of the reference solutions and approximate solutions for two gap sizes.

\pgfplotstableread{pmcPositiv.data}{\pmcPositivData}

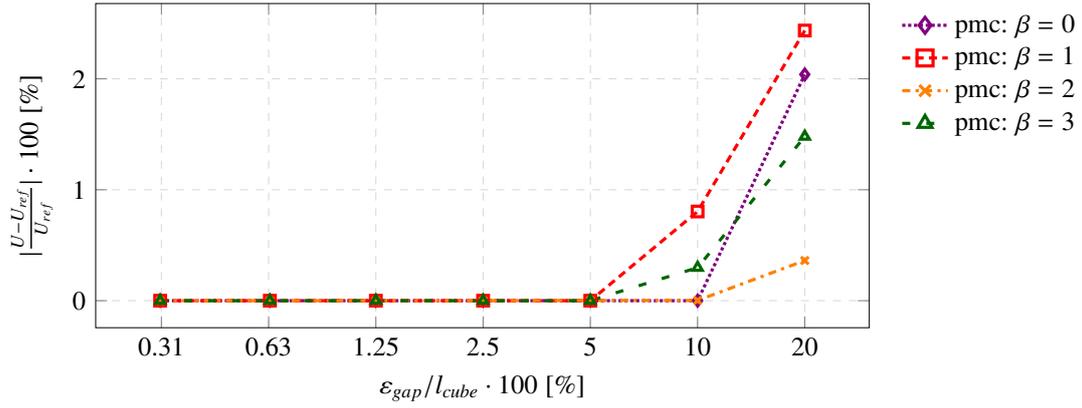
\begin{figure}[H]
  	\centering
	\begin{tikzpicture}
		\begin{axis}[ 
			width=0.8\textwidth,
			height=0.4\textwidth,
			xlabel={$\varepsilon_{gap} / l_{cube} \cdot 100\;[\%]$},
			ylabel={$\lvert \frac{U-U_{ref}}{U_{ref}}\rvert\cdot 100\;[\%]$},
			every axis plot/.append style={very thick},
			xmode=log,
			legend pos=outer north east,
			mark options={solid},
			grid=both,
       	 	grid style={dashed,gray!30},
       	 	xtick={0.3125,0.625,1.25,2.5,5,10,20},xticklabel=\pgfmathparse{exp(\tick)}\pgfmathprintnumber{\pgfmathresult}
 			]
			\addplot[color=violet,	densely	dotted,			mark=diamond	] table[x={gap}, y={b0}] 	{\pmcPositivData}; \addlegendentry{pmc: $\beta=0$}
			\addplot[color=red,	densely dashed,	mark=square		] table[x={gap}, y={b1}] 	{\pmcPositivData}; \addlegendentry{pmc: $\beta=1$}
			\addplot[color=orange,	dashdotted,		mark=x			] table[x={gap}, y={b2}] 	{\pmcPositivData}; \addlegendentry{pmc: $\beta=2$}
			\addplot[color=darkgreen,	loosely dashed, mark=triangle	] table[x={gap}, y={b3}] 	{\pmcPositivData}; \addlegendentry{pmc: $\beta=3$}
		\end{axis}
	\end{tikzpicture}
	\caption{Relative deviation of energies depending on the gap size for different positions of the cube.}	
	\label{fig:GapToErrorPMC}
\end{figure}

\noindent Although this study supports the quality of the presented approach, it cannot guarantee limitation of an error in energy of even of local solution quantities in general situations. They strongly depend on the complexity of the model and the amount and type of flaws. A crucial factor is also the location of the flaw. If it is located in highly stressed regions, the influence will be bigger than if it were located in regions of low stress. It remains to an engineer to judge the feasibility of the solution.
\section{Numerical examples} \label{sec:Examples}
To demonstrate the accuracy and robustness of the proposed approach, three examples are presented. The first simple example serves to verify the proposed method. To this end, a plate with a hole is simulated and compared to a flawless reference solution. The complex screw in the second example proves the applicability for sophisticated, defective CAD models. Again, a flawless reference model was available. The last example is an engine bracket taken directly from engineering practice. This model is a perfect example of a flawed geometry, as many NURBS-patches do not fit together. An attempt to mesh the model showed that 337.544 triangles had a free edge, i.e. are flawed.

\subsection{Example 1: Thick-walled plate with circular hole}\label{sec:Example1}
\begin{figure}[H]
  \centering
	\subfloat[Gap]{
	\includegraphics[width=0.3\textwidth]{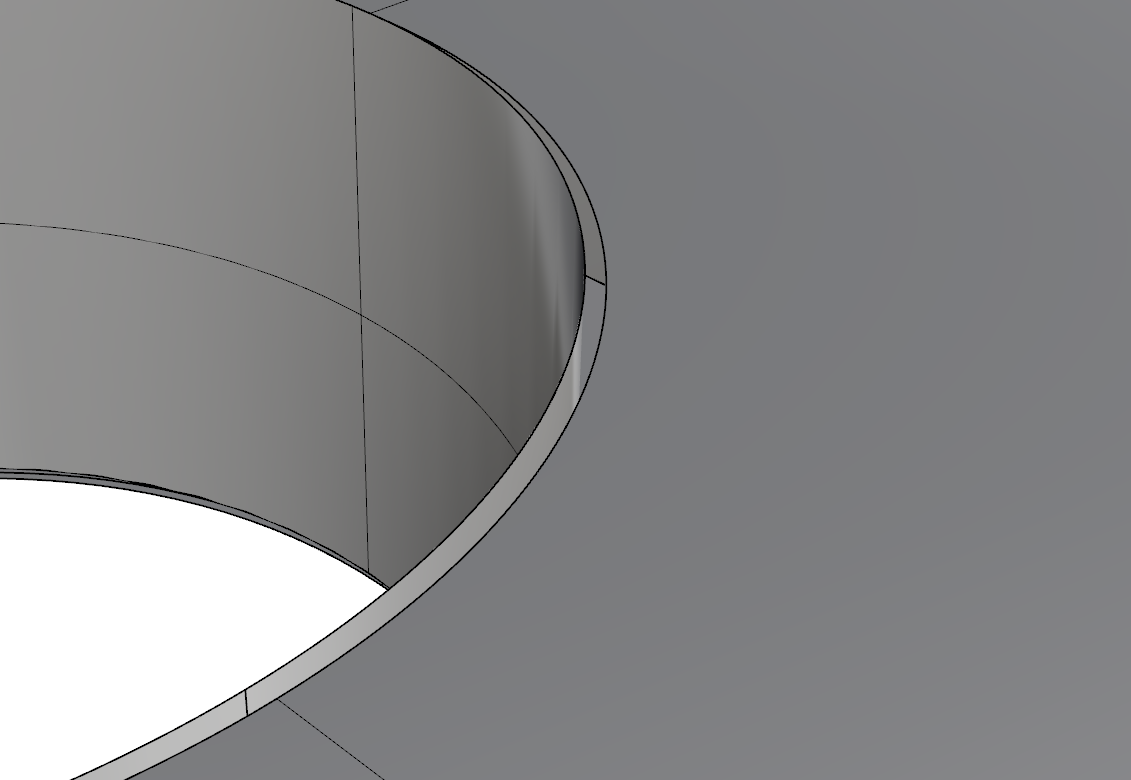}}
	\quad
	\subfloat[Intersection]{
	\includegraphics[width=0.3\textwidth]{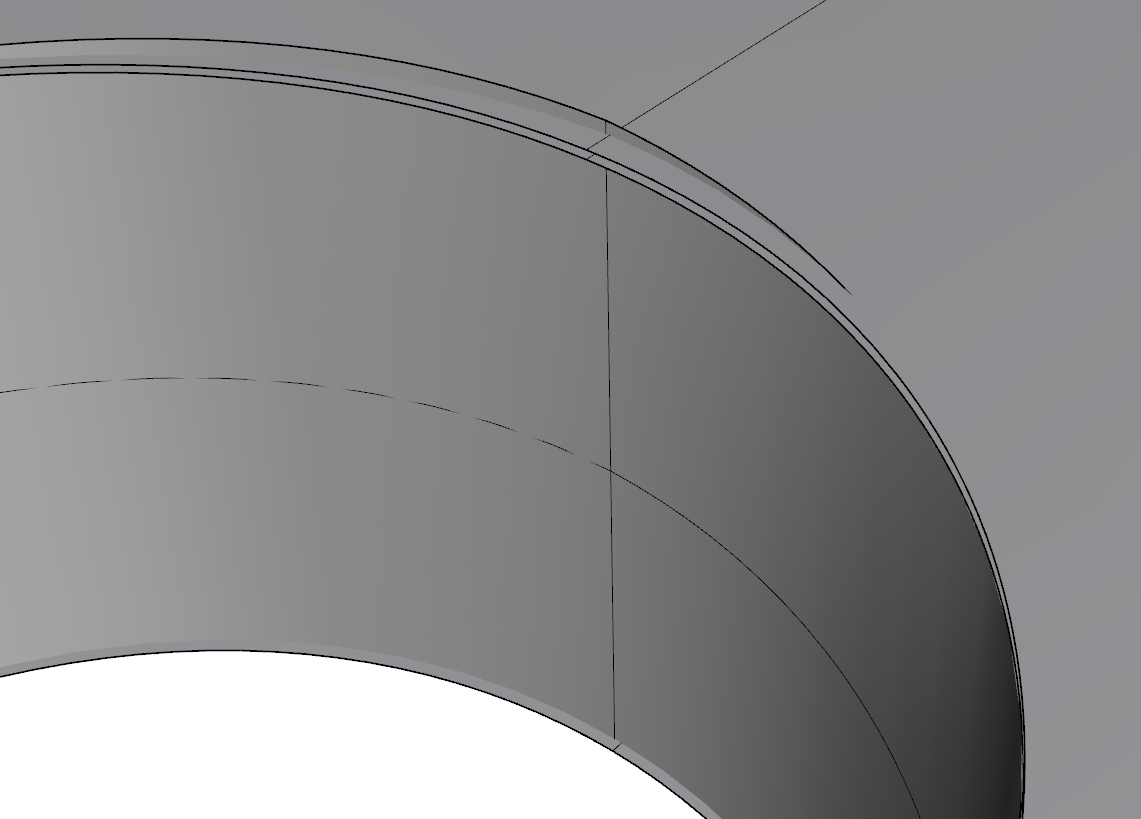}}
	\quad
	\subfloat[Offset]{
	\includegraphics[width=0.3\textwidth]{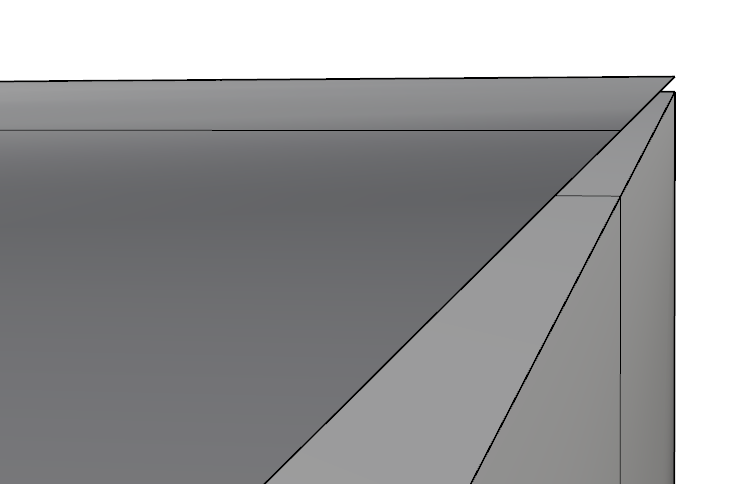}}
	\quad
	\caption{Flaw details: a) gap b) intersection c) offset of the top surface}
	\label{fig:DetailFlaws}
\end{figure}

As a classical benchmark for 3D problems, we choose the thick-walled plate with four circular holes~\cite{Parvizian2007}. The Young's modulus is set to $E = 10000.0 \,N/mm^2$ and the Poisson's ratio to $v = 0.30$. The plate is loaded with a surface traction $\overline{t}_{n} = 100.0 \,N/mm^2$. Symmetry boundary conditions are used, allowing to simulate only a quarter of the domain (see Fig.~\ref{fig:PlateWithHoleSketch}). The dimensions of the model are $b = h = 4.0 \,mm$ and $t = r = 1.0\, mm$. To show the robustness of the proposed method, several flaws -- namely intersections, gaps, double entities, and offsets -- are introduced on the surfaces (see Figs.~\ref{fig:FlawedPlateIso} and~\ref{fig:DetailFlaws}).

\begin{minipage}{.45\textwidth}
	\begin{figure}[H]
		\centering
  		\includegraphics[width=\textwidth]{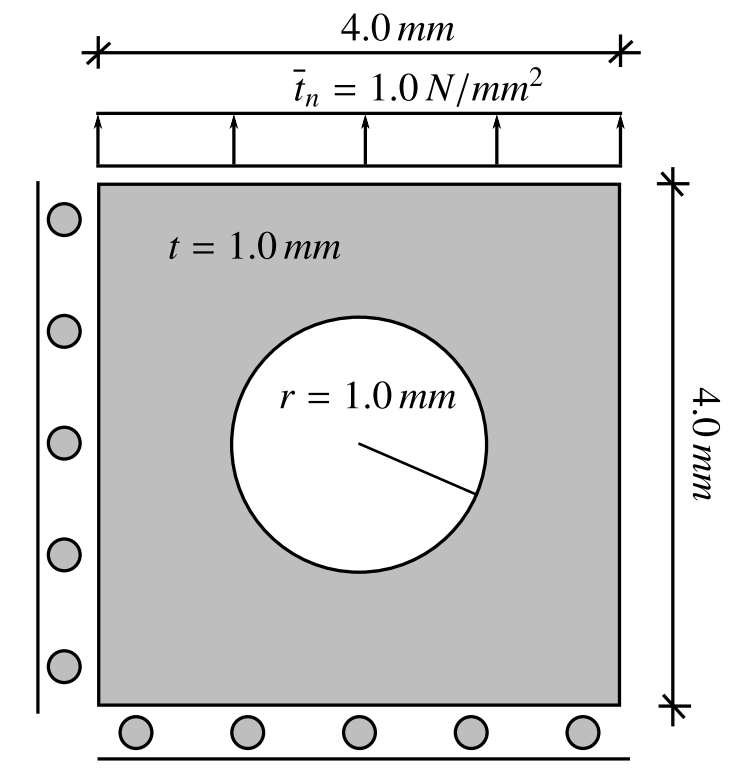}
		\caption{ Thick-walled plate with circular hole under surface load }
		\label{fig:PlateWithHoleSketch}	
	\end{figure}
\end{minipage}
\hfill
\begin{minipage}{.45\textwidth}
	\begin{figure}[H]
		\centering
		\includegraphics[width=\textwidth]{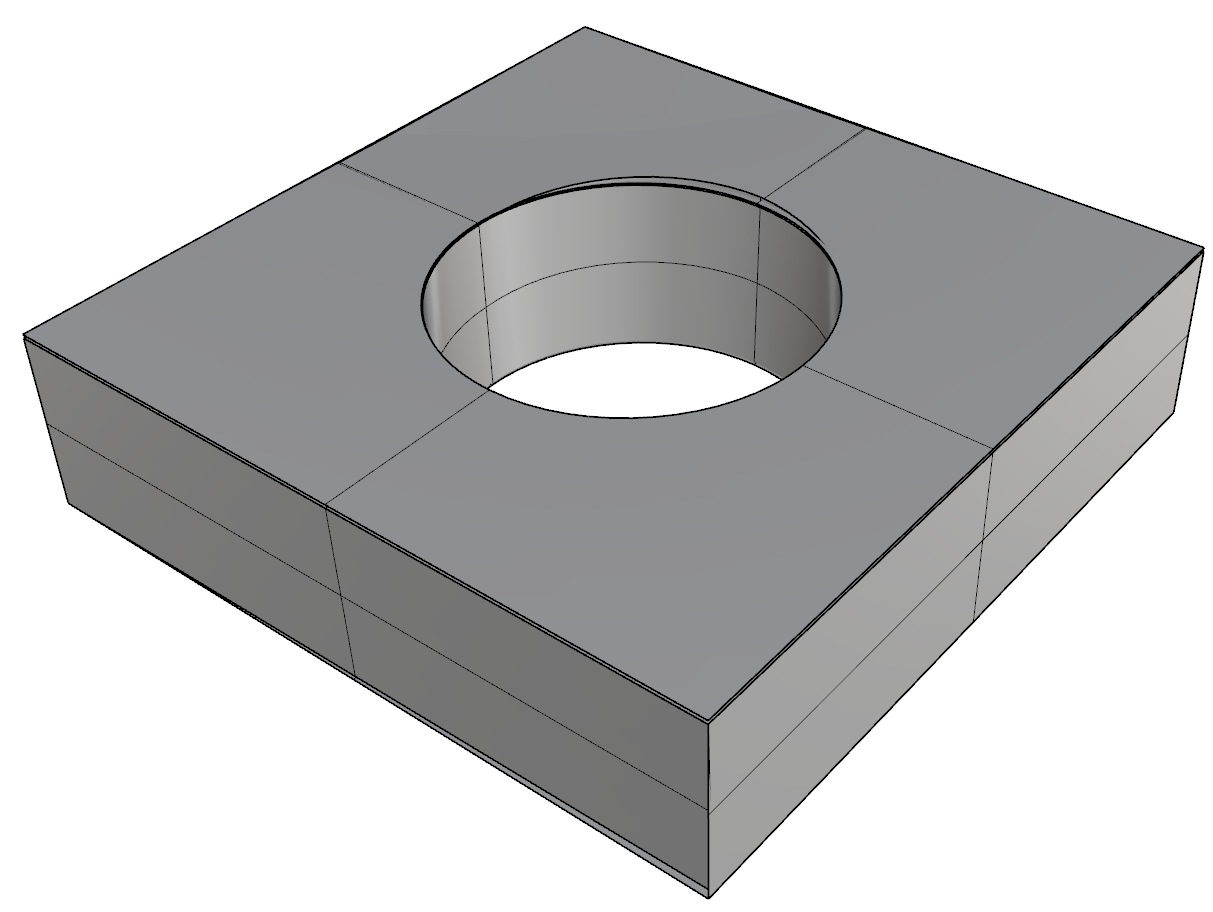}
		\caption{ Flawed B-Rep model containing several gaps, intersections, offsets, and multiple entities}
		\label{fig:FlawedPlateIso}
	\end{figure}
\end{minipage}

The domain is discretized into $10 \times 10 \times 1$ finite cells employing integrated Legendre polynomials as basis functions. The background grid and the qualitative displacement are depicted in Figure~\ref{fig:PlateWithHoleReconstruction}. A convergence study was carried out for $p$-refinement using $p = 1...6$. To measure the accuracy of the approach, the strain energy is computed and passed on to the reference solution $u_{ex}$, which was computed with an extensive boundary-conforming finite element analysis.
The minimal size of the cells $\hat{c}_{geo}$ of the geometric tree $\widehat{T\!R}_{geo}$ was limited by the size of the largest gap, allowing a maximum subdivision depth of $n_{max} = 5$. Note that the tree is also refined at the 'flat' surfaces of the plate. This results in $\sim$1.33 million cells, of which $\sim$1.15 million cells are located on the deepest level (see Fig.~\ref{fig:PlateWithHoleDisplacement}).\\
\begin{minipage}{.50\textwidth}
	\begin{figure}[H]
		\centering
		\includegraphics[width=0.9\textwidth]{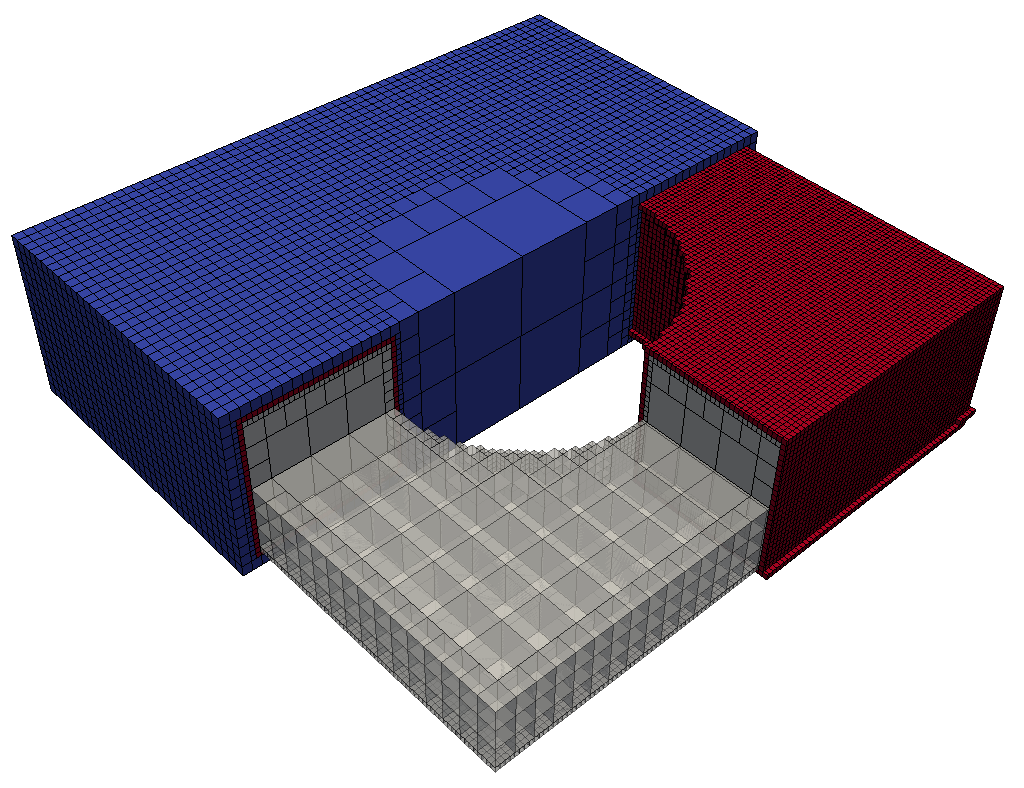}
		\caption{ Approximation tree $\widehat{T\!R}_{geo}$ with subdivision depth $n_{max} = 5$}
		\label{fig:PlateWithHoleReconstruction}	
	\end{figure}
\end{minipage}
\hfill
\begin{minipage}{.45\textwidth}
	\begin{figure}[H]
		\centering
		\includegraphics[width=0.8\textwidth]{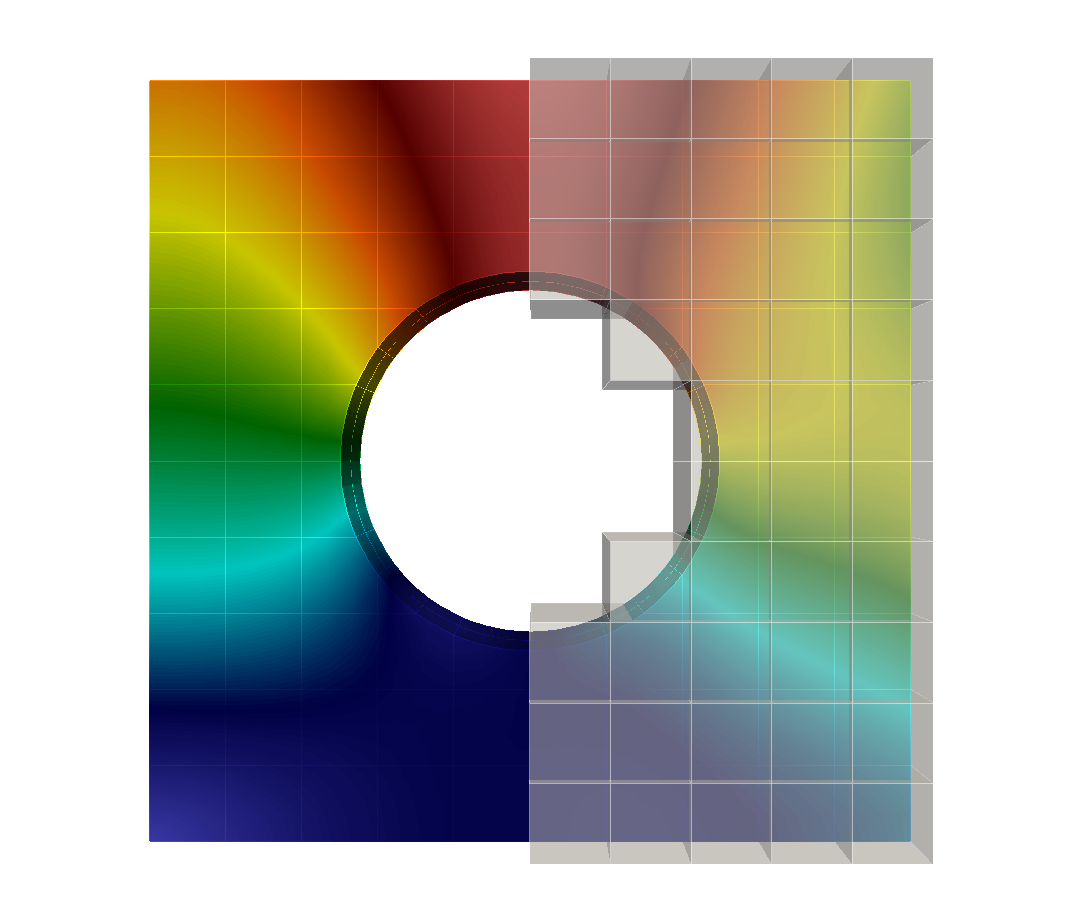}
		\caption{Displacement and finite cell discretization}
		\label{fig:PlateWithHoleDisplacement}
	\end{figure}
	
\end{minipage}
\vspace{0.5cm}

\begin{figure}[H]
	\centering
	\includegraphics[width=0.6\textwidth]{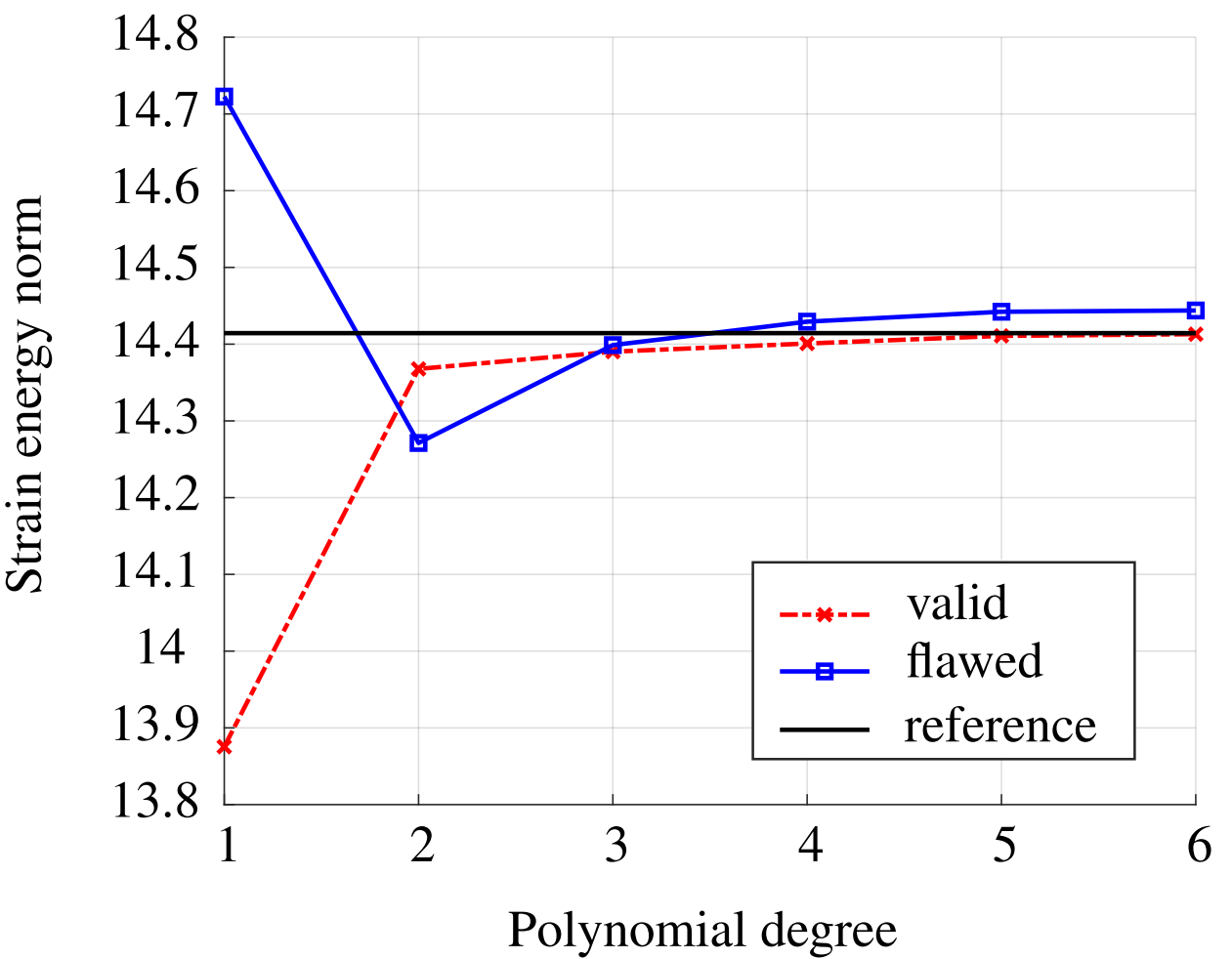}
	\caption{ Strain energy norm for polynomial degrees $p=1...6$}
	\label{fig:convergence}
\end{figure}

\Cref{fig:convergence} plots the strain energy for the different polynomial degrees. Note that the relative error in the strain energy can only be computed for the valid model, as this is the only possible basis to compute a reference solution.

\vspace{0.5cm}

\noindent In Figure~\ref{fig:convergence}, it can be seen, that both models converge to a slightly different value. This is, of course, to be expected -- as both models have a slightly different shape and volume. The good convergence of the flawed model is attributed to the fact that errors due to flaws are very localized. This is an inherent property of the proposed methodology. 

A detailed investigation of the flawed geometry can be carried out based on the ray-casting tests, which are applied on the integration cells (see Section~\ref{sec:PMC}). As an example, we consider a polynomial degree of $p=3$. For the integration of the system matrices, 6 406 920 points need to be evaluated. From these, a total number of 2 225 641 points ($\sim35\%$) are lying on cut cells. Typically, 12 to 18 ray-castings are carried out on each of these points. 1 503 636 points ($\sim$23\%) are ambiguous, i.e. at least one ray delivers a different result compared to the majority. In 656 009 cases ($\sim$10\%), a 'vote for the majority' is not possible, as the number of rays voting for inside and outside is equal. This large amount is mainly due to the many double entities. To compute the upper and lower boundaries of the energy norm, two additional simulations were carried out -- once with all ambiguous integration points counting as inside and once counting as outside. The energy norm for lower boundary was $\sim$0.4\% lower, and for the upper boundary $\sim$2.9\% larger compared to the simulation with 'vote for the majority'. Due to the fact that the error is restricted to the smallest geometrical leaves and the ray-casting errors occur only in the vicinity of the flaws, the deviation in the strain energy norm is rather small.

\subsection{Example 2: Screw}
\begin{figure}[H]
	\centering
	\subfloat[STL model]{
		\includegraphics[width=0.6\textwidth]{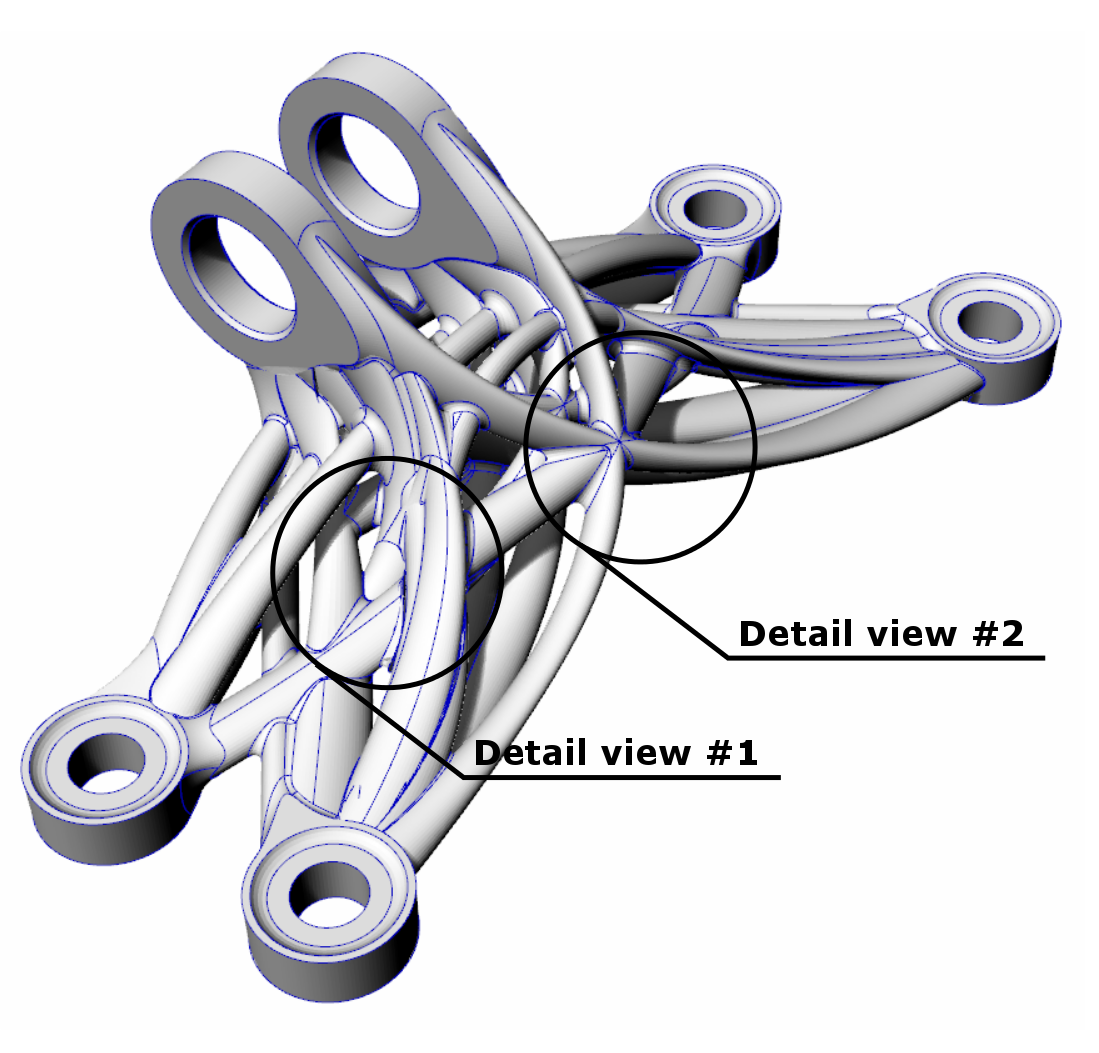}}
		\quad
	\subfloat[Details]{
        \includegraphics[width=0.35\textwidth]{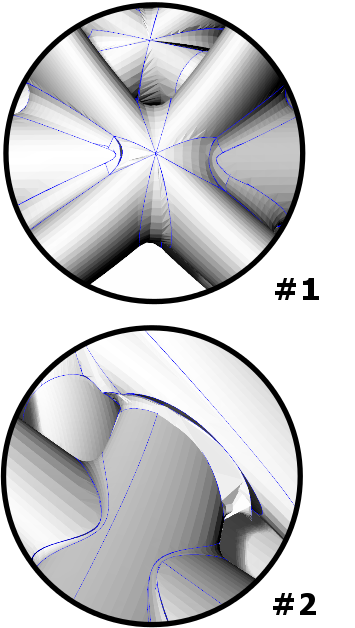}}
    \caption{ Blue lines denote open edges: In many cases the free edges are fairly close. As can be seen in the detailed views, however, some of the gaps and openings are quite large. }
	\label{fig:Flaws}
\end{figure}

This example demonstrates how the algorithm performs for a more complex geometry. To this end, we consider the potentially flawed CAD model of a screw, depicted in Fig.~\ref{fig:flaws}. The simulation was carried out on $10 \times 30 \times 10$ finite cells using trivariate B-Splines of polynomial degree $p = 3$ and the open knot vector  $U = [0,\, 0,\, 0,\, 0,\; 1,\; 2,\; 3,\, 3,\, 3,\, 3]$. The partitioning depth for the integration of cut FCM cells was set to $k = 3$. The tip surface was loaded with a constant pressure, and the bottom surface was clamped. At the top, several flaws were introduced, resulting in gaps, overlaps, and intersections. In the detailed view in Fig.~\ref{fig:Flaws}, the free edges (i.e. edges which have only one adjoined face) are highlighted in blue.

\begin{figure}[H]
	\centering
	\subfloat[$n_{max} = 5$]{
		\includegraphics[width=0.46\textwidth]{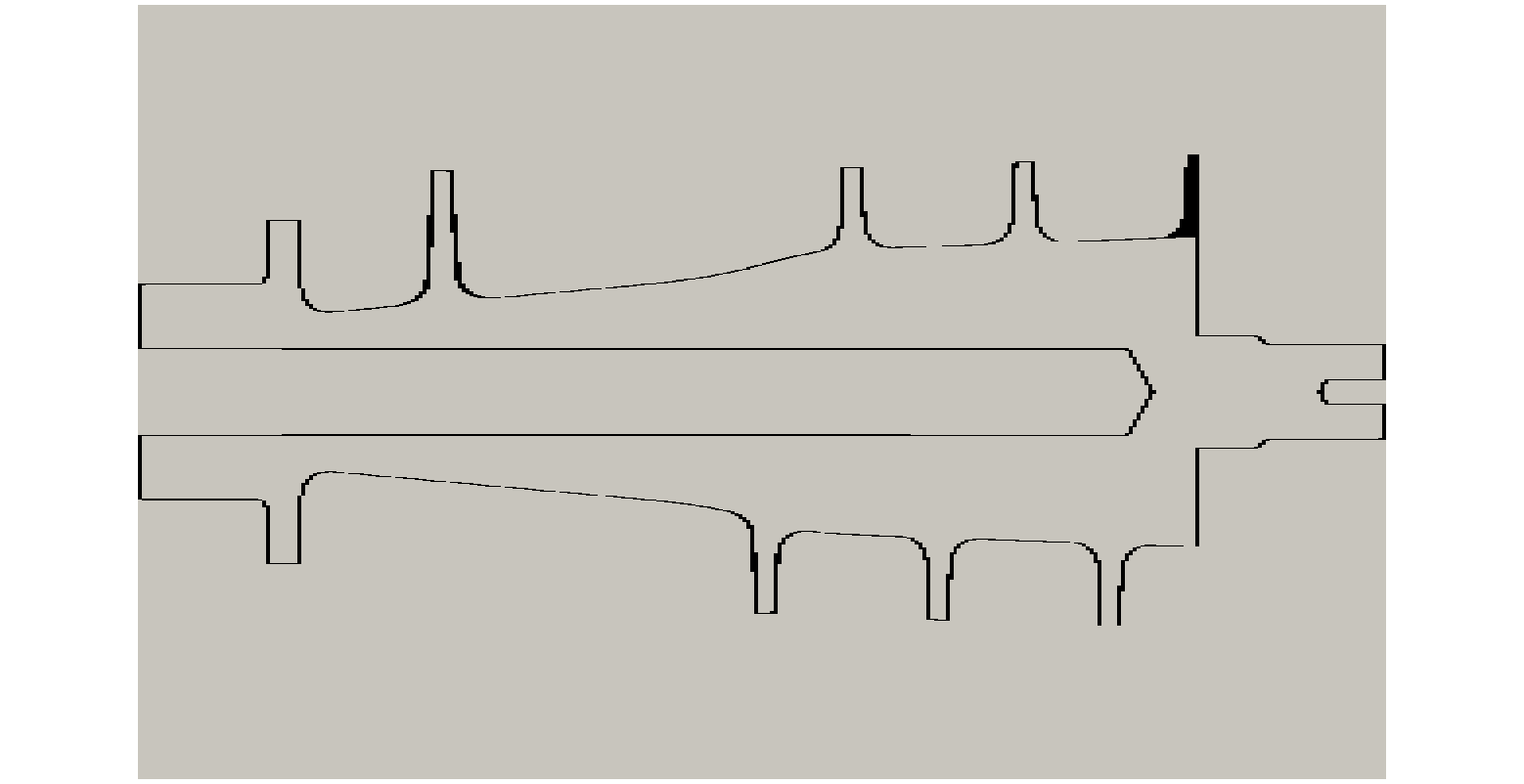}}
		\quad
	\subfloat[$n_{max} = 4$]{
		\includegraphics[width=0.46\textwidth]{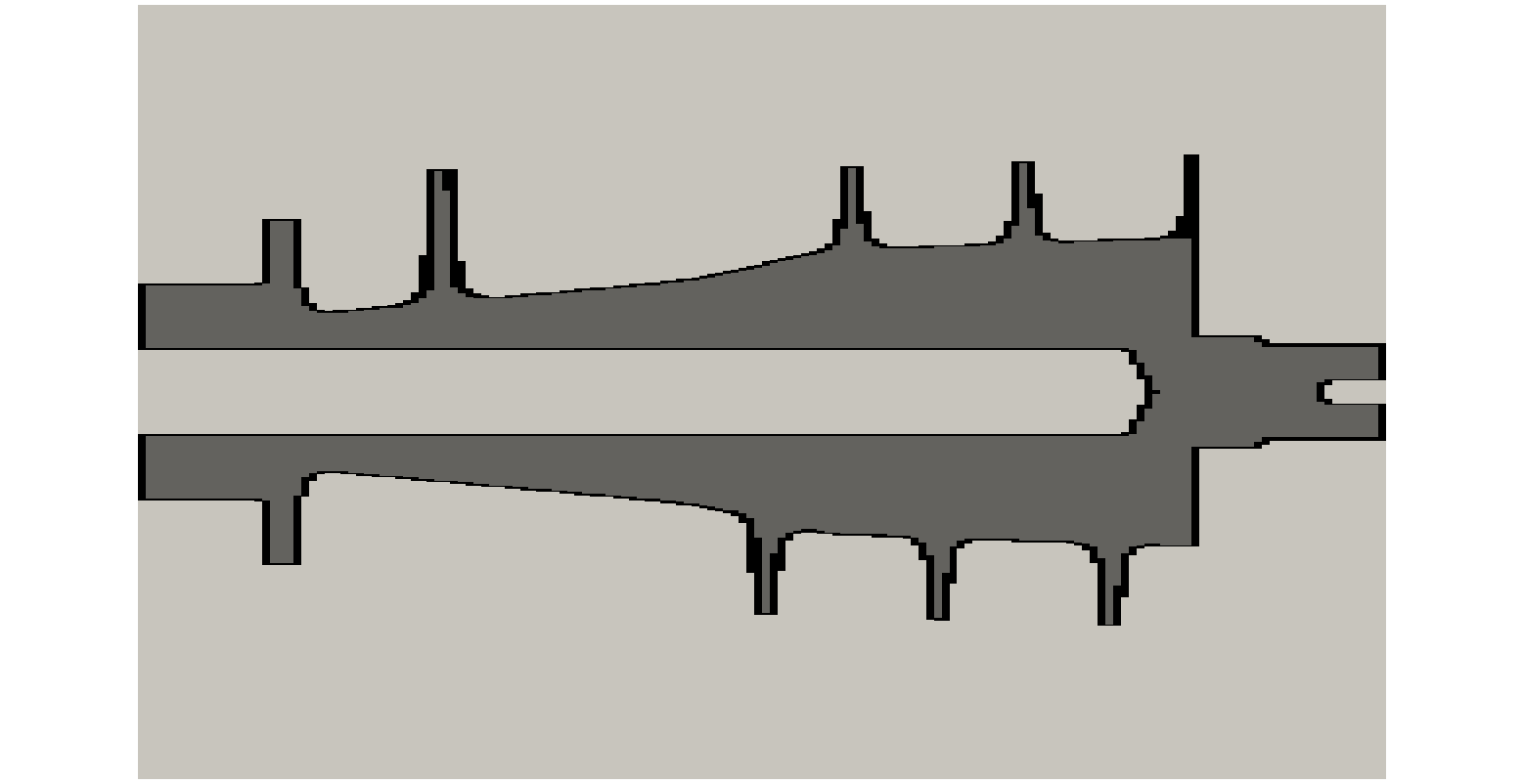}}
	\caption{Cut through an octree approximation $\widehat{T\!R}_{geo}$, with a) too small cut leaves $c_{geo}$ (black), so that all (non cut) leaves are marked as outside (light gray). b) with one subdivision level less leaves inside (dark gray) can be detected. }
	\label{fig:failedFloodfill}
\end{figure}

Fig.~\ref{fig:failedFloodfill} shows the effect of a too fine resolution of $\widehat{T\!R}_{geo}$. For a subdivision depth $n_{max} = 5$, the flood fill algorithm marks the entire domain as outside. 

\begin{figure}[H]
	\centering
	\subfloat[FCM mesh]{
		\includegraphics[width=0.31\textwidth]{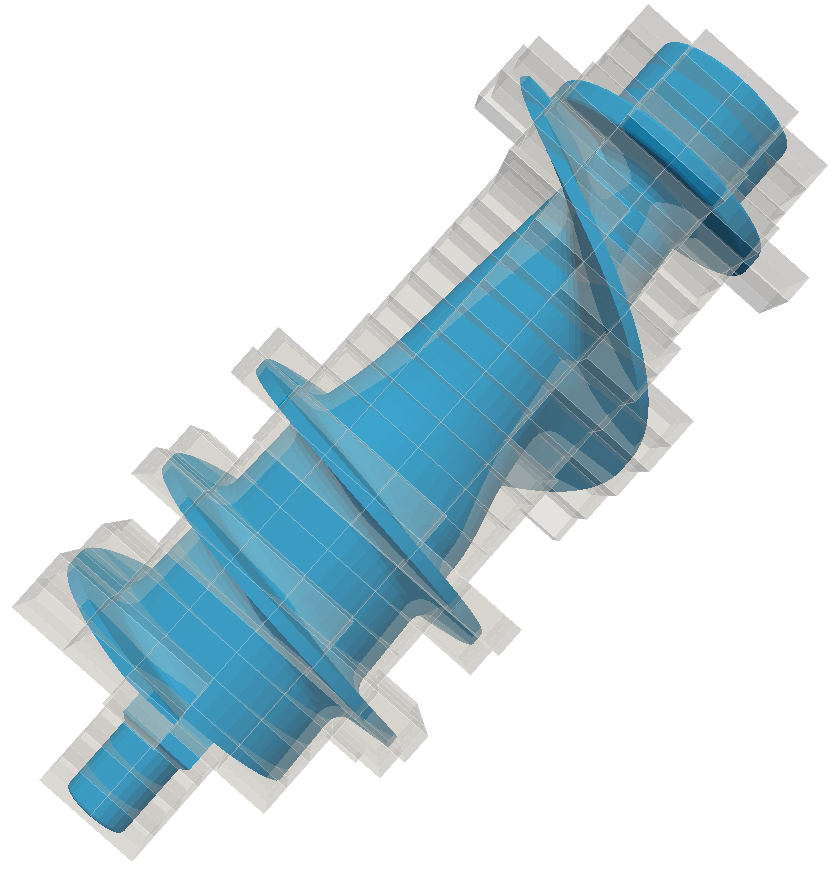}}
	\subfloat[Flawed]{
		\includegraphics[width=0.31\textwidth]{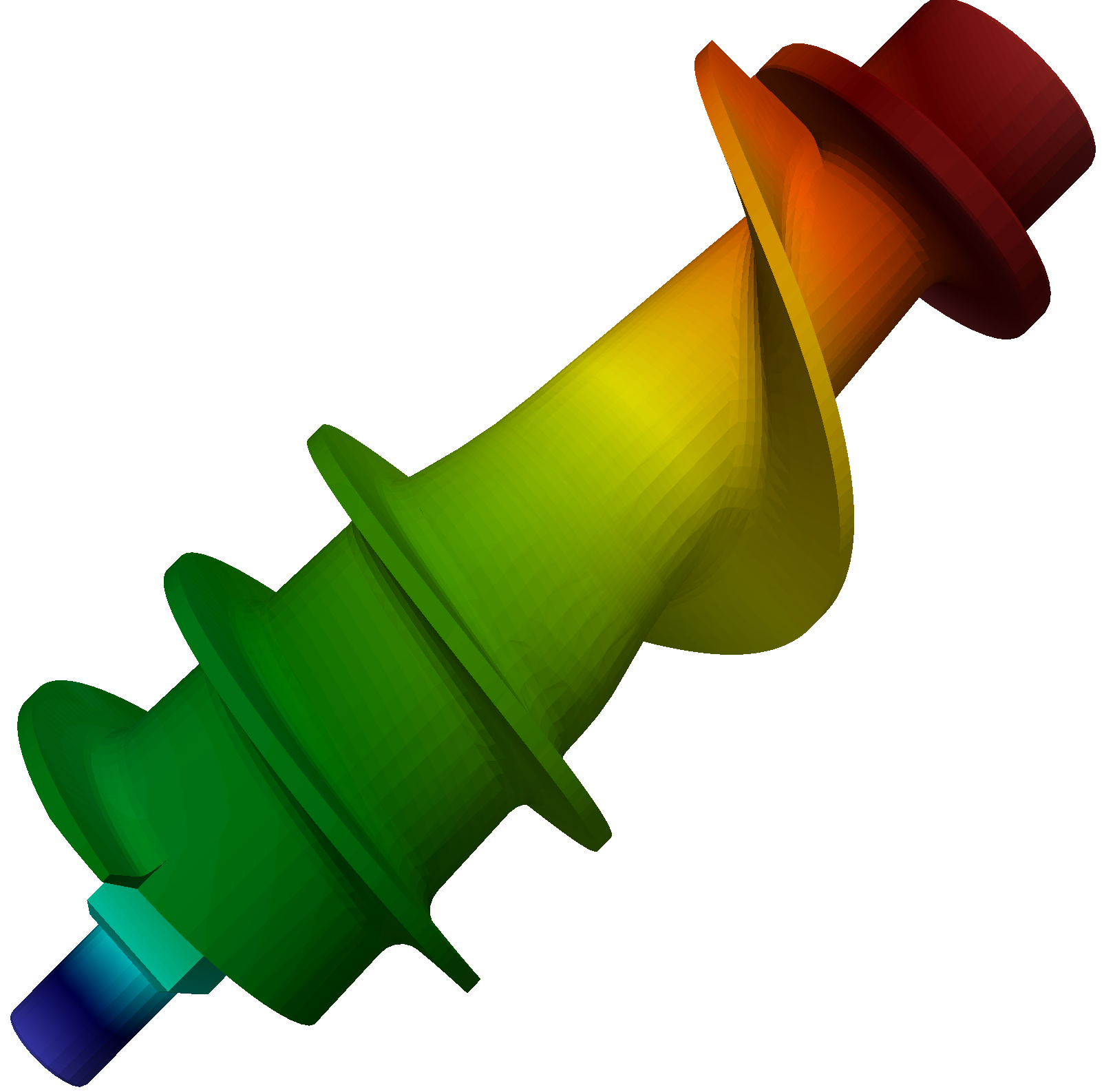}}
	\subfloat[Valid, including FCM mesh]{
		\includegraphics[width=0.31\textwidth]{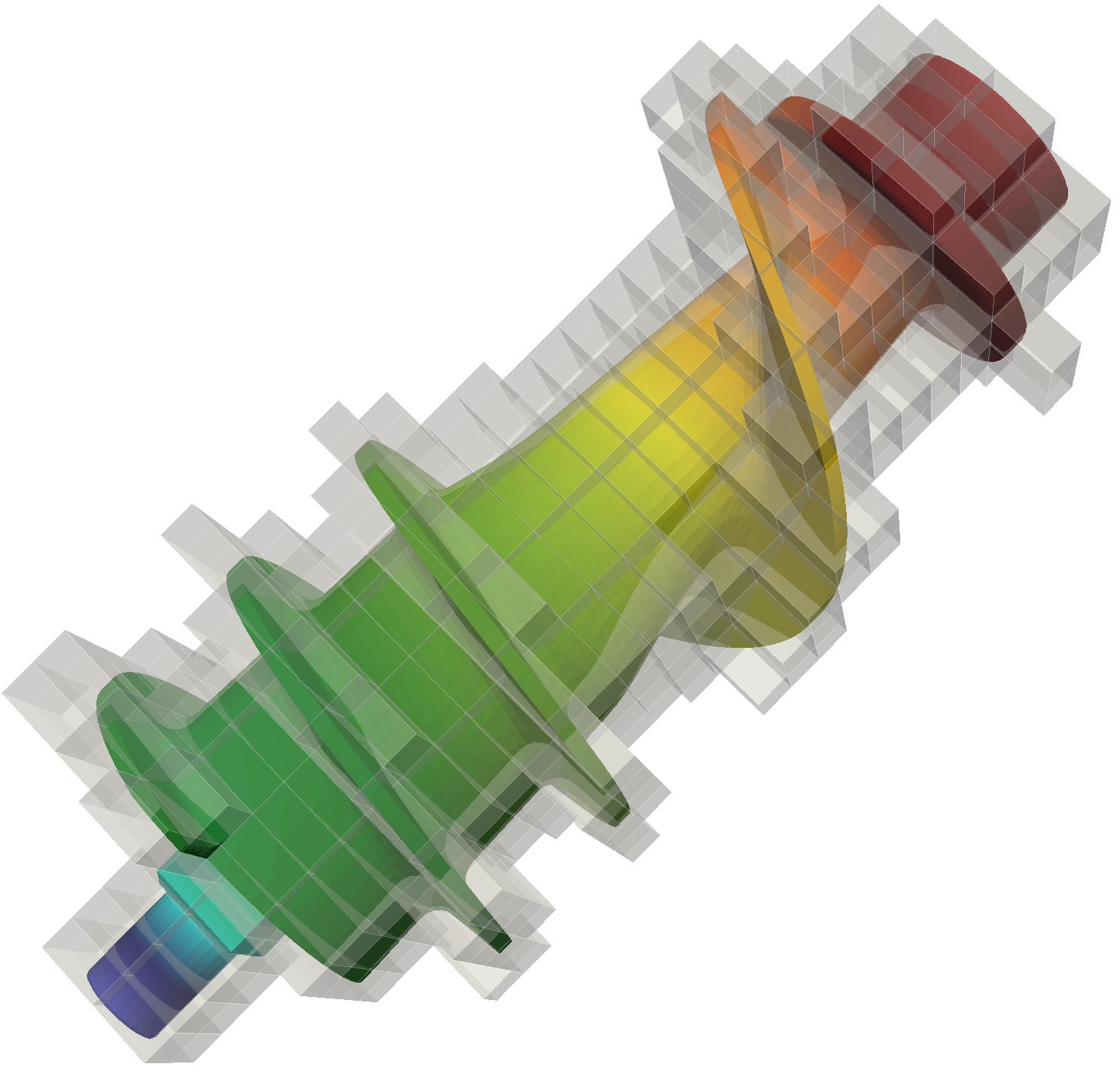}}
	\caption{ Finite Cell mesh and displacement of the flawed and the valid model}
	\label{fig:diplacementScrew}
\end{figure}

\begin{figure}[H]
	\centering
	\subfloat[Flawed]{
		\includegraphics[width=0.46\textwidth]{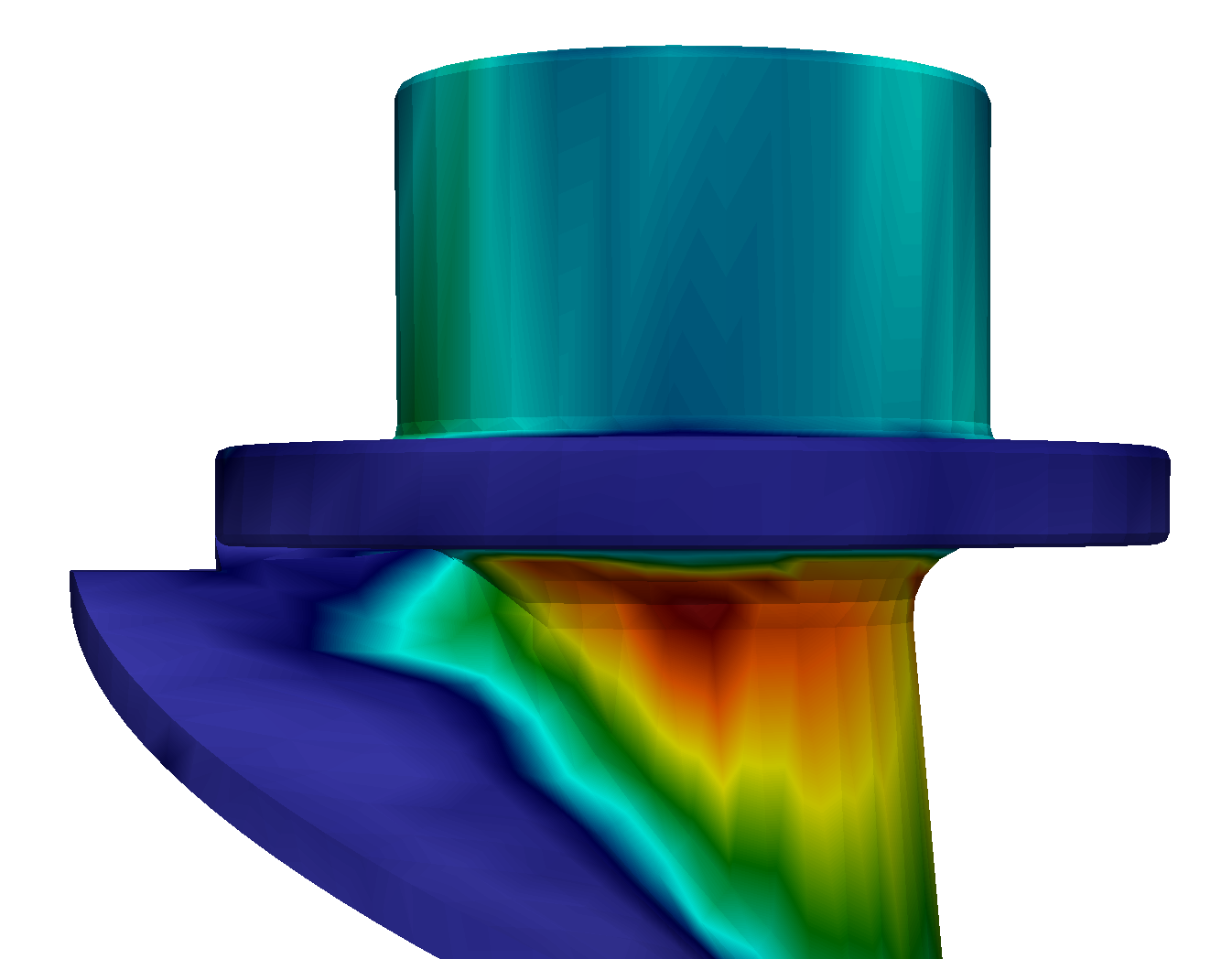}}
		\quad
	\subfloat[Valid]{
		\includegraphics[width=0.46\textwidth]{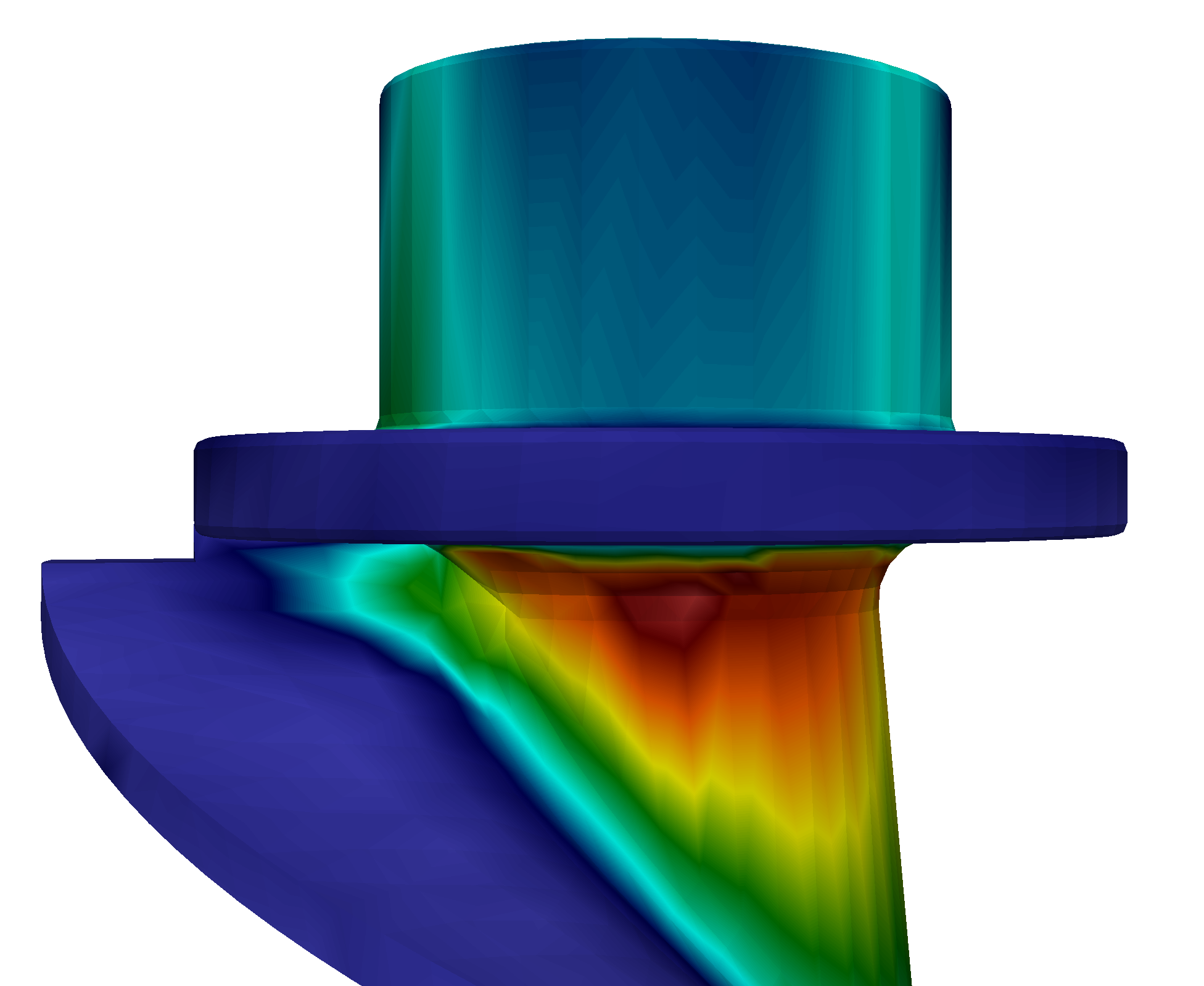}}
	\caption{Von Mises stresses around the flawed region}
	\label{fig:stressScrew}
\end{figure}

A visual inspection of the displacements of the flawed and the valid model shows no difference (see Fig.~\ref{fig:diplacementScrew}), whereas differences around the flaws can be detected for the von Mises stresses (see Fig.~\ref{fig:stressScrew}). The stresses at the flawed model are more noisy compared to the valid model.

\subsection{Example 3: Engine Brake}
\begin{figure}[H]
	\centering
	\subfloat[Initial design]{
		\includegraphics[width=0.3\textwidth]{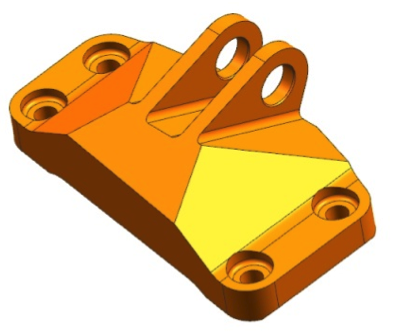}}
		\quad
	\subfloat[Optimized design]{
        \includegraphics[width=0.65\textwidth]{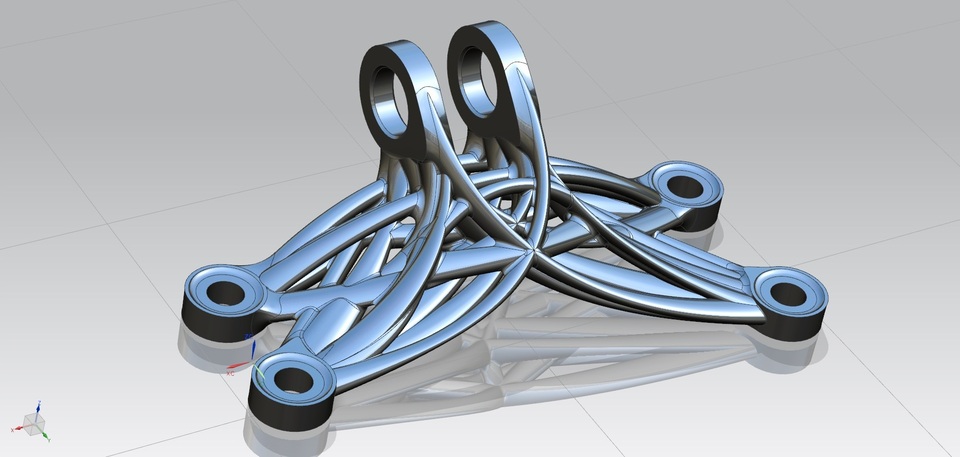}}
    \caption{General Electric design challenge for the optimal shape of a jet engine brake }
	\label{fig:DesignChallenge}
\end{figure}

In 2013, a collaboration of Grab Cad and General Electric arranged a competition to find the optimal design of an engine bracket for a General Electric turbofan \cite{GrabCad2013}. The submitted designs were then evaluated, and the top ten were produced using additive manufacturing. The model depicted in Fig.~\ref{fig:DesignChallenge} was designed by Sean Morrissey \footnote{https://grabcad.com/sean.morrissey-1}. 

In an attempt to perform a heat diffusion simulation motivated by a local heat source induced by a laser beam during additive manufacturing it turns out that 337.544 triangles have a free edge, indicating a gap/opening between the patches. 2324 triangles were oriented in the wrong direction and innumerable intersections occurred. Due to the immense amount  of flaws, geometry healing – and, thus, also the meshing -- is not applicable on this raw model. However, using the approach presented in this paper, we were able to immediately run a simulation without any further treatment of flaws. 

\begin{figure}[H]
	\centering
	\includegraphics[width=0.8\textwidth]{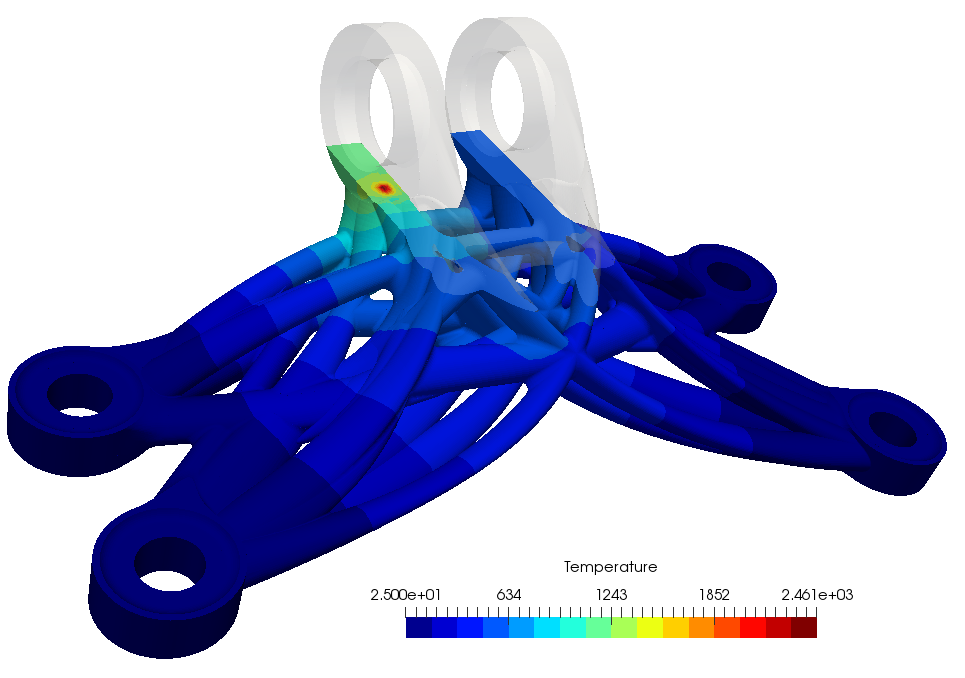}
	\caption{Temperature distribution in the bracket due to a point heat source.}
	\label{fig:brakeTemperature}
\end{figure}

We choose $18 \times 11 \times 6$ elements for the simulation. A partitioning depth for the geometry approximation tree $\widehat{T\!R}_{geo}$ of $n_{max} = 3$ on each finite cell was applicable. A laser beam is modeled by a small heat source where a local refinement of the finite cell grid was applied. Figure~\ref{fig:brakeTemperature} shows the resulting temperatures in the specimen.

\section{Conclusions} \label{sec:Conclusion}
This work presents a methodology to address challenges flawed CAD models pose to computational mechanics. Unlike other methods that rely on model reconstruction or geometry healing, the proposed approach herein allows for a numerical analysis directly on the corrupted 'dirty' geometry. Certainly, a simulation on broken geometries will inevitably lead to 
errors, which are yet of a similar nature to modeling errors due to a representation of a NURBS-based geometry by faceted surfaces. The size of this modeling error depends on the geometric size of the flaws, e.g., the width of a gap between patches. The influence of these errors remains local to the flaw itself. Moreover, the error can be bounded by performing bracketing simulations. Therein, the upper bound is delivered by a computation considering all ambiguous integration points to lie inside the physical domain and the lower bound is generated by considering the inverse situation. Several examples demonstrate the capability of the proposed method, showing that results of high accuracy can be obtained.

\newpage
\section*{Acknowledgements}
We gratefully acknowledge the support of the German Research Foundation under the Grant No. Ra 624/22-2.

Shuohui Yin from Hohai University, China as a visiting student at Technische Universit\"at M\"unchen, Germany, thanks Prof. Ernst Rank and Dr. Stefan Kollmannsberger at Technische Universit\"at M\"unchen for one-year academic guidance and discussion. He also thankfully acknowledges the support of the scholarship from China Scholarship Council (CSC).

\bibliographystyle{ieeetrDoi}
\bibliography{library}

\begin{thebibliography}{10}
\providecommand{\url}[1]{#1}
\csname url@samestyle\endcsname
\providecommand{\newblock}{\relax}
\providecommand{\bibinfo}[2]{#2}
\providecommand{\BIBentrySTDinterwordspacing}{\spaceskip=0pt\relax}
\providecommand{\BIBentryALTinterwordstretchfactor}{4}
\providecommand{\BIBentryALTinterwordspacing}{\spaceskip=\fontdimen2\font plus
\BIBentryALTinterwordstretchfactor\fontdimen3\font minus
  \fontdimen4\font\relax}
\providecommand{\BIBforeignlanguage}[2]{{%
\expandafter\ifx\csname l@#1\endcsname\relax
\typeout{** WARNING: IEEEtran.bst: No hyphenation pattern has been}%
\typeout{** loaded for the language `#1'. Using the pattern for}%
\typeout{** the default language instead.}%
\else
\language=\csname l@#1\endcsname
\fi
#2}}
\providecommand{\BIBdecl}{\relax}
\BIBdecl

\bibitem{Cottrell2009}
J.~A. Cottrell, T.~J.~R. Hughes, and Y.~Bazilevs,
  \emph{\BIBforeignlanguage{en}{Isogeometric {{Analysis}}: {{Toward
  Integration}} of {{CAD}} and {{FEA}}}}.\hskip 1em plus 0.5em minus
  0.4em\relax {John Wiley \& Sons}, Aug. 2009. ISBN 978-0-470-74909-8

\bibitem{Hughes2005}
T.~J.~R. Hughes, J.~A. Cottrell, and Y.~Bazilevs, ``Isogeometric analysis:
  {{CAD}}, finite elements, {{NURBS}}, exact geometry and mesh refinement,''
  \emph{Computer Methods in Applied Mechanics and Engineering}, vol. 194, no.
  39\textendash{}41, pp. 4135--4195, Oct. 2005. doi: 10.1016/j.cma.2004.10.008

\bibitem{Cirak2002}
F.~Cirak, M.~J. Scott, E.~K. Antonsson, M.~Ortiz, and P.~Schr\"oder,
  ``Integrated modeling, finite-element analysis, and engineering design for
  thin-shell structures using subdivision,'' \emph{Computer-Aided Design},
  vol.~34, no.~2, pp. 137--148, Feb. 2002. doi: 10.1016/S0010-4485(01)00061-6

\bibitem{Kagan2000}
P.~Kagan and A.~Fischer, ``Integrated mechanically based {{CAE}} system using
  {{B}}-{{Spline}} finite elements,'' \emph{Computer-Aided Design}, vol.~32,
  no.~8, pp. 539--552, Aug. 2000. doi: 10.1016/S0010-4485(00)00041-5

\bibitem{Yang2006}
J.~Yang and S.~Han, ``Repairing {{CAD}} model errors based on the design
  history,'' \emph{Computer-Aided Design}, vol.~38, no.~6, pp. 627--640, Jun.
  2006. doi: 10.1016/j.cad.2006.02.007

\bibitem{Mantyla1988}
M.~M\"antyl\"a, \emph{An Introduction to Solid Modeling}, ser. Principles of
  computer science series.\hskip 1em plus 0.5em minus 0.4em\relax Rockville:
  {Computer Science Press}, 1988, no.~13. ISBN 978-0-88175-108-6

\bibitem{Massarwi2016}
F.~Massarwi and G.~Elber, ``A {{B}}-spline based framework for volumetric
  object modeling,'' \emph{Computer-Aided Design}, vol.~78, pp. 36--47, Sep.
  2016. doi: 10.1016/j.cad.2016.05.003

\bibitem{Butlin1996}
G.~Butlin and C.~Stops, ``{{CAD Data Repair}},'' in \emph{Proceedings of the
  5th {{International Meshing Roundtable}}}, 1996, pp. 7--12.

\bibitem{Gu2001}
H.~Gu, T.~R. Chase, D.~C. Cheney, T.~T. Bailey, and D.~Johnson, ``Identifying,
  {{Correcting}}, and {{Avoiding Errors}} in {{Computer}}-{{Aided Design Models
  Which Affect Interoperability}},'' \emph{Journal of Computing and Information
  Science in Engineering}, vol.~1, no.~2, pp. 156--166, May 2001. doi:
  10.1115/1.1384887

\bibitem{Petersson2001}
N.~A. Petersson and K.~K. Chand, ``\BIBforeignlanguage{English}{Detecting
  {{Translation Errors}} in {{CAD Surfaces}} and {{Preparing Geometries}} for
  {{Mesh Generation}}},'' {Lawrence Livermore National Lab., CA (US)}, Newport
  Beach, CA, Tech. Rep. UCRL-JC-144019, Aug. 2001.

\bibitem{Yang2005a}
J.~Yang, S.~Han, and S.~Park, ``\BIBforeignlanguage{en}{A method for
  verification of computer-aided design model errors},''
  \emph{\BIBforeignlanguage{en}{Journal of Engineering Design}}, vol.~16,
  no.~3, pp. 337--352, Jun. 2005. doi: 10.1080/09544820500126565

\bibitem{Chong2007}
C.~S. Chong, A.~Senthil~Kumar, and H.~P. Lee, ``Automatic {{Mesh}}-healing
  {{Technique}} for {{Model Repair}} and {{Finite Element Model Generation}},''
  \emph{Finite Elements in Analysis and Design}, vol.~43, no.~15, pp.
  1109--1119, Nov. 2007. doi: 10.1016/j.finel.2007.06.009

\bibitem{Nooruddin2003}
F.~S. Nooruddin and G.~Turk, ``Simplification and repair of polygonal models
  using volumetric techniques,'' \emph{IEEE Transactions on Visualization and
  Computer Graphics}, vol.~9, no.~2, pp. 191--205, Apr. 2003. doi:
  10.1109/TVCG.2003.1196006

\bibitem{Bischoff2005a}
S.~Bischoff and L.~Kobbelt, ``\BIBforeignlanguage{en}{Structure {{Preserving
  CAD Model Repair}}},'' \emph{\BIBforeignlanguage{en}{Computer Graphics
  Forum}}, vol.~24, no.~3, pp. 527--536, Sep. 2005. doi:
  10.1111/j.1467-8659.2005.00878.x

\bibitem{Busaryev2009}
O.~Busaryev, T.~K. Dey, and J.~A. Levine, ``Repairing and {{Meshing Imperfect
  Shapes}} with {{Delaunay Refinement}},'' in \emph{2009 {{SIAM}}/{{ACM Joint
  Conference}} on {{Geometric}} and {{Physical Modeling}}}.\hskip 1em plus
  0.5em minus 0.4em\relax New York, NY, USA: {ACM}, 2009. doi:
  10.1145/1629255.1629259

\bibitem{Wang2002}
Z.~J. Wang and K.~Srinivasan, ``\BIBforeignlanguage{en}{An adaptive
  {{Cartesian}} grid generation method for `{{Dirty}}' geometry},''
  \emph{\BIBforeignlanguage{en}{International Journal for Numerical Methods in
  Fluids}}, vol.~39, no.~8, pp. 703--717, Jul. 2002. doi: 10.1002/fld.344

\bibitem{Lee2010}
Y.~K. Lee, C.~K. Lim, H.~Ghazialam, H.~Vardhan, and E.~Eklund, ``Surface {{Mesh
  Generation}} for {{Dirty Geometries}} by the {{Cartesian Shrink}}-wrapping
  {{Technique}},'' \emph{Engineering with Computers}, vol.~26, no.~4, pp.
  377--390, Aug. 2010. doi: 10.1007/s00366-009-0171-0

\bibitem{Gasparini2013}
R.~Gasparini, T.~Kosta, and I.~Tsukanov, ``Engineering analysis in imprecise
  geometric models,'' \emph{Finite Elements in Analysis and Design}, vol.~66,
  pp. 96--109, Apr. 2013. doi: 10.1016/j.finel.2012.10.011

\bibitem{Kantorovich1958}
L.~V. Kantorovich and V.~I. Krylov, \emph{\BIBforeignlanguage{en}{Approximate
  {{Methods}} of {{Higher Analysis}}}}.\hskip 1em plus 0.5em minus 0.4em\relax
  {Interscience Publishers}, 1958. ISBN 978-0-486-82160-3

\bibitem{IntactSolutions2013}
IntactSolutions, ``{{ScanAndSolve}},'' http://www.scan-and-solve.com/, 2013.

\bibitem{Parvizian2007}
J.~Parvizian, A.~D\"uster, and E.~Rank, ``Finite cell method,''
  \emph{Computational Mechanics}, vol.~41, no.~1, pp. 121--133, Apr. 2007. doi:
  10.1007/s00466-007-0173-y

\bibitem{Duster2008}
A.~D\"uster, J.~Parvizian, Z.~Yang, and E.~Rank, ``The finite cell method for
  three-dimensional problems of solid mechanics,'' \emph{Computer Methods in
  Applied Mechanics and Engineering}, vol. 197, no. 45\textendash{}48, pp.
  3768--3782, Aug. 2008. doi: 10.1016/j.cma.2008.02.036

\bibitem{Abedian2014}
A.~Abedian, J.~Parvizian, A.~D\"uster, and E.~Rank,
  ``\BIBforeignlanguage{en}{Finite cell method compared to h-version finite
  element method for elasto-plastic problems},''
  \emph{\BIBforeignlanguage{en}{Applied Mathematics and Mechanics}}, vol.~35,
  no.~10, pp. 1239--1248, Oct. 2014. doi: 10.1007/s10483-014-1861-9

\bibitem{Wassermann2017}
B.~Wassermann, S.~Kollmannsberger, T.~Bog, and E.~Rank, ``From geometric design
  to numerical analysis: {{A}} direct approach using the {{Finite Cell Method}}
  on {{Constructive Solid Geometry}},'' \emph{Computers \& Mathematics with
  Applications}, Mar. 2017. doi: 10.1016/j.camwa.2017.01.027

\bibitem{Cai2014}
S.~Cai, W.~Zhang, J.~Zhu, and T.~Gao, ``Stress constrained shape and topology
  optimization with fixed mesh: {{A B}}-spline finite cell method combined with
  level set function,'' \emph{Computer Methods in Applied Mechanics and
  Engineering}, vol. 278, pp. 361--387, Aug. 2014. doi:
  10.1016/j.cma.2014.06.007

\bibitem{Groen2016}
J.~P. Groen, M.~Langelaar, O.~Sigmund, and M.~Ruess,
  ``\BIBforeignlanguage{en}{Higher-order multi-resolution topology optimization
  using the finite cell method},'' \emph{\BIBforeignlanguage{en}{International
  Journal for Numerical Methods in Engineering}}, Jan. 2016. doi:
  10.1002/nme.5432

\bibitem{Joulaian2013}
M.~Joulaian and A.~D\"uster, ``\BIBforeignlanguage{en}{Local enrichment of the
  finite cell method for problems with material interfaces},''
  \emph{\BIBforeignlanguage{en}{Computational Mechanics}}, vol.~52, no.~4, pp.
  741--762, Oct. 2013. doi: 10.1007/s00466-013-0853-8

\bibitem{Joulaian2014}
M.~Joulaian, S.~Duczek, U.~Gabbert, and A.~D\"uster,
  ``\BIBforeignlanguage{en}{Finite and spectral cell method for wave
  propagation in heterogeneous materials},''
  \emph{\BIBforeignlanguage{en}{Computational Mechanics}}, vol.~54, no.~3, pp.
  661--675, Apr. 2014. doi: 10.1007/s00466-014-1019-z

\bibitem{Duczek2014}
S.~Duczek, M.~Joulaian, A.~D\"uster, and U.~Gabbert,
  ``\BIBforeignlanguage{en}{Numerical analysis of {{Lamb}} waves using the
  finite and spectral cell methods},''
  \emph{\BIBforeignlanguage{en}{International Journal for Numerical Methods in
  Engineering}}, vol.~99, no.~1, pp. 26--53, Jul. 2014. doi: 10.1002/nme.4663

\bibitem{Elhaddad2015}
M.~Elhaddad, N.~Zander, S.~Kollmannsberger, A.~Shadavakhsh, V.~N\"ubel, and
  E.~Rank, ``\BIBforeignlanguage{en}{Finite {{Cell Method}}: {{High}}-{{Order
  Structural Dynamics}} for {{Complex Geometries}}},''
  \emph{\BIBforeignlanguage{en}{International Journal of Structural Stability
  and Dynamics}}, vol.~15, no.~7, p. 1540018, Apr. 2015. doi:
  10.1142/S0219455415400180

\bibitem{Bog2017}
T.~Bog, N.~Zander, S.~Kollmannsberger, and E.~Rank,
  ``\BIBforeignlanguage{en}{Weak imposition of frictionless contact constraints
  on automatically recovered high-order, embedded interfaces using the finite
  cell method},'' \emph{\BIBforeignlanguage{en}{Computational Mechanics}}, Aug.
  2017. doi: 10.1007/s00466-017-1464-6

\bibitem{Mongeau2015}
A.~Mongeau, ``Large deformation two- and three- dimensional contact on embedded
  interfaces using the {{Finite Cell Method}},'' Master's {{Thesis}},
  Technische Universit\"at M\"unchen, Dec. 2015.

\bibitem{Kollmannsberger2015}
S.~Kollmannsberger, A.~\"Ozcan, J.~Baiges, M.~Ruess, E.~Rank, and A.~Reali,
  ``\BIBforeignlanguage{en}{Parameter-free, weak imposition of {{Dirichlet}}
  boundary conditions and coupling of trimmed and non-conforming patches},''
  \emph{\BIBforeignlanguage{en}{International Journal for Numerical Methods in
  Engineering}}, vol. 101, no.~9, pp. 670--699, Mar. 2015. doi:
  10.1002/nme.4817

\bibitem{Zander2015}
N.~Zander, T.~Bog, S.~Kollmannsberger, D.~Schillinger, and E.~Rank,
  ``\BIBforeignlanguage{en}{Multi-level hp-adaptivity: High-order mesh
  adaptivity without the difficulties of constraining hanging nodes},''
  \emph{\BIBforeignlanguage{en}{Computational Mechanics}}, vol.~55, no.~3, pp.
  499--517, Feb. 2015. doi: 10.1007/s00466-014-1118-x

\bibitem{Kudela2016}
L.~Kudela, N.~Zander, S.~Kollmannsberger, and E.~Rank, ``Smart octrees:
  {{Accurately}} integrating discontinuous functions in {{3D}},''
  \emph{Computer Methods in Applied Mechanics and Engineering}, vol. 306, pp.
  406--426, Jul. 2016. doi: 10.1016/j.cma.2016.04.006

\bibitem{Fries2015}
T.-P. Fries and S.~Omerovi\'c, ``\BIBforeignlanguage{en}{Higher-order accurate
  integration of implicit geometries},''
  \emph{\BIBforeignlanguage{en}{International Journal for Numerical Methods in
  Engineering}}, vol. 106, no.~5, pp. 323--371, Jan. 2015. doi:
  10.1002/nme.5121

\bibitem{Joulaian2016}
M.~Joulaian, S.~Hubrich, and A.~D\"uster, ``\BIBforeignlanguage{en}{Numerical
  integration of discontinuities on arbitrary domains based on moment
  fitting},'' \emph{\BIBforeignlanguage{en}{Computational Mechanics}}, vol.~57,
  no.~6, pp. 979--999, Jun. 2016. doi: 10.1007/s00466-016-1273-3

\bibitem{Hubrich2017}
S.~Hubrich, P.~D. Stolfo, L.~Kudela, S.~Kollmannsberger, E.~Rank,
  A.~Schr\"oder, and A.~D\"uster, ``\BIBforeignlanguage{en}{Numerical
  integration of discontinuous functions: Moment fitting and smart octree},''
  \emph{\BIBforeignlanguage{en}{Computational Mechanics}}, pp. 1--19, Jul.
  2017. doi: 10.1007/s00466-017-1441-0

\bibitem{Giraldo2017}
D.~Giraldo and D.~Restrepo, ``\BIBforeignlanguage{en}{The spectral cell method
  in nonlinear earthquake modeling},''
  \emph{\BIBforeignlanguage{en}{Computational Mechanics}}, pp. 1--21, Aug.
  2017. doi: 10.1007/s00466-017-1454-8

\bibitem{Schillinger2012b}
D.~Schillinger, M.~Ruess, N.~Zander, Y.~Bazilevs, A.~D\"uster, and E.~Rank,
  ``Small and large deformation analysis with the p- and {{B}}-spline versions
  of the {{Finite Cell Method}},'' \emph{Computational Mechanics}, vol.~50,
  no.~4, pp. 445--478, Feb. 2012. doi: 10.1007/s00466-012-0684-z

\bibitem{Rank2012}
E.~Rank, M.~Ruess, S.~Kollmannsberger, D.~Schillinger, and A.~D\"uster,
  ``Geometric modeling, isogeometric analysis and the finite cell method,''
  \emph{Computer Methods in Applied Mechanics and Engineering}, vol. 249-252,
  pp. 104--115, Dec. 2012. doi: 10.1016/j.cma.2012.05.022

\bibitem{Ruess2014}
M.~Ruess, D.~Schillinger, A.~I. \"Ozcan, and E.~Rank, ``Weak coupling for
  isogeometric analysis of non-matching and trimmed multi-patch geometries,''
  \emph{Computer Methods in Applied Mechanics and Engineering}, vol. 269, pp.
  46--71, Feb. 2014. doi: 10.1016/j.cma.2013.10.009

\bibitem{dePrenter2017a}
F.~{de Prenter}, C.~V. Verhoosel, G.~J. {van Zwieten}, and E.~H. {van
  Brummelen}, ``Condition number analysis and preconditioning of the finite
  cell method,'' \emph{Computer Methods in Applied Mechanics and Engineering},
  vol. 316, no. Supplement C, pp. 297--327, Apr. 2017. doi:
  10.1016/j.cma.2016.07.006

\bibitem{Burman2015a}
E.~Burman, P.~Hansbo, and M.~G. Larson, ``A stabilized cut finite element
  method for partial differential equations on surfaces: {{The
  Laplace}}\textendash{{Beltrami}} operator,'' \emph{Computer Methods in
  Applied Mechanics and Engineering}, vol. 285, pp. 188--207, Mar. 2015. doi:
  10.1016/j.cma.2014.10.044

\bibitem{Burman2010}
E.~Burman and P.~Hansbo, ``Fictitious domain finite element methods using cut
  elements: {{I}}. {{A}} stabilized {{Lagrange}} multiplier method,''
  \emph{Computer Methods in Applied Mechanics and Engineering}, vol. 199, no.
  41-44, pp. 2680--2686, Oct. 2010. doi: 10.1016/j.cma.2010.05.011

\bibitem{Patrikalakis2000}
N.~M. Patrikalakis, T.~Sakkalis, and G.~Shen,
  ``\BIBforeignlanguage{en}{Boundary {{Representation Models}}: {{Validity}}
  and {{Rectification}}},'' in \emph{\BIBforeignlanguage{en}{The
  {{Mathematics}} of {{Surfaces IX}}}}.\hskip 1em plus 0.5em minus 0.4em\relax
  {Springer, London}, 2000, pp. 389--409. ISBN 978-1-4471-1153-5
  978-1-4471-0495-7

\bibitem{Hoffmann1989}
C.~M. Hoffmann, \emph{Geometric and Solid Modeling: An Introduction}, ser. The
  Morgan Kaufmann series in computer graphics and geometric modeling.\hskip 1em
  plus 0.5em minus 0.4em\relax San Mateo, Calif: {Morgan Kaufmann}, 1989. ISBN
  978-1-55860-067-6

\bibitem{Bazilevs2010a}
Y.~Bazilevs, V.~M. Calo, J.~A. Cottrell, J.~A. Evans, T.~J.~R. Hughes,
  S.~Lipton, M.~A. Scott, and T.~W. Sederberg, ``Isogeometric analysis using
  {{T}}-splines,'' \emph{Computer Methods in Applied Mechanics and
  Engineering}, vol. 199, no. 5\textendash{}8, pp. 229--263, Jan. 2010. doi:
  10.1016/j.cma.2009.02.036

\bibitem{Sederberg2008}
T.~W. Sederberg, G.~T. Finnigan, X.~Li, H.~Lin, and H.~Ipson, ``Watertight
  {{Trimmed NURBS}},'' in \emph{{{ACM SIGGRAPH}} 2008 {{Papers}}}.\hskip 1em
  plus 0.5em minus 0.4em\relax New York, NY, USA: {ACM}, 2008. doi:
  10.1145/1399504.1360678

\bibitem{Rumpe2016}
B.~Rumpe, \emph{\BIBforeignlanguage{en}{Modeling with {{UML}}: {{Language}},
  {{Concepts}}, {{Methods}}}}.\hskip 1em plus 0.5em minus 0.4em\relax {Springer
  International Publishing}, 2016. ISBN 978-3-319-33932-0

\bibitem{Goldberg1983}
A.~Goldberg and D.~Robson, \emph{Smalltalk-80: {{The Language}} and {{Its
  Implementation}}}.\hskip 1em plus 0.5em minus 0.4em\relax Boston, MA, USA:
  {Addison-Wesley Longman Publishing Co., Inc.}, 1983. ISBN 978-0-201-11371-6

\bibitem{Schillinger2012}
D.~Schillinger, A.~D\"uster, and E.~Rank, ``\BIBforeignlanguage{en}{The
  hp-d-adaptive finite cell method for geometrically nonlinear problems of
  solid mechanics},'' \emph{\BIBforeignlanguage{en}{International Journal for
  Numerical Methods in Engineering}}, vol.~89, no.~9, pp. 1171--1202, 2012.
  doi: 10.1002/nme.3289

\bibitem{Zander2012}
N.~Zander, S.~Kollmannsberger, M.~Ruess, Z.~Yosibash, and E.~Rank, ``The
  {{Finite Cell Method}} for linear thermoelasticity,'' \emph{Computers \&
  Mathematics with Applications}, vol.~64, no.~11, pp. 3527--3541, Dec. 2012.
  doi: 10.1016/j.camwa.2012.09.002

\bibitem{Elhaddad2018}
M.~Elhaddad, N.~Zander, T.~Bog, L.~Kudela, S.~Kollmannsberger, J.~S. Kirschke,
  T.~Baum, M.~Ruess, and E.~Rank, ``\BIBforeignlanguage{en}{Multi-level
  hp-finite cell method for embedded interface problems with application in
  biomechanics},'' \emph{\BIBforeignlanguage{en}{International Journal for
  Numerical Methods in Biomedical Engineering}}, vol.~34, no.~4, p. e2951,
  2018. doi: 10.1002/cnm.2951

\bibitem{Kollmannsberger2018}
S.~Kollmannsberger, A.~\"Ozcan, M.~Carraturo, N.~Zander, and E.~Rank,
  ``\BIBforeignlanguage{en}{A hierarchical computational model for moving
  thermal loads and phase changes with applications to selective laser
  melting},'' \emph{\BIBforeignlanguage{en}{Computers \& Mathematics with
  Applications}}, vol.~75, no.~5, pp. 1483--1497, Mar. 2018. doi:
  10.1016/j.camwa.2017.11.014

\bibitem{Duster2017}
A.~D\"uster, E.~Rank, and B.~A. Szab\'o, ``\BIBforeignlanguage{English}{The
  p-version of the finite element method and finite cell methods},'' in
  \emph{\BIBforeignlanguage{English}{Encyclopedia of {{Computational}}
  Mechanics}}, E.~Stein, R.~Borst, and T.~J.~R. Hughes, Eds.\hskip 1em plus
  0.5em minus 0.4em\relax Chichester, West Sussex: {John Wiley \& Sons}, 2017,
  vol.~2, pp. 1--35. ISBN 978-1-119-00379-3

\bibitem{Dauge2015}
M.~Dauge, A.~D\"uster, and E.~Rank, ``\BIBforeignlanguage{en}{Theoretical and
  {{Numerical Investigation}} of the {{Finite Cell Method}}},''
  \emph{\BIBforeignlanguage{en}{Journal of Scientific Computing}}, vol.~65,
  no.~3, pp. 1039--1064, Mar. 2015. doi: 10.1007/s10915-015-9997-3

\bibitem{Schillinger2011}
D.~Schillinger and E.~Rank, ``An unfitted hp-adaptive finite element method
  based on hierarchical {{B}}-splines for interface problems of complex
  geometry,'' \emph{Computer Methods in Applied Mechanics and Engineering},
  vol. 200, no. 47-48, pp. 3358--3380, Nov. 2011. doi:
  10.1016/j.cma.2011.08.002

\bibitem{Zander2016}
N.~Zander, T.~Bog, M.~Elhaddad, F.~Frischmann, S.~Kollmannsberger, and E.~Rank,
  ``The multi-level hp-method for three-dimensional problems: {{Dynamically}}
  changing high-order mesh refinement with arbitrary hanging nodes,''
  \emph{Computer Methods in Applied Mechanics and Engineering}, vol. 310, pp.
  252--277, Oct. 2016. doi: 10.1016/j.cma.2016.07.007

\bibitem{Varduhn2016}
V.~Varduhn, M.-C. Hsu, M.~Ruess, and D.~Schillinger,
  ``\BIBforeignlanguage{en}{The tetrahedral finite cell method:
  {{Higher}}-order immersogeometric analysis on adaptive non-boundary-fitted
  meshes},'' \emph{\BIBforeignlanguage{en}{International Journal for Numerical
  Methods in Engineering}}, vol. 107, no.~12, pp. 1054--1079, Jan. 2016. doi:
  10.1002/nme.5207

\bibitem{Kamensky2015}
D.~Kamensky, M.-C. Hsu, D.~Schillinger, J.~A. Evans, A.~Aggarwal, Y.~Bazilevs,
  M.~S. Sacks, and T.~J.~R. Hughes, ``An immersogeometric variational framework
  for fluid\textendash{}structure interaction: {{Application}} to bioprosthetic
  heart valves,'' \emph{Computer Methods in Applied Mechanics and Engineering},
  vol. 284, pp. 1005--1053, Feb. 2015. doi: 10.1016/j.cma.2014.10.040 00015.

\bibitem{Duczek2016}
S.~Duczek and U.~Gabbert, ``\BIBforeignlanguage{en}{The finite cell method for
  polygonal meshes: Poly-{{FCM}}},''
  \emph{\BIBforeignlanguage{en}{Computational Mechanics}}, pp. 1--32, Jun.
  2016. doi: 10.1007/s00466-016-1307-x

\bibitem{Abedian2013}
A.~Abedian, J.~Parvizian, A.~D\"uster, H.~Khademyzadeh, and E.~Rank,
  ``Performance of {{Different Integration Schemes}} in {{Facing
  Discontinuities}} in the {{Finite Cell Method}},'' \emph{International
  Journal of Computational Methods}, vol.~10, no.~03, p. 1350002, Jun. 2013.
  doi: 10.1142/S0219876213500023

\bibitem{Ruess2012}
M.~Ruess, Y.~Bazilevs, D.~Schillinger, N.~Zander, and E.~Rank, ``Weakly
  enforced boundary conditions for the {{NURBS}}-based {{Finite Cell
  Method}},'' in \emph{European {{Congress}} on {{Computational Methods}} in
  {{Applied Sciences}} and {{Engineering}} ({{ECCOMAS}})}, Vienna, Austria,
  2012. ISBN 978-3-9502481-9-7

\bibitem{Ruess2013}
M.~Ruess, D.~Schillinger, Y.~Bazilevs, V.~Varduhn, and E.~Rank, ``Weakly
  enforced essential boundary conditions for {{NURBS}}-embedded and trimmed
  {{NURBS}} geometries on the basis of the finite cell method,''
  \emph{International Journal for Numerical Methods in Engineering}, vol.~95,
  no.~10, pp. 811--846, Sep. 2013. doi: 10.1002/nme.4522

\bibitem{Guo2015a}
Y.~Guo and M.~Ruess, ``Nitsche's method for a coupling of isogeometric thin
  shells and blended shell structures,'' \emph{Computer Methods in Applied
  Mechanics and Engineering}, vol. 284, pp. 881--905, Feb. 2015. doi:
  10.1016/j.cma.2014.11.014

\bibitem{Preparata1985}
F.~P. Preparata and M.~I. Shamos, \emph{Computational {{Geometry}}: {{An
  Introduction}}}.\hskip 1em plus 0.5em minus 0.4em\relax New York, NY, USA:
  {Springer-Verlag New York, Inc.}, 1985. ISBN 978-0-387-96131-6

\bibitem{Zalik2001}
B.~{\v Z}alik and I.~Kolingerova, ``A cell-based point-in-polygon algorithm
  suitable for large sets of points,'' \emph{Computers \& Geosciences},
  vol.~27, no.~10, pp. 1135--1145, Dec. 2001. doi:
  10.1016/S0098-3004(01)00037-1

\bibitem{Sitek2006}
A.~Sitek, R.~H. Huesman, and G.~T. Gullberg,
  ``\BIBforeignlanguage{eng}{Tomographic reconstruction using an adaptive
  tetrahedral mesh defined by a point cloud},''
  \emph{\BIBforeignlanguage{eng}{IEEE transactions on medical imaging}},
  vol.~25, no.~9, pp. 1172--1179, Sep. 2006. doi: 10.1109/TMI.2006.879319

\bibitem{Taylor1994}
G.~Taylor, ``Point in {{Polygon Test}},'' \emph{Survey Review}, vol.~32, no.
  254, pp. 479--484, Oct. 1994. doi: 10.1179/sre.1994.32.254.479

\bibitem{Salomon1978}
K.~B. Salomon, ``An efficient point-in-polygon algorithm,'' \emph{Computers \&
  Geosciences}, vol.~4, no.~2, pp. 173--178, Jan. 1978. doi:
  10.1016/0098-3004(78)90085-7

\bibitem{Foley1997}
J.~D. Foley, A.~V. Dam, S.~K. Feiner, J.~F. Hughes, and R.~L. Phillips,
  \emph{\BIBforeignlanguage{en}{Introduction to {{Computer Graphics}}}}.\hskip
  1em plus 0.5em minus 0.4em\relax {Addison-Wesley}, 1997. ISBN
  978-0-201-60921-9

\bibitem{Strang1973}
G.~Strang, \emph{An Analysis of the Finite Element Method}.\hskip 1em plus
  0.5em minus 0.4em\relax Englewood Cliffs, N.J: {Prentice-Hall}, 1973. ISBN
  0-13-032946-0

\bibitem{GrabCad2013}
GrabCad, ``General {{Electric}} jet engine bracket challenge - {{GrabCAD}},''
  \url{https://grabcad.com/challenges/ge-jet-engine-bracket-challenge}, 2013.

\end{thebibliography}

\end{document}